\documentclass[12pt]{article}
\usepackage{graphicx}
\usepackage{amsfonts}
\usepackage{amssymb,amsmath}
\usepackage{latexsym}
\usepackage{color}
\usepackage{cite}
\input{colordvi.tex}

\begin{document}

\begin{titlepage}

\begin{center}

{\Large \bf  
On Classification of Models of Large Local-Type Non-Gaussianity
}

\vskip .45in

{\large
Teruaki Suyama$^1$,
Tomo Takahashi$^2$, 
Masahide Yamaguchi$^3$,\\
and 
Shuichiro Yokoyama$^4$
}

\vskip .45in

{\em
$^1$Research Center for the Early Universe, Graduate School
  of Science, \\ The University of Tokyo, Tokyo 113-0033, Japan
 \vspace{0.2cm}\\
$^2$Department of Physics, Saga University, Saga 840-8502, Japan 
 \vspace{0.2cm} \\
$^3$Department of Physics, Tokyo Institute of Technology, Tokyo
152-8551, Japan
 \vspace{0.2cm} \\
$^4$Department of Physics and Astrophysics, Nagoya University,
Aichi 464-8602, Japan
}

\end{center}

\vskip .4in

\begin{abstract}
  We classify models generating large local-type non-Gaussianity into
  some categories by using some ``consistency relations'' among the
  non-linearity parameters $f_{\rm NL}^{\rm local}, \tau_{\rm NL}^{\rm
  local}$ and $g_{\rm NL}^{\rm local}$, which characterize the size of
  bispectrum for the former and trispectrum for the later two. Then we
  discuss how one can discriminate models of large local-type
  non-Gaussianity with such relations.  We first classify the models by
  using the ratio of $\tau_{\rm NL}^{\rm local}/(6 f_{\rm NL}^{\rm
  local}/5)^2$, which is unity for ``single-source'' models and deviates
  from unity for ``multi-source'' ones.  We can make a further
  classification of models in each category by utilizing the relation
  between $f_{\rm NL}^{\rm local}$ and $g_{\rm NL}^{\rm local}$.  Our
  classification suggests that observations of trispectrum would be very
  helpful to distinguish models of large non-Gaussianity and may reveal
  the generation mechanism of primordial fluctuations.
\end{abstract}

\end{titlepage}

\tableofcontents

\clearpage

\setcounter{page}{1}

\section{Introduction}

Non-Gaussianity of primordial fluctuations is now one of the important
observables to probe the physics of the early Universe.  Although the
inflaton, which acquires quantum fluctuations during inflation and
generates (almost) scale-invariant and adiabatic primordial
fluctuations, has been considered to be responsible for the origin of
density fluctuations in the Universe, this scenario might not be the one
realized in the nature.  Primordial fluctuations generated from the
inflaton during slow-roll inflation are expected to be almost Gaussian,
however, current observations such as WMAP indicate that the primordial
fluctuations might deviate from Gaussian ones and the deviation could be
relatively large compared to the one expected from the ``slow-rolling''
inflation models.  (For recent analyses, see
\cite{Komatsu:2010fb,Xia:2010yu,DeBernardis:2010kc,Xia:2010pe}.) The
degrees of non-Gaussianity are often represented by a so-called
non-linearity parameter $f_{\rm NL}$, which characterizes the size of
bispectrum of the curvature perturbation.  Depending on the momentum
distribution of the bispectrum or the shape of three point function,
three types of $f_{\rm NL}$ have been discussed in the
literature~\cite{Komatsu:2010fb,Senatore:2009gt}: local, equilateral
and orthogonal types.  The limits on these $f_{\rm NL}$s have been
obtained as \cite{Komatsu:2010fb}: $ -10 < f_{\rm NL}^{\rm local} < 74 $
for the local type, $ -214 < f_{\rm NL}^{\rm equil} < 266 $ for the
equilateral type and $ - 410< f_{\rm NL}^{\rm orthog} < 6 $ for the
orthogonal type (95\,\% C.L.).  Although all of these are still
consistent with Gaussian fluctuations, they may give some hint of
non-Gaussian ones since the central value of $f_{\rm NL}$ for some types
is away from the zero.  If the primordial fluctuations are confirmed to
be deviated from Gaussian ones in the future, which may well be probed
with more precise observations such as Planck \cite{:2006uk},
fluctuations from the ``slow-rolling'' inflaton would be excluded
at least as a dominant mechanism of the generation of density
fluctuations.  However, many other mechanisms have also been discussed
in the literature and quite a few of them can generate large
non-Gaussianity.

Now we have plenty of models generating large non-Gaussian primordial
fluctuations. Thus the question we should ask next is ``how can we
differentiate these models?''  In this paper, we discuss this issue by
using bispectrum and trispectrum of the curvature
perturbation.  Although, as mentioned above, we can divide models into
some categories depending on the shape of the three point functions
(local, equilateral and orthogonal types)\footnote{
  There are various inflationary models that can produce large $f_{\rm
    NL}^{\rm equil}$, where scalar fields have some non-canonical
  kinetic terms~\cite{Alishahiha:2004eh, ArkaniHamed:2003uz,
    Silverstein:2003hf, Cheung:2007st, Chen:2006nt, Seery:2005wm,
    Li:2008qc}.  The orthogonal-type $f_{\rm NL}^{\rm orthog}$
  approximately arises from a linear combination of higher-derivative
  scalar-field interaction terms~\cite{Cheung:2007st,Senatore:2009gt}.
}, the shape is not enough to differentiate models since there remain
many models for each type of the shape.  Furthermore, 
$f_{\rm NL}$ predicted in those models can fall onto almost the same
value by tuning some model parameters.  Thus, obviously, the
determination of $f_{\rm NL}$ is not enough to pin down the models of
large non-Gaussianity even if $f_{\rm NL}$ is found to be large in the
future.  The purpose of this paper is to pursue the strategy of how
one can differentiate models of large non-Gaussianity.  To this end,
we consider higher order statistics such as the trispectrum in
addition to the bispectrum\footnote{
  Another possible direction may be the one using the scale dependence
  of $f_{\rm NL}$
  \cite{LoVerde:2007ri,Sefusatti:2009xu,Byrnes:2009pe,Byrnes:2010ft,Byrnes:2010xd,Huang:2010cy}.
}. The size of the trispectrum can be parametrized by other 
non-linearity parameters $\tau_{\rm NL}$ and $g_{\rm NL}$\footnote{
Current observational limits for $\tau_{\rm NL}^{\rm local}$ and
$g_{\rm NL}^{\rm local}$ are given by $-0.6 \times 10^4 < \tau^{\rm local}_{\rm
  NL} < 3.3 \times 10^4$ ($95 \%$ C.L.) and $-7.4 \times 10^5 < g^{\rm
  local}_{\rm NL} < 8.2 \times 10^5$ ($95 \%$ C.L.) from cosmic
microwave background observations~\cite{Smidt:2010sv}. The authors of \cite{Desjacques:2009jb} also obtained $ -3.5
\times 10^5 < g^{\rm local}_{\rm NL} < 8.2 \times 10^5$ ($95 \%$ C.L.)
from large scale structure observations 
by assuming that $f_{\rm NL}^{\rm local}=0$ and $\tau_{\rm NL}^{\rm local}=0$.
}
 and 
the importance of the trispectrum has been emphasized in some literature
\cite{Enqvist:2008gk,Enqvist:2009eq,Smidt:2010sv,Smidt:2010ra,Byrnes:2006vq,Desjacques:2009jb}. However, 
here we give a systematic study of the bispectrum and the trispectrum
of models with large non-Gaussianity and make some classifications by
using the ``consistency relations'' between the non-linearity
parameters $f_{\rm NL}, \tau_{\rm NL}$ and $g_{\rm NL}$.

Among three types mentioned above (local, equilateral and orthogonal types), 
we focus on local-type models in this paper.  As will be shown, by using the 
``consistency relations,''  we
may be able to discriminate models of large non-Gaussianity.

The organization of this paper is as follows. In the next section, we
summarize the formalism for the discussion of local-type
non-Gaussianity of primordial fluctuations. Then in
Section~\ref{sec:local}, we first make a classification of models by
using some relation between the non-linearity parameters $f_{\rm NL},
\tau_{\rm NL}$ and $g_{\rm NL}$ in a systematic manner.  Some example
models will also be discussed.  The final section is devoted to summary
and conclusion of this paper.

\section{Formalism}
\label{sec:formalism}

First we summarize the formalism to discuss local-type non-Gaussianity
of primordial fluctuations, which is expected to be generated from
super-horizon dynamics of curvature fluctuations.  For this purpose, we
adopt the $\delta N$ formalism
\cite{Starobinsky:1986fxa,Sasaki:1995aw,Sasaki:1998ug,Lyth:2004gb}.  In
this formalism, the super-horizon scale curvature perturbation on the
uniform energy density hypersurface $\zeta$ at some time $t=t_f$ is
given by the perturbation in the number of $e$-folding measured from the
initial time $t_\ast$ to the time $t_f$. Here we take the initial time
$t_\ast$ to be the time shortly after the horizon crossing and the
initial hypersurface to be a flat one.  Then the curvature perturbation
is given, up to the third order, as
\begin{eqnarray}
\zeta (t_f) \simeq N_a \delta \varphi^a_\ast 
+ {1 \over 2}N_{ab}\delta \varphi^a_\ast \delta \varphi^b_\ast
 + {1 \over 6} N_{abc} 
 \delta  \varphi^a_\ast \delta \varphi^b_\ast \delta \varphi_\ast^c~,
\label{eq:deltaN}
\end{eqnarray}
where a subscript $a,b$ and $c$ labels a scalar field, which is
assumed to be Gaussian fluctuations $\delta \varphi^a$ at $t=t_*$
in the following discussion, and $N_a = d N / d \varphi_\ast^a$ and so
on. The summation is implied for repeated indices.

The power spectrum $P_\zeta$, bispectrum $B_\zeta$ and trispectrum
$T_\zeta$ of the curvature perturbation are given by
\begin{equation}
\label{eq:power}
\langle \zeta_{\vec k_1} \zeta_{\vec k_2} \rangle
=
{(2\pi)}^3 P_\zeta (k_1) \delta ({\vec k_1}+{\vec k_2}),
\end{equation}
\begin{eqnarray}
\langle \zeta_{\vec k_1} \zeta_{\vec k_2} \zeta_{\vec k_3} \rangle
&=&
{(2\pi)}^3 B_\zeta (k_1,k_2,k_3) \delta ({\vec k_1}+{\vec k_2}+{\vec k_3}),
\label{eq:bi}
\end{eqnarray}
\begin{eqnarray}
\langle
\zeta_{\vec k_1} \zeta_{\vec k_2} \zeta_{\vec k_3} \zeta_{\vec k_4}
\rangle
&=&
{(2\pi)}^3 T_\zeta (k_1,k_2,k_3,k_4) \delta ({\vec k_1}+{\vec k_2}+{\vec k_3}+{\vec k_4}),
\label{eq:tri}
\end{eqnarray}
where $B_\zeta$ and $T_\zeta$ can be written as
\begin{eqnarray}
B_\zeta (k_1,k_2,k_3)
&=&
\frac{6}{5} f_{\rm NL}^{\rm local}
\left(
P_\zeta (k_1) P_\zeta (k_2)
+ P_\zeta (k_2) P_\zeta (k_3)
+ P_\zeta (k_3) P_\zeta (k_1)
\right), \\
\label{eq:def_f_NL}
T_\zeta (k_1,k_2,k_3,k_4)
&=&
\tau_{\rm NL}^{\rm local} \left(
P_\zeta(k_{13}) P_\zeta (k_3) P_\zeta (k_4)+11~{\rm perms.}
\right) \nonumber \\
&&
+ \frac{54}{25} g_{\rm NL}^{\rm local} \left( P_\zeta (k_2) P_\zeta (k_3) P_\zeta (k_4)
+3~{\rm perms.} \right),
\label{eq:def_tau_g_NL}
\end{eqnarray}
with $k_{13} = |{\vec k_1} + {\vec k_3}|$.  Here $f_{\rm NL}^{\rm local}, \tau_{\rm
  NL}^{\rm local}$ and $g_{\rm NL}^{\rm local}$ are non-linearity
parameters of the local type. From Eq.~(\ref{eq:deltaN}), the power spectrum of the curvature
perturbation is given by
\begin{eqnarray}
\label{eq:P_zeta}
P_\zeta(k) = N_a N^a  P_\delta (k)~,
\end{eqnarray}
where $P_\delta(k)$ is the power spectrum for fluctuations of a scalar field:
\begin{eqnarray}
\label{eq:powerdelta}
&&
\langle \delta \varphi^a_{*\vec{k}_1}\delta \varphi^b_{*\vec{k}_2}\rangle
\equiv (2\pi)^3 \delta^{ab} \delta \left( \vec{k}_1 + \vec{k}_2\right)P_\delta(k_1)
=  (2\pi)^3 \delta^{ab} \delta \left( \vec{k}_1 + \vec{k}_2\right)
\frac{2\pi^2}{k_1^3} \mathcal{P}_\delta (k_1),
\end{eqnarray}
with $\mathcal{P}_\delta (k) = (H_\ast / 2\pi)^2$ and $H_\ast$ being
the Hubble parameter at $t=t_*$.

Since $\delta \varphi_a$ are  Gaussian fields,  the leading order contributions to the bispectrum are given by four-point functions of $\delta \varphi_a$. The next leading order contributions are given by six-point functions. As is explained in detail in \cite{Byrnes:2007tm,Yokoyama:2008by}, we can systematically classify those contributions by assigning a diagram to each contribution. In terms of the diagram approach, the four-point function of $\delta \varphi_a$ is represented by the tree diagram and the six-point function by the diagram containing a single loop. This can be easily generalized to higher order correlation functions of $\zeta$ (and also to 
the power spectrum). The leading order contributions to the $n$-point function of $\zeta$ are given by $2(n-1)$-point function of $\delta \varphi_a$, which are represented by the tree diagrams. The next leading order contributions are given by $2n$-point function, which are represented by the one-loop diagrams.

In many cases, the contributions from the loop diagrams are tiny and can
be safely neglected. Therefore, we first drop the loop corrections and
consider only tree contributions.  We will discuss a case in which
effects from loop corrections cannot be neglected in a later section and
an appendix.

By calculating the bispectrum and the trispectrum at the tree level,
we obtain the non-linearity parameters $f_{\rm NL}, \tau_{\rm NL}$ and
$g_{\rm NL}$ as \cite{Lyth:2005fi,Alabidi:2005qi,Byrnes:2006vq}
\begin{eqnarray}
{6 \over 5}f_{\rm NL}^{\rm local}
= 
\frac{N_a N_b N^{ab}}
{\left( N_c N^c \right)^2},
\label{eq:fNL}
\end{eqnarray}
\begin{eqnarray}
\tau_{\rm NL}^{\rm local}
 = 
 \frac{N_a N_{b} N^{ac} N_c^{~b}} 
{\left( N_d N^d \right)^3},
\label{eq:tauNL}
\end{eqnarray}
\begin{eqnarray}
\frac{54}{25}g_{\rm NL}^{\rm local}
=  
\frac{
N_{abc} N^a N^b N^c
}
{\left( N_d N^d \right)^3}.
\label{eq:gNL}
\end{eqnarray}
In the following, since we concentrate on non-Gaussianity of the local-type models,
we omit a superscript ``local'' unless some confusions
arise.
%

\section{Models of local-type non-Gaussianity}
\label{sec:local}

\subsection{A classification of local type}

Even if we limit ourselves to models generating large local-type
non-Gaussianity, there still remain many possibilities.  To discuss
how we discriminate those models, here we classify the local-type
models into some categories. For this purpose, we first give a particular attention to the
relation between $f_{\rm NL}$ and $\tau_{\rm NL}$.  Notice that, when
a single scalar field $\sigma$ is only responsible for density
fluctuations in the Universe, $\delta N$ formalism indicates:
\begin{eqnarray}
\frac{6}{5}f_{\rm NL}
=
 \frac{N_{\sigma\sigma}}{ N_\sigma^2 }, 
 ~~~~~~~~
\tau_{\rm NL}
=
 \frac{N_{\sigma\sigma}^2}{ N_\sigma^4 },
  ~~~~~~~~
\frac{54}{25} g_{\rm NL}
=
 \frac{N_{\sigma\sigma\sigma}}{ N_\sigma^3 }.
\label{eq:non_linear_param}
\end{eqnarray}
From these equations, we can easily find the following relation between $f_{\rm NL}$ and 
$\tau_{\rm NL}$:
\begin{eqnarray}
\label{eq:fNL_tauNL}
\tau_{\rm NL} = \left( \frac65 f_{\rm NL} \right)^2.
\end{eqnarray}
Since some simple models such as the pure curvaton and the pure
modulated reheating scenarios fall onto this type, we use the above
relation to make a classification.  We call models in which
Eq.~\eqref{eq:fNL_tauNL} is satisfied as ``single-source model.''
Notice that $\tau_{\rm NL}$ is determined once $f_{\rm NL}$ is given in
models of this class.

The next category is for models which do not have a universal relation
between $f_{\rm NL}$ and $\tau_{\rm NL}$, although there exists an
inequality as given below. This is a general situation where there are
multiple sources of density fluctuations.  Thus we call this category
``multi-source model.''  An example of this type includes mixed
fluctuation models, where fluctuations from both of the inflaton
and another scalar field such as the curvaton can be responsible for
density fluctuations. Since the inflaton gives almost Gaussian
fluctuations, non-Gaussianity mostly originates from fluctuations of the
other source in such a case.

The third category would be for models where there is some universal
relation between $f_{\rm NL}$ and $\tau_{\rm NL}$.  In fact, this
definition includes models of the first class discussed above as a
special case. However, the ``single source model'' is somewhat a very
special type, and thus we treat them as a separate category. One of
examples of this third category is so-called ``ungaussiton'' model
\cite{Linde:1996gt,Boubekeur:2005fj,Suyama:2008nt}, in which the
relation $\tau_{\rm NL} \propto f_{\rm NL}^{4/3}$ holds.  In this model,
the second order perturbation of the ``ungaussiton'' field dominates the
linear order one.  Since the dominance of the second order perturbations
indicates that the fluctuations are completely non-Gaussian, we need
another source which gives a Gaussian perturbation at linear order.
Thus this model would require multi-sources of fluctuations in this
sense.  Hence we call models in this third category ``constrained
multi-source model.''

Here it is worth noting that the following inequality holds
in the $f_{\rm NL}$--$\tau_{\rm NL}$ plane:
\begin{equation}
\tau_{\rm NL} \geq \left(\frac65  f_{\rm NL} \right)^2,
\label{eq:inequality}
\end{equation}
which can be obtained by using Cauchy-Schwarz inequality and was first
found by two of the present authors (T.S. and M.Y.)
\cite{Suyama:2007bg}.  The derivation of this
inequality~\eqref{eq:inequality} in \cite{Suyama:2007bg} was based on
the expressions for $f_{\rm NL}$ and $\tau_{\rm NL}$ given in
Eqs.~\eqref{eq:fNL} and \eqref{eq:tauNL}, which are valid at the tree
level.  In this paper, we further show that, even when some loop corrections
are dominant in $f_{\rm NL}$ and $\tau_{\rm NL}$, the
inequality~\eqref{eq:inequality} holds as far as loop contributions are
subdominant in the power spectrum, which is required from current
observations.   Since
all models generating large local-type non-Gaussianity known to
date satisfy the above assumptions, the inequality~\eqref{eq:inequality}
is very important to test the local-type models. Thus we call the
inequality~\eqref{eq:inequality} ``local-type inequality'' in the
following. We discuss this issue in detail again in
Section~\ref{subsec:constrained}.  Also notice that single-source models
correspond to the boundary of this local-type inequality.

\begin{figure}[htbp]
  \begin{center}
    \resizebox{120mm}{!}{
    \includegraphics{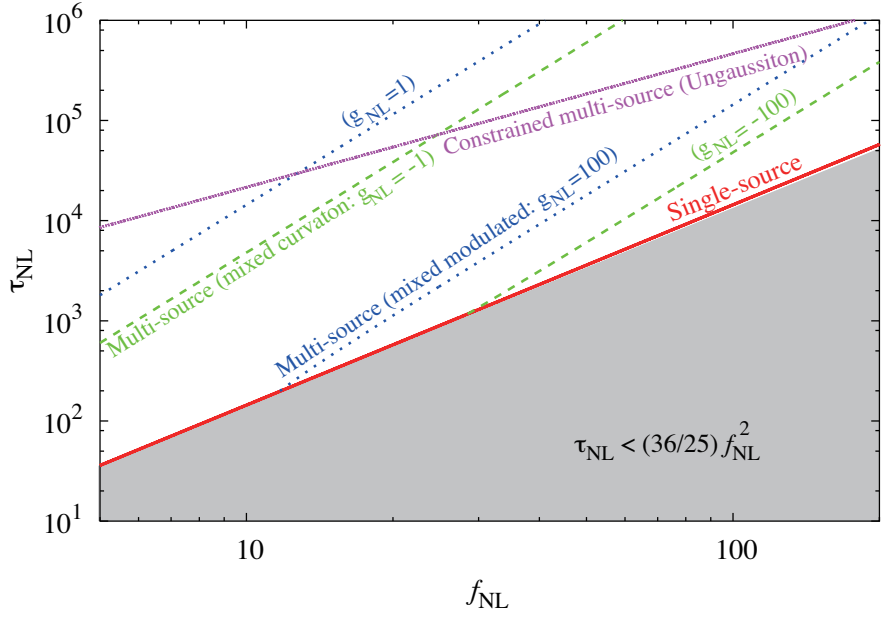}
    }
  \end{center}
  \caption{$f_{\rm NL}$--$\tau_{\rm NL}$ diagram. The relation between
  $f_{\rm NL}$ and $\tau_{\rm NL}$ is shown for three categories:
  single-source, multi-source and constrained multi-source models. For
  multi-source and constrained multi-source models, the cases for some
  representative explicit models (mixed curvaton and inflaton,
  mixed modulated reheating and inflaton, and ungaussiton models) are
  plotted. All the three categories satisfy the inequality $\tau_{\rm
  NL} \gtrsim (6 f_{\rm NL} /5)^2$ as far as loop contributions are
  subdominant in the power spectrum.  The region with $\tau_{\rm NL} <
  (6 f_{\rm NL}/5)^2$ is shaded with gray.}
 \label{fig:fNL_tauNL}
\end{figure}

In the following sections, we discuss various models for each
category. We start with ``single-source model,'' then discuss
``multi-source model'' and ``constrained multi-source model''
afterwards.  In Fig.~\ref{fig:fNL_tauNL}, some models of these
categories are shown in the ``$f_{\rm NL}$--$\tau_{\rm NL}$'' diagram,
from which we can have some idea of 
to what extent models of each category can give different predictions on
$f_{\rm NL}$ and $\tau_{\rm NL}$.  We also give a summary of the
categories and their examples in Table~\ref{tab:summary}.

\begin{figure}[htbp]
  \begin{center}
    \resizebox{120mm}{!}{
    \includegraphics{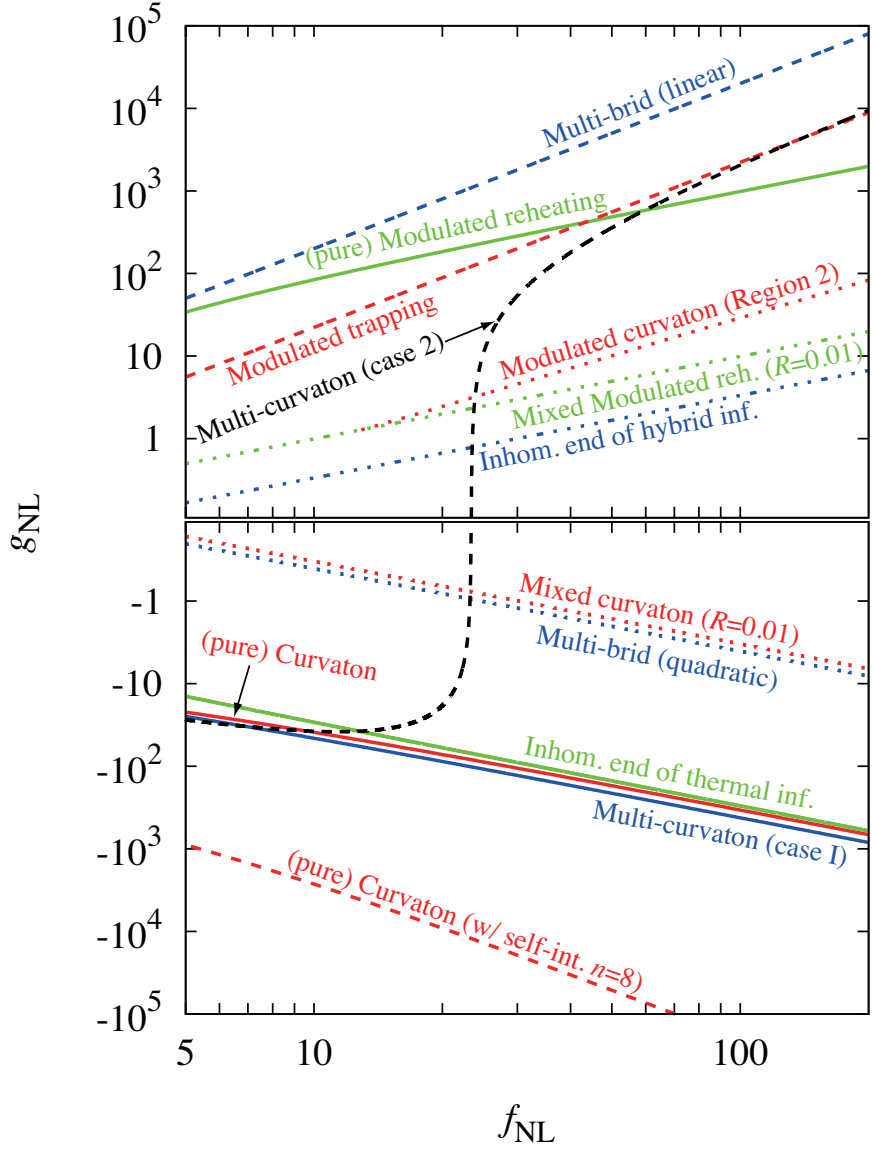}
    }
  \end{center}
  \caption{$f_{\rm NL}$--$g_{\rm NL}$ diagram. The relation between
    $g_{\rm NL}$ and $f_{\rm NL}$ is plotted for models given in
    Table~\ref{tab:summary}.  }
  \label{fig:fNL_gNL_diagram}
\end{figure}

Now we have categorized models of local-type into three classes by
using the key quantity $\tau_{\rm NL}/(6 f_{\rm NL}/5)^2$.  However,
since each category still includes some (or many) possible models, we
may need another quantity to discriminate them. For this purpose, we
can further utilize the relation between $f_{\rm NL}$ and $g_{\rm NL}$.
Although  the $f_{\rm NL}$--$g_{\rm NL}$ relation can change depending
on the model parameters, we can roughly divide models into three types
further  by looking at their relative size.
 As we will argue in the following sections, the relation $|g_{\rm NL}|
\sim |f_{\rm NL}|$ holds in some models,  then we call such models 
``linear $g_{\rm NL}$ type.''  In other models,
$g_{\rm NL}$ could be suppressed compared to $f_{\rm NL}$,
i.e. $g_{\rm NL} \sim ~{\rm (suppression~factor)} \times f_{\rm NL}$,
which we denote this type of models as ``suppressed $g_{\rm NL}$
type.''  The other type is ``enhanced $g_{\rm NL}$ type'' in which the
relation between $f_{\rm NL}$ and $g_{\rm NL}$ can be given as $g_{\rm
  NL} \sim f_{\rm NL}^n$ with $n >1$ (but in most models discussed in this paper, $n=2$).  By
using the $f_{\rm NL}$--$\tau_{\rm NL}$ and $f_{\rm NL}$--$g_{\rm NL}$
relations, we may be able to discriminate models well.  In
Table~~\ref{tab:summary}, the $f_{\rm NL}$--$g_{\rm NL}$ relations for
example models are shown and depicted in
Fig.~\ref{fig:fNL_gNL_diagram}.

\begin{table}
\vspace{-1cm}
  \centering 
  \begin{tabular}{c|c|c}
\hline
Category & $f_{\rm NL}$--$\tau_{\rm NL}$ relation & Examples and  $f_{\rm NL}$--$g_{\rm NL}$ relation \\
\hline 
Single-source  
& $ \tau_{\rm NL} = \left( 6 f_{\rm NL} /5 \right)^2$ 
& (pure) curvaton (w/o self-interaction) \\ 
&&$\left[ g_{\rm NL} = -(10/3) f_{\rm NL} - (575/108) \right]^{(a)}$ \\  \cline{3-3}
&& (pure) curvaton (w/ self-interaction) \\
&& $\left[ g_{\rm NL} = A_{\rm NQ} f_{\rm NL}^2  + B_{\rm NQ} f_{\rm NL} + C_{\rm NQ}\right]^{(b)}$ \\  \cline{3-3}
&& (pure) modulated reheating \\
&& $\left[ g_{\rm NL} = 10 f_{\rm NL}  -(50/3) \right]^{(c)}$ \\  \cline{3-3}
&& modulated-curvaton scenario \\
&& $\left[ g_{\rm NL} = 3  r_{\rm dec}^{1/2} f_{\rm NL}^{3/2}\right]^{(d)}$ \\  \cline{3-3}
&& Inhomogeneous end of hybrid inflation\\
&& $\left[ g_{\rm NL} = (10/3) \eta_{\rm cr}  f_{\rm NL} \right]^{}$ \\  \cline{3-3}
&& Inhomogeneous end of thermal inflation\\
&& $\left[ g_{\rm NL} = -(10/3)   f_{\rm NL} - (50/27) \right]^{(e)}$ \\  \cline{3-3}
&& Modulated trapping\\
&& $\left[ g_{\rm NL} = (2/9)   f_{\rm NL}^2 \right]^{(f)}$ \\  \cline{3-3}
\hline 
Multi-source  
& $ \tau_{\rm NL} > \left( 6 f_{\rm NL} / 5 \right)^2$ 
& mixed curvaton and inflaton \\
&&$\left[ g_{\rm NL} =  -(10/3) (R/(1+R) )f_{\rm NL}  - (575/108)   (R/(1+R) )^3 \right]^{(g)}$ \\  \cline{3-3}
&& mixed modulated and inflaton \\
&&$\left[ g_{\rm NL} = 10 (R/(1+R) )f_{\rm NL} - (50/3)  (R/(1+R) )^3
 \right]^{(h)}$ \\  \cline{3-3}
&& mixed modulated trapping and inflaton \\
&&$\left[ g_{\rm NL} = (2/9)( (1+R)/R )f_{\rm NL}^2 = (25/162) \tau_{\rm NL}
\right]^{(i)}$ \\  \cline{3-3}
&& multi-curvaton \\
&&$\left[g_{\rm NL} =   C_{\rm mc} f_{\rm NL}, ~~g_{\rm NL} = (4/15) f_{\rm NL}^2 \right]^{(j)}$ \\ \cline{3-3}
&& Multi-brid inflation (quadratic potential) \\
&&$\left[ g_{\rm NL} = -(10/3) \eta f_{\rm NL}, ~~ g_{\rm NL} =  2 f_{\rm NL}^2   \right]^{(k)}$ \\  \cline{3-3}
&& Multi-brid inflation (linear potential) \\
&&$\left[ g_{\rm NL} =  2 f_{\rm NL}^2 \right]^{(l)}$ \\  \cline{3-3}
\hline 
Constrained & & \\
 multi-source
& $ \tau_{\rm NL} = C f_{\rm NL}^n$ 
& ungaussiton ($C\simeq 10^3,\ n=4/3$)\\
\hline 
\end{tabular}
\begin{flushleft}
{\scriptsize
$^{(a)}$For the case with $r_{\rm dec} \ll 1$. \\
$^{(b)}$$A_{\rm NQ}, B_{\rm NQ}$ and $C_{\rm NQ}$ are given in 
Eqs.~\eqref{eq:curvaton_ANQ}-\eqref{eq:curvaton_CNQ} and this expression is for 
$r_{\rm dec} \ll 1$. \\
$^{(c)}$$\Gamma_{\sigma\sigma\sigma}=0$ is assumed.\\
$^{(d)}$This relation holds in the Region 2. For other cases, see text.\\
$^{(e)}$$g^{\prime\prime\prime}=0$ is assumed.\\
$^{(f)}$$\lambda=\sigma / M$ and $m = g\sigma$ are assumed.\\
$^{(g)}$A quadratic potential and $r_{\rm dec} \ll 1$ are assumed for the curvaton sector. 
$R\equiv P_\zeta^{(\sigma)} / P_\zeta^{(\phi)}$ is the ratio of the power spectra. This 
relation can also be written as 
$g_{\rm NL} \simeq -(24/5) (f_{\rm NL}^3/\tau_{\rm NL} ) - (9936/625)(f_{\rm NL}^6/\tau_{\rm NL}^3)$.  \\
$^{(h)}$$\Gamma_{\sigma\sigma\sigma}=0$ is assumed for the modulated reheating sector.
This relation can also be written as $g_{\rm NL} \simeq
(72/5) (f_{\rm NL}^3/\tau_{\rm NL})  - (31104/625)(f_{\rm NL}^6/\tau_{\rm NL}^3)$.  \\
$^{(i)}$$\lambda=\sigma / M$ and $m = g\sigma$ are assumed for the modulaton sector.\\
$^{(j)}$The former and the latter relations are for the cases where both curvatons are subdominant 
and dominant at their decay, respectively. $C_{\rm mc}$ 
is $\mathcal{O}(1)$ coefficient and always negative. \\
$^{(k)}$The former and the latter relations are for the equal mass
 and the large mass ratio cases, 
respectively. \\
$^{(l)}$For the equal mass case with $g_1 = g_2$.\\

}
\end{flushleft}
  \caption{Summary of the categories and their examples. }
  \label{tab:summary}
\end{table}

\subsection{Single-source model}
\label{subsec:single_source}

If future observations confirm the relation $\tau_{\rm NL} = (6
f_{\rm NL}/5)^2$, models of large non-Gaussianity we should pursue
would be the ones in this category. Models categorized in this class
include the pure curvaton and the pure modulated reheating scenarios,
and so on.  In the following, we look at some of models in this class
more closely.

\subsubsection{Pure curvaton model}
\label{subsec:pure_curvaton}

To make a prediction on the non-linearity parameters in the pure
curvaton model \cite{Enqvist:2001zp, Lyth:2001nq, Moroi:2001ct}\footnote{
  We later also consider a mixed model where density fluctuations originate
  both from the curvaton and the inflaton.  To be definite, we call
  the model where only the curvaton generates density fluctuations
  the pure curvaton model.
}, first we need to specify the curvaton potential.  Although in most
study of the curvaton~\cite{Moroi:2002rd,Lyth:2002my,
  Lyth:2003ip,Bartolo:2003jx, Malik:2006pm, Sasaki:2006kq,
  Huang:2008ze, Multamaki:2008yv,
  Beltran:2008ei,Moroi:2008nn,Takahashi:2009cx}, a quadratic potential
is adopted, it is also possible to have a self-interaction term.  Thus
we assume the following potential
\cite{Enqvist:2005pg,Enqvist:2008gk,Huang:2008zj,Enqvist:2009eq,Enqvist:2009zf,Enqvist:2009ww}\footnote{
  For the analyses of other non-quadratic type potentials, see
  \cite{Huang:2008bg,Kawasaki:2008mc,Chingangbam:2009xi,Choi:2010re}.
}:
\begin{equation}
\label{eq:V}
V(\sigma)
=
\frac{1}{2} m_\sigma^2 \sigma^2
+
\lambda m_\sigma^4 \left( \frac{\sigma}{m_\sigma} \right)^n~,
\end{equation}
where $m_\sigma$ is the mass of the curvaton, $\lambda$ is the
coupling and $n$ is the power for a self-interaction term. In the
following discussion, we assume that the self-interaction term is
subdominant compared to the quadratic one.  In general, it is possible
that a self-interaction term dominates first, and then the quadratic
term becomes dominant at later time. However, in such a case, rigorous
numerical calculations will be needed. Thus we restrict ourselves to
the ``nearly'' quadratic potential case when we include a
self-interaction term\footnote{
For a general case, see \cite{Enqvist:2009zf,Enqvist:2009ww}.
}.

In fact, to make a concrete prediction on non-Gaussianity in this model,
we need to specify the decay rate $\Gamma_\sigma$, the mass $m_\sigma$
and the initial value for the curvaton $\sigma_\ast$.  However, as far as the non-linearity parameters are
concerned, the information on these quantities is encoded into the
parameter $r_{\rm dec}$, which corresponds to the ratio of the energy
density of the curvaton to the total one at the decay\footnote{
  When the initial amplitude for the curvaton is large enough, the
  curvaton can drive the second inflation after the standard
  inflation. In this case, this statement does not hold true. However,
  the value of $f_{\rm NL}$ is $\mathcal{O}(1)$ in such a situation,
  hence we do not discuss such a case here.
}.  The definition of $r_{\rm dec}$ and its approximate expression using 
$\Gamma_\sigma, m_\sigma$ and $\sigma_\ast$ 
are given by
\begin{equation}
\left.
r_{\rm dec}
\equiv \frac{3 \rho_\sigma}{4 \rho_{\rm rad} + 3\rho_\sigma}\right|_{\rm dec} 
\sim 
\frac{\sigma_\ast^2}{M_{\rm pl}^2 \sqrt{\Gamma_\sigma / m_\sigma}}\,,
\end{equation}
where $M_{\rm pl} \simeq 2.4 \times 10^{18}$~GeV is the reduced
Planck mass. Thus in the following, we do not specify explicit values
for $\Gamma_\sigma, m_\sigma$ and $\sigma_\ast$.

In the curvaton model, adopting the sudden-decay approximation and
assuming radiation-dominated background from the period of the start
of its oscillation to its decay\footnote{
  In Ref.~\cite{Enqvist:2009eq}, it has been shown that the background
  equation of state also affects the curvature perturbation generated
  from the curvaton, in particular, its non-Gaussianity.
},  the curvature perturbation can be given by\footnote{
  In the curvaton model, isocurvature fluctuations can also be
  generated. However, we do not consider such a case here. For the
  study of isocurvature fluctuations in the model, see
  \cite{Moroi:2002rd,Lyth:2003ip,Beltran:2008ei,Moroi:2008nn}.
},
\begin{eqnarray}
\label{eq:zeta_cur}
\zeta_{\rm cur} &=&
\frac{2}{3} r_{\rm dec} \frac{\sigma'_{\rm osc}}{\sigma_{\rm osc}}  \delta \sigma_\ast
+
\frac{1}{9} \left[ 3r_{\rm dec}\left(
1 +
\frac{\sigma_{\rm osc} \sigma_{\rm osc}^{\prime\prime}}{\sigma_{\rm osc}^{\prime 2}}
\right)
- 4 r^2_{\rm dec} -2  r^3_{\rm dec}
\right]
\left( \frac{\sigma'_{\rm osc}}{\sigma_{\rm osc}} \right)^2  (\delta \sigma_\ast )^2 \notag \\
&&+
\frac{4}{81} \left[
\frac{9r_{\rm dec}}{4}  \left(
\frac{\sigma_{\rm osc}^2 \sigma_{\rm osc}^{\prime\prime\prime}}
{\sigma_{\rm osc}^{\prime 3}}
+
3\frac{\sigma_{\rm osc} \sigma_{\rm osc}^{\prime\prime}}{\sigma_{\rm osc}^{\prime 2}}
\right)
-9r^2_{\rm dec}
\left(
1
+
\frac{\sigma_{\rm osc} \sigma_{\rm osc}^{\prime\prime}}{\sigma_{\rm osc}^{\prime 2}}
\right)
\right. \notag \\
&&
\left.
+\frac{r^3_{\rm dec}}{2} \left(
1
-
9\frac{\sigma_{\rm osc} \sigma_{\rm osc}^{\prime\prime}}{\sigma_{\rm osc}^{\prime 2}}
\right)
+10r^4_{\rm dec} + 3r^5_{\rm dec}
\right]
\left( \frac{\sigma'_{\rm osc}}{\sigma_{\rm osc}} \right)^3  (\delta \sigma_\ast )^3~,
\end{eqnarray}
where $\sigma_{\rm osc}$ denotes the value of the curvaton field at
the time of the start of oscillation and the prime indicates a
derivative with respect to $\sigma_\ast$, namely $\sigma_{\rm
  osc}^\prime = d \sigma_{\rm osc} / d \sigma_\ast$ and so on.
$\sigma_{\rm osc}$ and its derivatives represent the non-linear
evolution of the curvaton after the horizon exit.

From the expression of the curvature perturbation given above, the
non-linearity parameters $f_{\rm NL}$ and $g_{\rm NL}$ can be
evaluated as
\begin{eqnarray}
\label{eq:fNL_cur}
\frac65 f_{\rm NL} &=& \frac{3}{2 r_{\rm dec}} \left(
1 +
\frac{\sigma_{\rm osc} \sigma_{\rm osc}^{\prime\prime}}{\sigma_{\rm osc}^{\prime 2}}
\right)
- 2
 -r_{\rm dec}, \\
\label{eq:gNL_cur}
\frac{54}{25} g_{\rm NL} &=& 
\frac{9}{4r^2_{\rm dec}}  \left(
\frac{\sigma_{\rm osc}^2 \sigma_{\rm osc}^{\prime\prime\prime}}
{\sigma_{\rm osc}^{\prime 3}}
+
3\frac{\sigma_{\rm osc} \sigma_{\rm osc}^{\prime\prime}}{\sigma_{\rm osc}^{\prime 2}}
\right)
-\frac{9}{r_{\rm dec}}
\left(
1
+
\frac{\sigma_{\rm osc} \sigma_{\rm osc}^{\prime\prime}}{\sigma_{\rm osc}^{\prime 2}}
\right) \notag \\
&& \qquad 
+\frac{1}{2} \left(
1
-
9\frac{\sigma_{\rm osc} \sigma_{\rm osc}^{\prime\prime}}{\sigma_{\rm osc}^{\prime 2}}
\right)
+10r_{\rm dec} + 3r^2_{\rm dec}.
\end{eqnarray}

When the potential has a pure quadratic form, we have
\begin{equation}
\frac{\sigma'_{\rm osc}}{\sigma_{\rm osc}} = \frac{1}{\sigma_\ast}, ~~
\sigma_{\rm osc}^{\prime\prime} =0,~~
\sigma_{\rm osc}^{\prime\prime\prime} =0.
\end{equation}
Thus, in this case with $r_{\rm dec} \ll 1$, we can find the following relation between
$f_{\rm NL}$ and $g_{\rm NL}$ \cite{Enqvist:2008gk}:
\begin{equation}
\label{eq:fNL_gNL_pure_cur}
g_{\rm NL} = -\frac{10}{3} f_{\rm NL}-{575 \over 108} + O(r_{\rm dec}).
\end{equation}
On the other hand, when we cannot neglect the contribution from a
self-interaction term, the non-linear evolution of the curvaton field
after the horizon exit  changes the relation between
$f_{\rm NL}$ and $g_{\rm NL}$, for $r_{\rm dec} \ll 1$, as
\begin{equation}
\label{eq:nonQ_relation}
g_{\rm NL} \simeq A_{\rm NQ} f_{\rm NL}^2  + B_{\rm NQ} f_{\rm NL} + C_{\rm NQ},
\end{equation}
where the coefficients are given by
\begin{eqnarray}
\label{eq:curvaton_ANQ}
A_{\rm NQ} & = & \frac23  \left(
\frac{\sigma_{\rm osc}^2 \sigma_{\rm osc}^{\prime\prime\prime}}
{\sigma_{\rm osc}^{\prime 3}}
+
3\frac{\sigma_{\rm osc} \sigma_{\rm osc}^{\prime\prime}}{\sigma_{\rm osc}^{\prime 2}}
\right)
\left(
1 +
\frac{\sigma_{\rm osc} \sigma_{\rm osc}^{\prime\prime}}{\sigma_{\rm osc}^{\prime 2}}
\right)^{-2}, \\
\label{eq:curvaton_BNQ}
B_{\rm NQ} & = & -\frac{10}{3} \left( 
1 
- 5 \frac{\sigma_{\rm osc} \sigma_{\rm osc}^{\prime\prime}}{\sigma_{\rm osc}^{\prime 2}}
- 2 \frac{\sigma_{\rm osc}^2 \sigma_{\rm osc}^{\prime\prime\prime}}
{\sigma_{\rm osc}^{\prime 3}}
\right)
\left(
1 +
\frac{\sigma_{\rm osc} \sigma_{\rm osc}^{\prime\prime}}{\sigma_{\rm osc}^{\prime 2}}
\right)^{-2}, \\
\label{eq:curvaton_CNQ}
C_{\rm NQ} 
&=&
- \frac{50}{27} \left( 3 -  \frac{\sigma_{\rm osc}^2 \sigma_{\rm osc}^{\prime\prime\prime}}
{\sigma_{\rm osc}^{\prime 3}} 
\right) \left(
1 +
\frac{\sigma_{\rm osc} \sigma_{\rm osc}^{\prime\prime}}{\sigma_{\rm osc}^{\prime 2}}
\right)^{-2} - \frac{25}{108} 
\left( 
1- 9 \frac{\sigma_{\rm osc} \sigma_{\rm osc}^{\prime\prime}}{\sigma_{\rm osc}^{\prime 2}}
\right).
\end{eqnarray}
These coefficients depend on the power of a self-interaction term and
its relative size to the mass term, which can be calculated
numerically \cite{Enqvist:2008gk,Enqvist:2009eq}.
There is a clear distinction in the relation between $f_{\rm NL}$ and
$g_{\rm NL}$ depending on the contribution from a self-interaction term.
From the viewpoint of the $f_{\rm NL}$--$g_{\rm NL}$ relation, the model
with and without a self-interaction term can be regarded as ``enhanced
$g_{\rm NL}$" and ``linear $g_{\rm NL}$" types, respectively.  Thus once
$f_{\rm NL}$ and $g_{\rm NL}$ are determined by observations with some
accuracy, it may be even possible to probe the form of the potential
\cite{Enqvist:2008gk,Enqvist:2009eq} as well.  In
Fig.~\ref{fig:fNL_gNL_diagram}, we show the case with $V(\sigma) = (1/2)
m_\sigma^2 \sigma^2$ (pure quadratic potential) and $V(\sigma) = (1/2)
m_\sigma^2 \sigma^2 + \lambda \sigma^8 / m_\sigma^4$ (with a
self-interaction term). As seen from the figure, we can easily see the
difference in the $f_{\rm NL}$--$g_{\rm NL}$ relation between the cases
with and without a self-interaction.

\subsubsection{Pure modulated reheating model}
\label{sec:pure_modulated}

In the modulated reheating scenario \cite{Dvali:2003em,Kofman:2003nx},
the decay rate of the inflaton $\Gamma$ fluctuates in
space\footnote{
Thus, the reheating temperature after inflation also
fluctuates in space in this model, which may generate significant isocurvature
fluctuations and give severe constraints in some settings
\cite{Takahashi:2009dr,Kamada:2010yz}.
} because of its dependence on a
modulus field $\sigma$, which acquires quantum fluctuations during
inflation, and large non-Gaussianity can be generated
\cite{Zaldarriaga:2003my,Suyama:2007bg,Ichikawa:2008ne}.  In this
scenario, the level of non-Gaussianity is highly dependent on the
assumption of the decay rate.  Thus we need to specify the dependence of
$\Gamma$ on $\sigma$ to give a concrete prediction.  Furthermore, in
fact, we also need to specify the form of the inflaton potential and the
interaction for the decay.  But we first give a general expression for
the curvature perturbations $\zeta$ in this scenario, taking into
account the above mentioned uncertainties.  $\zeta$ in this scenario can
be given by \cite{Suyama:2007bg,Ichikawa:2008ne}
\begin{eqnarray}
\label{eq:zeta_mod}
\zeta_{\rm mod} & = &  
A(x) \frac{\Gamma_{\sigma}}{\Gamma} \delta \sigma_\ast
+ \frac{1}{2} \left( 
A(x) \frac{\Gamma_{\sigma \sigma}}{\Gamma} 
+ 
B(x) \frac{\Gamma_{\sigma}^2}{\Gamma^2} 
 \right) \delta \sigma_\ast^2 \notag \\
 &&
+ \frac{1}{6} \left( 
A(x) \frac{\Gamma_{\sigma \sigma \sigma}}{\Gamma} 
+  3B(x) \frac{\Gamma_{\sigma} \Gamma_{\sigma \sigma}}{\Gamma^2}
+ C(x) \frac{ \Gamma_{\sigma}^3}{\Gamma^3}
\right)\delta \sigma_\ast^3.
\end{eqnarray}
Here $\Gamma_\sigma = d \Gamma / d\sigma, $ $\Gamma_{\sigma\sigma} =
d^2 \Gamma /d \sigma^2$ and so on.  $A(x), B(x)$ and $C(x)$ are
functions of $x \equiv \Gamma / H_c$ with $H_c$ being the Hubble
parameter after several oscillations of the inflaton field.  Since the
number of $e$-folds after inflation only depends on the quantity $x
= \Gamma / H_c$, which can be confirmed by a dimensional analysis of
the background evolution equations, the coefficients such as $A, B$
and $C$ depend only on $x$.  From the above expression, we obtain the
non-linearity parameters for this model as
\begin{eqnarray}
&&\frac{6}{5}f_{\rm NL} = 
 \frac{B(x)}{A(x)^2}
 +
 \frac{1}{A(x)}\frac{\Gamma \Gamma_{\sigma \sigma}}{\Gamma_\sigma^2}, 
\label{eq:fNL_puremod}\\
&&\frac{54}{25}g_{\rm NL}
=
 \frac{C(x)}{A(x)} 
 + \frac{3 B(x)}{A(x)} \frac{ \Gamma \Gamma_{\sigma \sigma}}{\Gamma_\sigma^2}
 +\frac{\Gamma^2 \Gamma_{\sigma \sigma \sigma}}{\Gamma_\sigma^3}.
 \label{eq:gNL_puremod}
\end{eqnarray}
To make some explicit predictions for these parameters, we need to
know the values of $A(x), B(x)$ and $C(x)$ for a given $x$, which
requires us to specify the interaction responsible for the decay and
the potential under which the inflaton oscillates. Furthermore,  some numerical analysis may be
 needed.  However, some
approximations are available which can give quite accurate results, in
particular, for the case of $x \ll 1$
\cite{Suyama:2007bg,Ichikawa:2008ne}.

To evaluate to what extent the non-linearity parameters can
be large, here we assume that the potential of the inflaton is $V
\propto \phi^2$ and the inflaton decays through $\mathcal{L}_{\rm int}
= -y \phi \bar{\psi}\psi$. Then the coefficients are evaluated as $A =
-1/6, B = 1/6$ and $C= -1/3$ (for other types of potentials and
interactions, see \cite{Ichikawa:2008ne}).  In this case, 
the non-linearity parameters are given by 
\begin{eqnarray}
&&\frac{6}{5}f_{\rm NL} = 
6 
 -6 \frac{\Gamma \Gamma_{\sigma \sigma}}{\Gamma_\sigma^2}, 
\label{eq:fNL_mod}\\
\notag  \\
&&\frac{54}{25}g_{\rm NL}
=
36 \left(
2 -3  \frac{ \Gamma \Gamma_{\sigma \sigma}}{\Gamma_\sigma^2}
 + \frac{\Gamma^2 \Gamma_{\sigma \sigma \sigma}}{\Gamma_\sigma^3}
 \right).
 \label{eq:gNL_mod}
\end{eqnarray}

To get some numbers for $f_{\rm NL}$ and $g_{\rm NL}$, we further need
to assume an explicit form for $\Gamma$. Here we take the following
form:
\begin{equation}
\label{eq:Gamma_sigma}
\Gamma = \Gamma_0 \left( 
1 + \alpha \frac{\sigma}{M} + \beta \frac{\sigma^2}{M^2} 
\right),
\end{equation}
where $\alpha$ and $\beta$ are some coefficients, $M$ is some scale
and we assume $|\sigma| \ll M$ to justify the expansion in terms of
$\sigma /M$.  With this form of $\Gamma$, the non-linearity parameters
are given by\footnote{
Although here we assume that the curvature perturbation is totally 
generated from this mechanism, in general,
fluctuations from the inflaton would also contribute, in particular 
to the power spectrum. In order that
the contribution from the inflaton is negligible at linear order, $M /
M_{\rm pl} \lesssim 0.2 \alpha \sqrt{\epsilon}$ should be
satisfied with $\epsilon$ being the slow-roll parameter for the inflaton, 
which is obtained by requiring that $\zeta_{\rm mod} >
\zeta_{\rm inf}$ with $\zeta_{\rm inf}$ being the curvature
perturbation generated from the inflaton.  
Another requirement  comes from the constraint 
on the tensor-to-scalar ratio $r \lesssim 0.1$, from which 
we obtain $M/M_{\rm pl} < 0.01 \alpha$.
These relations can be satisfied by taking  the scale $M$ appropriately.
}
\begin{equation}
\frac65 f_{\rm NL} \simeq 6 \left( 1 - \frac{2\beta}{\alpha^2} \right),
~~~~~
\frac{54}{25} g_{\rm NL} \simeq  36  \left( 2 - \frac{6\beta}{\alpha^2} \right).
\end{equation}
Thus, for example, if we take $\alpha = 1/2$ and $\beta = -1$, then
one can have $f_{\rm NL} \sim 45$ and $g_{\rm NL} \sim 430$.

In fact, we can find a relation between $f_{\rm NL}$ and $g_{\rm NL}$
independent of the form of $\Gamma$ as far as the third derivative
$\Gamma_{\sigma\sigma\sigma}$ is negligible.
  After some algebra, one can derive the relation between
$f_{\rm NL}$ and $g_{\rm NL}$ as \cite{Suyama:2007bg}
\begin{equation}
\label{eq:fNL_gNL_modulated}
g_{\rm NL} = 10 f_{\rm NL} - \frac{50}{3}.
\end{equation}
Notice that the signs of $f_{\rm NL}$ and $g_{\rm NL}$ are the
same for large values of $f_{\rm NL}$, which is different from the one
in the curvaton model. Thus once the signs of $f_{\rm NL}$ and $g_{\rm
  NL}$ are determined from observations, we can discriminate between
the curvaton and the modulated reheating models (see also
Fig.~\ref{fig:fNL_gNL_diagram}). In both models, the simple cases predict
that $g_{\rm NL}$ is of the same order of $f_{\rm NL}$ as given in
Eqs. (\ref{eq:fNL_gNL_pure_cur}) and (\ref{eq:fNL_gNL_modulated}),
which is what we call ``linear $g_{\rm NL}$'' type.  Thus, 
as far as the modulated reheating scenario is concerned,
larger $g_{\rm NL}$ may suggest a non-negligible third derivative
of the decay rate $\Gamma_{\sigma\sigma\sigma}$.

\subsubsection{Modulated-curvaton model}

In the modulated reheating scenario, density fluctuations originate
from those of a modulus field. Such a field may also play a role of
the curvaton at later time after it ``modulates'' the reheating. In
this case, the fluctuations become a mixture of those generated from
the modulated reheating and the curvaton mechanism although the source
of fluctuations is a single field $\sigma$.  In fact, this kind of
possibility has not been well explored in the literature.  Therefore,
we here discuss this scenario in some detail.

\begin{figure}[htbp]
  \begin{center}
    \resizebox{100mm}{!}{
    \includegraphics{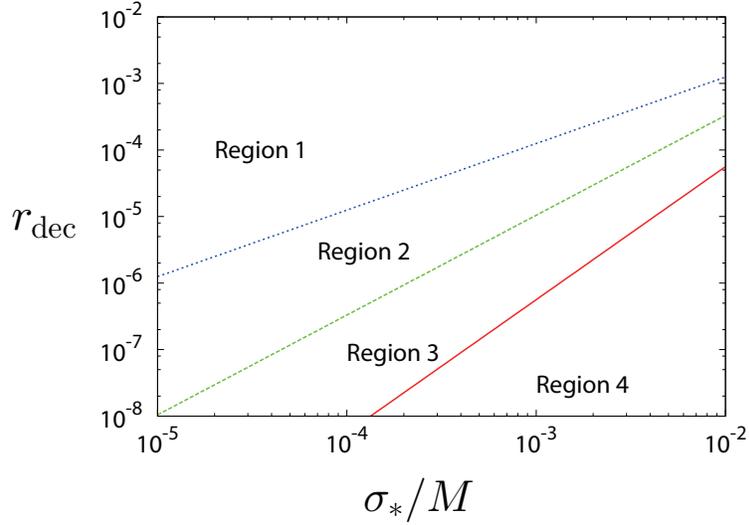}
    }
  \end{center}
  \caption{
Parameter regions where the curvaton fluctuations dominate over those
from the modulated reheating in all orders (Region 1), and the
fluctuations from modulated reheating dominate the curvaton ones, in
the first and the third orders (Region 2), only in the second order
(Region 3), and in all orders (Region 4). Note that here ``all orders"
just means ``up to the third order" since we consider fluctuations
only up to this order. We used
    $\alpha=\frac{1}{2}$ and $\beta=-1$ for illustrative purpose.} \label{fig:mod_cur_region} 
\end{figure}

The curvature perturbation $\zeta$ in this scenario can be written as
the sum of the contributions from the modulated reheating and the
curvaton mechanism:
\begin{eqnarray}
\zeta 
&=& 
\zeta_{\rm cur} + \zeta_{\rm mod} \notag \\
&=&
\left( 
\frac{2r_{\rm dec}}{ 3\sigma_\ast}
-\frac{\Gamma_{\sigma}}{ 6 \Gamma}
\right ) \delta \sigma_\ast
+
\left[ 
\frac{1}{9 \sigma_\ast^2} \left( 3r_{\rm dec} - 4 r_{\rm dec}^2 -2  r_{\rm dec}^3
\right)
- \frac{1}{12} \left(
\frac{\Gamma_{\sigma \sigma}}{\Gamma} 
-
 \frac{\Gamma_{\sigma}^2}{\Gamma^2} 
 \right)
\right] \delta \sigma_\ast^2 \notag \\
&& 
+
\left[ 
\frac{4}{81\sigma_\ast^3} \left(
-9r_{\rm dec}^2
+\frac{r_{\rm dec}^3}{2} 
+10r_{\rm dec}^4 + 3r_{\rm dec}^5
\right)
-\frac{1}{36} \left( 
\frac{2 \Gamma_{\sigma}^3}{\Gamma^3}
- \frac{3 \Gamma_{\sigma} \Gamma_{\sigma \sigma}}{\Gamma^2}
+ \frac{\Gamma_{\sigma \sigma \sigma}}{\Gamma} 
\right)
\right]
\delta \sigma_*^3, \notag \\
\end{eqnarray}
where $\zeta_{\rm cur}$ and $\zeta_{\rm mod}$ are the curvature
perturbations generated from the curvaton mechanism and the modulated
reheating, and are given in Eqs.\,\eqref{eq:zeta_cur} and
\eqref{eq:zeta_mod}.  Here we assume that the curvaton potential is
quadratic and the potential for the inflaton is $V \propto \phi^2$.
The non-linearity parameters are then given by
\begin{eqnarray}
\frac65 f_{\rm NL}
&=&
2{\left( 
-\frac{4r_{\rm dec}}{\sigma_\ast} +\frac{\Gamma_{\sigma}}{\Gamma} 
\right)}^{-2} 
\left[ 3 \left( \frac{\Gamma_{\sigma}^2}{\Gamma^2} 
-\frac{\Gamma_{\sigma \sigma}}{\Gamma} \right)
+\frac{4r_{\rm dec}}{\sigma_\ast^2} (3 - 4r_{\rm dec} - 2r^2_{\rm dec}) 
\right], \\ \notag \\
\frac{54}{25} g_{\rm NL}
&=& 
- 36 {\left(  
-\frac{4r_{\rm dec}}{\sigma_\ast} 
+\frac{\Gamma_{\sigma}}{\Gamma} \right)}^{-3} \notag \\
&& \times
\left[ 
-\frac{2 \Gamma_{\sigma}^3}{\Gamma^3}
+ \frac{3 \Gamma_{\sigma} \Gamma_{\sigma \sigma}}{\Gamma^2}
- \frac{\Gamma_{\sigma \sigma \sigma}}{\Gamma} +\frac{8r^2_{\rm dec}}{9 \sigma_\ast^3} \left( -18+r_{\rm dec}+20r^2_{\rm dec}+6r^3_{\rm dec}\right)
\right]. \notag \\
\end{eqnarray}
To discuss the prediction of this model more explicitly, we need to
specify the form of the decay rate. Here we again adopt the form given
by Eq.~\eqref{eq:Gamma_sigma}.  Since this model includes both cases
of the curvaton and the modulated reheating, some limits should be
identical to those models.  However, notice that which part gives a
dominant contribution to the total curvature perturbation $\zeta$ can be different order by
order. Assuming $\sigma_*/ M \ll 1$ and $r_{\rm dec} \ll 1$, the
condition in which $\zeta_{\rm cur} > \zeta_{\rm mod}$ at the first
order can be given by
\begin{equation}
\label{eq:region_1st}
r_{\rm dec} > |C_1| \frac{\sigma_*}{M},
\end{equation}
where $C_1 = \alpha/4$ is a constant.  In the second order, the
corresponding condition is
\begin{equation}
\label{eq:region_2nd}
r_{\rm dec} > |C_2| \left( \frac{\sigma_*}{M} \right)^2,
\end{equation}
where $C_2 = (2 \beta - \alpha^2 )/4$.  For the third order, it is given by
\begin{equation}
\label{eq:region_3rd}
r_{\rm dec} > \sqrt{\left| C_3 \right|} \left( \frac{\sigma_*}{M} \right)^{3/2},
\end{equation}
where $C_3 = (\alpha^3 - 3 \alpha \beta)/8$.  As far as the parameters
$\alpha, \beta$ are $\mathcal{O}(1)$, the constants $C_1, C_2$ and $C_3$
are also $\mathcal{O}(1)$. From the inequalities above, one can
easily notice that there can exist, for example, the case where the
curvaton part dominates over the modulated reheating part in the second
order, however, in the third order, the modulated reheating part gives a
more contribution.  When $\alpha, \beta = \mathcal{O}(1)$, the
parameters $r_{\rm dec}$ in the curvaton and $\sigma_* / M$ in the
modulated reheating totally determine the size of $\zeta$ in each order.
From the conditions Eqs.~\eqref{eq:region_1st}, \eqref{eq:region_2nd}
and \eqref{eq:region_3rd}, we can divide the parameter space in the
$\sigma_*/M$--$r_{\rm dec}$ plane into four regions, which is shown in
Fig.~\ref{fig:mod_cur_region}.  Since we consider the case of $\sigma_*
/ M \ll 1$, for a fixed value of $\sigma_* / M$, the corresponding
values of $r_{\rm dec}$ for each region are:
\begin{equation}
\label{eq:mod_cur_region}
r_{{\rm dec},4} 
< |C_2| \left( \frac{\sigma_*}{M} \right)^2 <
r_{{\rm dec},3} 
< \sqrt{| C_3 | } \left( \frac{\sigma_*}{M} \right)^{3/2} <
r_{{\rm dec},2} 
< |C_1| \left( \frac{\sigma_*}{M} \right) <
r_{{\rm dec},1},
\end{equation}
where the subscript $i$ which appears in $r_{{\rm dec}, i}$ indicates
the region shown in Fig.~\ref{fig:mod_cur_region}.  Region~1
corresponds to the case where the curvaton fluctuations always
dominate over those from the modulated reheating.  In Region~2, only
at linear order $\zeta_{\rm mod}^{(1)} > \zeta_{\rm cur}^{(1)}$ holds,
but in the second and third orders, $\zeta_{\rm cur}^{(2)} >
\zeta_{\rm mod}^{(2)}$ and $\zeta_{\rm cur}^{(3)} > \zeta_{\rm
  mod}^{(3)}$ are satisfied.  In Region~3, only in the second order,
$\zeta_{\rm cur}^{(2)} > \zeta_{\rm mod}^{(2)}$ holds, but in linear
and the third orders, fluctuations from the modulated reheating give
dominant contributions. Region~4 corresponds to the pure modulated
reheating case.

Now we look at each case in order.  In Region~1, $r_{\rm dec}$ should
satisfy $r_{\rm dec} > |C_1| \sigma_*/M$, which is required to have
$\zeta_{\rm cur} > \zeta_{\rm mod}$ at all (up to the third) orders.
Here we assume that $\sigma_\ast > \delta \sigma_\ast = H_\ast /2\pi$,
which guarantees the linear order perturbation always dominates over
the second order one. Since $\zeta_{\rm cur}$ should be equal to the
observed amplitude of fluctuation, we require $\zeta_{\rm
  cur} \sim 10^{-5}$, which yields another condition $r_{\rm
  dec} \sim 10^{-5}\sigma_\ast/H_\ast$.  By eliminating $r_{\rm dec}$ from
the above two conditions, we find that $M$ must satisfy a bound $M >
10^5 |C_1| H_\ast$.

In Region 2, although the curvature fluctuations from the
modulated reheating give more contribution than those from the
curvaton at linear order, the curvaton dominates over the modulated
reheating both in the second and third orders.  In this case, the
relation between $f_{\rm NL}$ and $g_{\rm NL}$ can be given by
\begin{equation}
g_{\rm NL} \sim 3  r_{\rm dec}^{1/2} f_{\rm NL}^{3/2}.
\end{equation}
Although $g_{\rm NL}$ is proportional to $f_{\rm NL}^{3/2}$, the size of $g_{\rm NL}$
cannot be very large, comparing to that of $f_{\rm NL}$, due to the suppression by $r_{\rm dec}$. 
 (Notice that $r_{\rm dec} \lesssim \sigma_\ast / M$ in this region.) 
Furthermore, parameter sets in Region 2 generally give too large $f_{\rm
NL}$.  The expression for $f_{\rm NL}$ in Region 2 is approximately
given by
\begin{equation}
\frac65 f_{\rm NL} \simeq \frac{24 r_{\rm dec}}{\alpha^2} \left( \frac{\sigma_\ast}{M} \right)^{-2}.
\end{equation}
$r_{\rm dec}$ should satisfy the inequality of Eq.~\eqref{eq:mod_cur_region}, from which 
we can find 
\begin{equation}
\frac{6}{\sqrt{\sigma_\ast / M}} \lesssim \frac65 f_{\rm NL} \lesssim \frac{6}{\sigma_\ast / M}.
\end{equation}
Since here we assume that $\sigma_\ast / M  \ll 1$, $f_{\rm NL}$ is expected to be large.
For example, if we take $\sigma_\ast / M < 10^{-3}$, $f_{\rm NL} \sim 200$, which 
is already outside the current limit.
But, in fact, by fine-tuning some parameters such as $\alpha$ and $\beta$, 
we can have the situation where $f_{\rm NL} < 100$. For example, if one takes $\beta \simeq \alpha^2/ 3$,
this choice of parameters effectively enlarges Region 2 since $C_3 \sim 0$ (see Eq.~\eqref{eq:mod_cur_region}). We show such a case in Fig.~\ref{fig:fNL_gNL_diagram}.

In Region 3, the relation between $f_{\rm NL}$ and $g_{\rm NL}$
can be written as
\begin{equation}
g_{\rm NL} \sim  \left( \frac{\sigma_*}{M} \right)^{2}
 \frac{f_{\rm NL}}{r_{\rm dec}}.
\end{equation}
In this region, the inequality $(\sigma_\ast/ M)^2 /r_{\rm dec} <
1$ should be satisfied (see Eq.~\eqref{eq:mod_cur_region}), thus
$g_{\rm NL}$ tends to be smaller than $f_{\rm NL}$ as in the case of Region 2, which can be
regarded as ``suppressed $g_{\rm NL}$'' type.

Region~4 corresponds to the limit of the pure modulated reheating
case. Thus this region follows the argument in 
the previous subsection~\ref{sec:pure_modulated}.

As a final remark on this scenario, it should be noted that the linear
order term can be canceled to vanish for some parameter values of
$r_{\rm dec}$ and $\sigma_*/M$ even if $\sigma_\ast > \delta
\sigma_\ast$ is satisfied.  In this situation, by introducing the
fluctuations from the inflaton, this scenario becomes like the
ungaussiton model which is discussed later.

\subsubsection{Inhomogeneous end of hybrid inflation}
\label{subsubsec:inhomo}

In this subsection, we discuss a hybrid inflation model, where a light
scalar field other than inflaton exists and fluctuations of such a
field give cosmic density
perturbations~\cite{Bernardeau:2002jf,Bernardeau:2004zz,Lyth:2005qk,Salem:2005nd,Alabidi:2006wa}.
In usual hybrid inflation models, the inflationary phase ends due to
the tachyonic instability of a waterfall field and the end of
inflation homogeneously occurs.  If, however, such a waterfall field
is coupled with a light scalar field, then the fluctuations of the
light scalar field can drive the inhomogeneous end of inflation, from
which the curvature perturbation can be generated.

Here let us consider one of the simplest models introduced in
Ref.~\cite{Lyth:2005qk}, where the potential for the
model is given by
\begin{eqnarray}
V={\lambda \over 4}\left( \frac{v^2}{\lambda} - \chi^2\right)^2 + {1 \over 2}g^2\phi^2\chi^2 + {1 \over 2}m_\phi^2 \phi^2
+ {1 \over 2}f^2\sigma^2\chi^2 + {1 \over 2}m_\sigma^2 \sigma^2~,
\label{eq:simplepotential}
\end{eqnarray}
where $g$, $f$ and $\lambda$ are some coupling constants and  $v$ is 
a vacuum expectation value (VEV). 
 $m_\phi$ and $m_\sigma$ are the masses for the
inflaton $\phi$ and a light scalar field $\sigma$, respectively.  $\chi$ is the
waterfall field.  Here we assume that only $\sigma$ acquires
fluctuations\footnote{
For the case where fluctuations of $\phi$ also contribute to $\zeta$, which becomes a
multi-source case, see the discussion in Sec.~\ref{subsec:multi_brid}.
}.  The effective mass squared of the waterfall field is
\begin{eqnarray}
m_\chi^2 = -  v^2 + g^2 \phi^2 + f^2 \sigma^2~,
\end{eqnarray}
and then a critical value of the inflaton field corresponding to $m_\chi^2 =
0$ is given by
\begin{eqnarray}
\phi_{\rm cr} = {\sqrt{v^2 - f^2 \sigma^2} \over g}~.
\label{eq:critical}
\end{eqnarray}
Here, we have assumed that $ v^2 > f^2 \sigma^2$.  Notice that
the critical value of the inflaton can fluctuate due to the
fluctuations of the light field $\sigma$.  Assuming that the inflation
ends at $\phi = \phi_{\rm cr}$ because of the tachyonic instability of
the waterfall field (hence we adopt a sudden-end approximation), the
total $e$-folding number during inflation can be estimated as
\begin{eqnarray}
N = - {1 \over M_{\rm Pl}^2}\int^{\phi_{\rm cr}}_{\phi_*} {V \over V_\phi} d\phi~,
\label{eq:efoldsmultibrid}
\end{eqnarray}
where we have used the slow-roll approximation.  Since the
fluctuations of the light field $\sigma$ affect the perturbation of
the $e$-folding number, which corresponds to the total curvature
perturbation $\zeta$, through the fluctuation of the critical value of
the inflaton $\phi_{\rm cr}$, we have
\begin{eqnarray}
\zeta &=& {\partial N \over \partial \phi_{\rm cr}} {d \phi_{\rm cr} \over d\sigma}
\delta \sigma_*+{1 \over 2}\left[ 
{\partial^2 N \over \partial \phi_{\rm cr}^2} \left({d \phi_{\rm cr} \over d\sigma}\right)^2 +
{{\partial N \over \partial \phi_{\rm cr}} {d^2 \phi_{\rm cr} \over d\sigma^2}}\right] \delta \sigma_*^2~\nonumber\\
&&+{1 \over 6}\left[
{\partial^3 N \over \partial \phi_{\rm cr}^3} \left({d \phi_{\rm cr} \over d\sigma}\right)^3
+ 3 {\partial^2 N \over \partial \phi_{\rm cr}^2} \left({d \phi_{\rm cr} \over d\sigma}\right)\left({d^2 \phi_{\rm cr} \over d\sigma^2} \right)
+
{{\partial N \over \partial \phi_{\rm cr}} {d^3 \phi_{\rm cr} \over d\sigma^3}}\right] \delta \sigma_*^3
~.
\end{eqnarray}
Hence, the non-linearity parameters are given by
\begin{eqnarray}
{6 \over 5}f_{\rm NL} &\! = \!&  {N_{\phi\phi} \over N_\phi^2}\biggr|_{\phi=\phi_{\rm cr}} 
 + {1 \over N_{\phi}}\biggr|_{\phi=\phi_{\rm cr}} {\phi''_{\rm cr} \over {\phi'_{\rm cr}}^2} 
~,\label{eq:generalfnl}\\
{54 \over 25}g_{\rm NL} &\! = \!&  {N_{\phi\phi\phi} \over N_\phi^3}\biggr|_{\phi=\phi_{\rm cr}} 
+ 3 {N_{\phi\phi} \over N_{\phi}^3}\biggr|_{\phi=\phi_{\rm cr}}
{\phi''_{\rm cr} \over {\phi'_{\rm cr}}^2}
+ {1 \over N_{\phi}^2}\biggr|_{\phi=\phi_{\rm cr}} {\phi'''_{\rm cr} \over {\phi'_{\rm cr}}^3}
~,\label{eq:generalgnl}
\end{eqnarray}
where the prime denotes the derivative with respect to $\sigma$.  Since
in the ordinary hybrid inflation model the slow-roll conditions are
satisfied until $\phi = \phi_{\rm cr}$, we find that the first
terms in Eqs.~(\ref{eq:generalfnl}) and (\ref{eq:generalgnl}) are
suppressed by the slow-roll parameters.  In order to realize the large
non-Gaussianity in this model,  large contributions from the second
terms in Eqs.~(\ref{eq:generalfnl}) and (\ref{eq:generalgnl}) or the
third term in Eq. (\ref{eq:generalgnl}) are needed.  Neglecting the
terms suppressed by slow-roll parameters, the expressions for
non-linearity parameters are reduced to
\begin{eqnarray}
{6 \over 5}f_{\rm NL} &\! \simeq \! &  - \sqrt{2\epsilon_{\rm cr}} {\phi''_{\rm cr} \over {\phi'_{\rm cr}}^2}~,\label{eq:slorollfnl} \\
{54 \over 25}g_{\rm NL} &\! \simeq \!&
-(2\epsilon_{\rm cr} - \eta_{\rm cr}) {18 \over 5}f_{\rm NL}
+ 2\epsilon_{\rm cr} {\phi'''_{\rm cr} \over {\phi'_{\rm cr}}^3}\label{eq:slowrollgnl}
~,
\end{eqnarray}
where $\epsilon$ and $\eta$ are the slow-roll parameters defined by
\begin{equation}
\label{eq:slow_roll}
\epsilon \equiv \frac{1}{2} M_{\rm Pl}^2 \left( \frac{V_\phi}{V} \right)^2, \quad  \quad
\eta \equiv M_{\rm Pl}^2 \frac{V_{\phi\phi}}{V},
\end{equation}
and we denote the quantities evaluated at $\phi = \phi_{\rm cr}$ by a
subscript ``${\rm cr}$.''

From Eq.~(\ref{eq:critical}), we have
\begin{eqnarray}
{d \phi_{\rm cr} \over d\sigma} &\! =\!&- {f^2 \sigma \over g^2 \phi_{\rm cr}}~,\\
{d^2 \phi_{\rm cr} \over d\sigma^2} &\! =\!& - {f^2 \over g^2 \phi_{\rm cr}} \left(  { v^2 \over g^2 \phi_{\rm cr}^2}\right)~,\\ 
{d^3 \phi_{\rm cr} \over d\sigma^3} &\! =\!& -3 {f^4 \sigma \over g^4 \phi_{\rm cr}^3} \left( { v^2 \over g^2 \phi_{\rm cr}^2}\right)~.
\end{eqnarray}
Then, the non-linearity parameters can be written as
\begin{eqnarray}
{6 \over 5}f_{\rm NL} &\! = \!& \eta_{\rm cr} { v^2 \over f^2 \sigma^2 }~,\\
{54 \over 25}g_{\rm NL} &\! = &\! 6 \eta_{\rm cr}^2 {v^2 \over f^2 \sigma^2 }~,
\end{eqnarray}
where we have used $\eta = M_{\rm Pl} \sqrt{2\epsilon} / \phi$ since $
\sqrt{2 \epsilon} = M_{\rm pl} m^2_\phi \phi/ V$ and $\eta = M_{\rm
  pl}^2 m^2_\phi / V$ during inflation in this model. Furthermore we have also
adopted the approximation $\epsilon_{\rm cr} \ll \eta_{\rm cr}
\ll 1$, which is satisfied as in a usual hybrid inflation.  Hence, when
$ v^2 \gg f^2 \sigma^2 / \eta_{\rm cr}$, large $f_{\rm NL}$ can
be generated in this scenario.  Notice that, in general, fluctuations
from the inflaton would also exist, which gives another condition for
neglecting the inflaton fluctuations: $f^2 \sigma > g^2 \phi_{\rm cr}
\left( \epsilon_{\rm cr} / \epsilon_{*}\right)^{1/2}$.  This relation
can hold by taking the value of $\sigma$ appropriately.  Then, we can
find the following relation between $f_{\rm NL}$ and $g_{\rm NL}$ :
\begin{eqnarray}
\label{eq:gNL_fNL_lyth}
g_{\rm NL} = \eta_{\rm cr} {10 \over 3}f_{\rm NL}~.
\end{eqnarray}
Compared to the modulated reheating scenario, the ratio of the
non-linearity parameters $g_{\rm NL} / f_{\rm NL}$ is suppressed by the
slow-roll parameter $\eta$ in this scenario and thus this model can be regarded as
``suppressed $g_{\rm NL}$'' type.  Another simple model introduced by
Alabidi and Lyth~\cite{Alabidi:2006wa} also predicts that $g_{\rm NL} /
f_{\rm NL}$ is suppressed by the slow-roll parameter.

\subsubsection{Inhomogeneous end of thermal inflation}\label{subsubsec:thermal_inf}

Primordial fluctuations can also be generated by modulating the end of
thermal inflation \cite{Kawasaki:2009hp} (see also
\cite{Matsuda:2009yt} for the discussion on the generation of
primordial fluctuation from inhomogeneous cosmological phase
transition).  Here we briefly review the mechanism following
\cite{Kawasaki:2009hp}.

The thermal inflation \cite{Lyth:1995ka} can be realized by using a
flaton field $\phi$, a flat direction in supersymmetric theories,
whose effective potential can be given by
\begin{eqnarray}
V_{\rm eff} = V_{\rm cons} + {1 \over 2}\left( g T^2 - m^2 \right)\phi^2
 + {\lambda \over 6}{1 \over M_{\rm Pl}^2}\phi^6~,
\end{eqnarray}
where $(1/2) gT^2$ is a thermal correction to the potential due to the
interaction of the flaton field with some particles in the thermal
bath with $g$ being an effective coupling between them and $T$ being
the cosmic temperature.  When $\sqrt{g}~T > m$, the flaton field is
trapped at the minimum $\phi=0$, then at some time when the energy density of
the flaton given by $V_{\rm cons}$ becomes
larger than the background energy density, the Universe is dominated
by the false vacuum in the potential, which drives a mini-inflation.
However, when the temperature decreases down to $T_c = m/\sqrt{g}$,
the flaton rolls down to the VEV, then the mini-inflation ends.  The
total number of $e$-folding 
from the time $t_i$ when the mini-inflation begins to the time $t_f$ when the flaton 
rolls down to the VEV is given by
\begin{eqnarray}
N(t_c, t_{\rm in}) = \int^{t_c}_{t_{\rm in}} H dt
  = 
  \ln \left( {a_c \over a_{\rm in}} \right) = - \ln \left( {T_c \over T_{\rm in}} \right) ~,
\label{eq:thermalefold1}
\end{eqnarray}
where $T_{\rm in}$ is the temperature when the mini-inflation begins.
If the coupling $g$ depends on a light scalar field $\sigma$ as
$g=g(\sigma)$, the coupling $g$ can fluctuate since a light scalar
field $\sigma$ can acquire quantum fluctuations during inflation,
which gives rise to fluctuation in the number of $e$-folding, that is,
the curvature fluctuation. By utilizing the $\delta N$ formalism, the
curvature perturbation can be given by
\begin{eqnarray}
\zeta = \delta N &\! = &\! - {\delta T_c \over T_c} + {1 \over 2} \left( {\delta T_c \over T_c} \right)^2
- {1 \over 3} \left( {\delta T_c \over T_c} \right)^3
\nonumber \\
&\! = \!&{1 \over 2}\left\{
{g' \over g} \delta \sigma_*
+ {1 \over 2}\left[ {g'' \over g} - \left({g' \over g}\right)^2\right]\delta \sigma_*^2
+ {1 \over 6}\left[ 
{g^{\prime\prime\prime} \over g} - 3 {g'' g'\over g^2} 
+ 2 \left({g' \over g}\right)^3 \right] \delta \sigma_*^3  \right \}~,
\end{eqnarray}
from which we can derive the non-linearity parameters in this model as
\begin{eqnarray}
{6 \over 5}f_{\rm NL} &=& 
2 \left[ { g'' / g \over \left( g' / g \right)^2} - 1\right]~, \\
{54 \over 25} g_{\rm NL} &=& 4
\left[ 
{ g^{'''} / g \over \left( g' / g \right)^3} - 3 {g^{''} / g \over \left( g' / g \right)^2} + 
2 \right]~,
\end{eqnarray}
where the prime represents the derivative with respect to
$\sigma_\ast$.  When $g^{'''}$ is negligible, the following relation
holds between $f_{\rm NL}$ and $g_{\rm NL}$:
\begin{eqnarray}
g_{\rm NL} = -\frac{10}{3} f_{\rm NL} - \frac{50}{27}~.
\end{eqnarray}
Thus this model can be considered as ``linear $g_{\rm NL}$'' type in
terms of the $f_{\rm NL}$--$g_{\rm NL}$ relation.  Although the
mechanism is quite similar to the modulated reheating scenario and the
inhomogeneous end of hybrid inflation discussed in the previous subsection, the relation
between $f_{\rm NL}$ and $g_{\rm NL}$ is different.  In particular,
the relative sign between $f_{\rm NL}$ and $g_{\rm NL}$ differs from
the one in the modulated reheating although the size of $f_{\rm NL}$
and $g_{\rm NL}$ is almost the same order. On the other hand, the size
of $g_{\rm NL}$  is larger than that for the case of the
inhomogeneous end of hybrid inflation, where $g_{\rm NL} / f_{\rm NL}$
is suppressed by the slow-roll parameter.

\subsubsection{Modulated trapping mechanism}\label{subsubsec:modulated_trap}

During inflation, the resonant particle production can occur because
of the coupling of the inflaton to other particles
\cite{Chung:1999ve,Elgaroy:2003hp,Romano:2008rr}. If the relevant
coupling $\lambda$ depends on a light scalar field $\sigma$, dubbed as
modulaton in \cite{Langlois:2009jp}, the curvature perturbation can be
generated through fluctuations of $\sigma$ which originate from quantum
fluctuations during inflation.  Such a mechanism is called modulated
trapping, which was proposed in \cite{Langlois:2009jp}.

Here we briefly discuss this mechanism, following closely
\cite{Langlois:2009jp}, and give some relation between $f_{\rm NL}$ and
$g_{\rm NL}$ in the model. Let us consider the following
interaction for the inflaton $\phi$:
\begin{equation}
\mathcal{L}_{\rm int} = -\frac{1}{2} \mathcal{N} (m - \lambda \phi ) \bar{\chi} \chi,
\end{equation}
where $\chi$ is a fermion coupled to the inflaton and $\mathcal{N}$ is
the number of species of particles with the same mass.  When the
inflaton crosses the value $\phi = \phi_{\rm pp} = m /\lambda$, $\chi$
field becomes effectively massless, then they are resonantly produced.
At this moment, the occupation number for $\chi$ abruptly increases
from zero to
\begin{equation}
n_{\rm pp} = \frac{\lambda^{3/2}}{2 \pi^3} | \dot{\phi}_{\rm pp} |^{3/2},
\end{equation}
where ``${\rm pp}$'' indicates that the quantities are evaluated at
the time of particle production.  Once the particles are produced, its
number density just decreases with the cosmic expansion, thus after its
production, the number density can be written as
\begin{equation}
n(t) = n_{\rm pp} \left( \frac{a}{a_{\rm pp}} \right)^{-3} \Theta (t  - t_{\rm pp}).
\end{equation}
By using the above expression and adopting the Hartree approximation
in the equation of motion for $\phi$, we can evaluate the effect of
the backreaction of the particle production to $\phi$. The equation of
motion for $\phi$ can be written as
\begin{equation}
\ddot{\phi} + 3 H \dot{\phi} + \frac{dV (\phi) }{d \phi} 
= 
\mathcal{N} \lambda n_{\rm pp}  \left( \frac{a}{a_{\rm pp}} \right)^{-3} \Theta (t  - t_{\rm pp}).
\end{equation}
Assuming that the particle production and its dilution occur in a
short time scale compared to the cosmic expansion during inflation,
$H_\ast$ and $d V(\phi_\ast) /d\phi$ can be regarded as constants
during the particle production. Defining
\begin{equation}
\Delta \phi (t) \equiv \phi (t, \lambda \ne 0)  - \phi (t, \lambda=0), 
\end{equation}
then, from the equation of motion above, we obtain
\begin{equation}
\label{eq:delta_phidot}
\Delta \dot{\phi} (t > t_{\rm pp}) = 
\mathcal{N} \lambda n_{\rm pp} \exp [ - 3 H_{\rm pp} (t - t_{\rm pp}) ] ( t - t_{\rm pp}).
\end{equation}

Now we evaluate the curvature perturbation by using the $\delta N$
formalism.  The number of $e$-folding which is attributed to the particle
production can be given as
\begin{equation}
\Delta N^{(\lambda \ne 0 )} = - H_{\rm pp} \frac{\Delta \phi}{\dot{\phi}_{\rm pp}}.
\end{equation}
By integrating $\Delta \dot{\phi}$ given in
Eq.~\eqref{eq:delta_phidot} from the time of particle production, we
obtain $\Delta \phi$ as
\begin{equation}
\Delta \phi = \int_{t_\ast}^{\infty} \Delta \dot{\phi} dt 
= \frac{\mathcal{N} \lambda n_{\rm pp}}{9 H_{\rm pp}^2}.
\end{equation}
Thus $\Delta N^{(\lambda \ne 0 )}$ can be evaluated as
\begin{equation}
\Delta N^{(\lambda \ne 0 )} = - H_\ast \frac{\Delta \phi}{\dot{\phi}_{\rm pp}} 
= \frac{\lambda^{5/2} \mathcal{N} |\dot{\phi}_{\rm pp} |^{1/2}}{18 \pi^3 H_{\rm pp}},
\end{equation}
from which the curvature perturbation due to the fluctuations of
$\sigma$ is given by the formula in the $\delta N$ formalism:
\begin{equation}
\zeta 
= 
\left( \Delta N^{(\lambda \ne 0 )} \right)_{, \sigma} \delta \sigma_\ast 
+\frac{1}{2} \left( \Delta N^{(\lambda \ne 0 )} \right)_{, \sigma\sigma} (\delta \sigma_\ast)^2
+\frac{1}{6} \left( \Delta N^{(\lambda \ne 0 )} \right)_{, \sigma\sigma\sigma} (\delta \sigma_\ast)^3.
\end{equation}

To give some explicit expressions for the non-linearity parameters
such as $f_{\rm NL}$ and $g_{\rm NL}$, we need to assume the
functional form or $\sigma$ dependence of $\lambda$ and $m$, which is
somewhat model-dependent.  As discussed in \cite{Langlois:2009jp}, in
the case where $\lambda$ is independent of $\sigma$, but the mass
depends on $\sigma$ as $m= g \sigma$, the non-linearity parameters
cannot be much larger than unity. Thus we do not consider such a case
here.  When the coupling $\lambda$ and the mass $m$ depend on $\sigma$
as $\lambda = \sigma / M$ and $m = g\sigma$ with $M$ being some energy
scale, $f_{\rm NL}$ and $g_{\rm NL}$ are evaluated as
\begin{eqnarray}
\frac65 f_{\rm NL} &=& \frac{9}{5 e \beta}, \\ \notag \\
\frac{54}{25} g_{\rm NL} &=& \frac{27}{25 e^2 \beta^2}, 
\end{eqnarray}
where $\beta$ is the ``efficiency factor'', which is defined and given as
\begin{equation}
\beta \equiv 
\frac{ {\rm Max} (\Delta \dot{\phi}) }{ | \dot{\phi}_{\rm pp}|}
=
\frac{ \mathcal{N} \lambda^{5/2} |\dot{\phi}_{\rm pp} |^{1/2} } {6\pi^3 e H_{\rm pp}}.
\end{equation}
Since the particle production occurs at the cost of reducing the
kinetic energy of the inflaton, $\beta$ cannot exceed 1. If we take
$\beta \simeq 0.01$, $f_{\rm NL}$ can be as large as $f_{\rm NL} \simeq
55$ \cite{Langlois:2009jp}.  Furthermore, from the above expression,
we can find the relation between $f_{\rm NL}$ and $g_{\rm NL}$ as
\begin{equation}
\label{eq:mod_trap_relation}
g_{\rm NL} = \frac{2}{9} f_{\rm NL}^2,
\end{equation}
which is ``enhanced $g_{\rm NL}$'' type.  In fact, the relation
$g_{\rm NL} \sim f_{\rm NL}^2$ holds even if we assume somewhat
general type of functional form for $\lambda = (\sigma/M)^p$.  Thus
this mechanism predicts a large values for $g_{\rm NL}$ relative to
$f_{\rm NL}$ compared to some other ``modulated coupling'' scenarios
such as modulated reheating and inhomogeneous end of thermal
inflation, which give $|g_{\rm NL}| \sim |f_{\rm NL}|$.

\subsection{Multi-source model} \label{sec:multi_source}

Although the origin of density fluctuations is usually assumed to be a
single source, in general, it is possible that multiple sources can be
simultaneously responsible for density fluctuations ~\cite{
  Ichikawa:2008iq, Ichikawa:2008ne, Takahashi:2009cx, Langlois:2004nn,
  Moroi:2005kz, Moroi:2005np }. For example, in the curvaton scenario,
the curvaton field alone is usually supposed to generate the curvature
perturbation, however, even in this scenario, the inflaton should
exist to drive the superluminal expansion at the early epoch and it
can acquire quantum fluctuations. Thus in general, both the inflaton
and the curvaton fields can be responsible for cosmic density
fluctuations today. In a case where fluctuations from multiple sources
can simultaneously give a sizable contribution to observed ones, the
relations among the non-linearity parameters are different from those
discussed in Section~\ref{subsec:single_source}.

Here we discuss the case where two scalar fields $\phi$ and $\sigma$
generate density fluctuations. In this case, the curvature perturbation
can be generally written as
\begin{eqnarray}
\label{eq:multizeta}
\zeta
&=& N_\phi \delta \phi_\ast + N_\sigma \delta \sigma_\ast \nonumber \\
&& + \frac{1}{2} \biggl( N_{\phi\phi} (\delta \phi_\ast)^2 
   + 2 N_{\phi\sigma} \delta \phi_\ast \delta \sigma_\ast   
   + N_{\sigma\sigma} (\delta \sigma_\ast)^2 \biggr) \nonumber \\
&& + \frac{1}{6} \biggl(N_{\phi\phi\phi} (\delta \phi_\ast)^3 
   + 3 N_{\phi\phi\sigma} (\delta \phi_\ast)^2 \delta \sigma_\ast   
   + 3 N_{\phi\sigma\sigma} \delta \phi_\ast (\delta \sigma_\ast)^2   
   + N_{\sigma\sigma\sigma} (\delta \sigma_\ast)^3 \biggr).
\notag \\
\end{eqnarray}
Then the non-linearity parameters can be given by
\begin{eqnarray}
\label{eq:multinon}
f_{\rm NL} &=&
 \left( \frac{1}{1+R} \right)^2 
\left[ f_{\rm NL}^{(\phi)} + 2 R f_{\rm NL}^{(\phi\sigma)} + R^2 f_{\rm NL}^{(\sigma)} \right], \\
\tau_{\rm NL} &=&
 \left( \frac{1}{1+R} \right)^3 
\left[   \left( \frac{6}{5}f_{\rm NL}^{(\phi)} + R \frac{6}{5}f_{\rm NL}^{(\phi\sigma)} \right)^2
 + R^3  
\left( \frac{6}{5}f_{\rm NL}^{(\sigma)} + \frac{1}{R} \frac{6}{5}f_{\rm NL}^{(\phi\sigma)} \right)^2 \right], \\
g_{\rm NL} &=&
 \left( \frac{1}{1+R} \right)^3 
\left[ g_{\rm NL}^{(\phi)} + 3 R g_{\rm NL}^{(\phi\phi\sigma)} + 3
 R^2 g_{\rm NL}^{(\phi\sigma\sigma)} + R^3 g_{\rm NL}^{(\sigma)} \right],
\end{eqnarray}
where we have defined  non-linearity parameters for each
contribution as
\begin{eqnarray}
\label{eq:multinondef}
&&\frac{6}{5}f_{\rm NL}^{(\phi)} =
 \frac{N_{\phi\phi}}{ N_\phi^2 }, \quad
\frac{6}{5}f_{\rm NL}^{(\phi\sigma)} =
 \frac{N_{\phi\sigma}}{ N_\phi N_\sigma }, \quad
\frac{6}{5}f_{\rm NL}^{(\sigma)} =
 \frac{N_{\sigma\sigma}}{ N_\sigma^2 }, \nonumber \\
&&\frac{54}{25} g_{\rm NL}^{(\phi)} =
 \frac{N_{\phi\phi\phi}}{ N_\phi^3 }, \quad
\frac{54}{25} g_{\rm NL}^{(\phi\phi\sigma)} =
 \frac{N_{\phi\phi\sigma}}{ N_\phi^2 N_\sigma }, \quad
\frac{54}{25} g_{\rm NL}^{(\phi\sigma\sigma)} =
 \frac{N_{\phi\sigma\sigma}}{ N_\phi N_\sigma^2 }, \quad
\frac{54}{25} g_{\rm NL}^{(\sigma)} =
 \frac{N_{\sigma\sigma\sigma}}{ N_\sigma^3 }.
\end{eqnarray}
$R$ represents the ratio of the power spectrum for the curvature
perturbation from fluctuations of $\phi$ and $\sigma$:
\begin{equation}
\label{eq:def_R}
R \equiv  \frac{P_\zeta^{(\sigma)} (k_{\rm ref})}{P_\zeta^{(\phi)} (k_{\rm ref})},
\end{equation}
where 
\begin{equation}
P_\zeta^{(\phi)}  (k_{\rm ref}) = N_\phi^2 P_\delta  (k_{\rm ref}), ~~~~~~ 
P_\zeta^{(\sigma)}  (k_{\rm ref})= N_\sigma^2 P_\delta  (k_{\rm ref}), 
\end{equation}
with $P_\delta$ being defined in Eq.~\eqref{eq:powerdelta} and the power spectra 
are evaluated at some reference scale $k_{\rm ref}$. 
 Then, we obtain the following relation,
\begin{eqnarray}
\label{eq:multiineq}
  \tau_{\rm NL} = \left( \frac{1+\overline{R}}{\overline{R}} \right)
                  \left( \frac65 f_{\rm NL} \right)^2 
                  \ge \left( \frac65 f_{\rm NL} \right)^2,
\end{eqnarray}
where the ratio $\overline R$ is defined as
\begin{eqnarray}
\label{eq:overRdef}
  \overline{R} \equiv \frac{1}{R} 
     \left( \frac{R_{f_{\rm NL}}+R^2}{R_{f_{\rm NL}}-R} \right)^2 \ge 0,
\end{eqnarray}
with       
\begin{eqnarray}
\label{eq:RfNLdef}
   R_{f_{\rm NL}} \equiv \frac{f_{\rm NL}^{(\phi)}+Rf_{\rm NL}^{(\phi\sigma)}}
                      {f_{\rm NL}^{(\sigma)}+ (1/R)f_{\rm NL}^{(\phi\sigma)}}.
\end{eqnarray}
These formulae can be applied for a general two-field case.  

\subsubsection{Mixed model with inflaton fluctuations} \label{sec:mixed_inf}

In a special case where $\phi$ is the inflaton field and $\sigma$ is
some other light field, $\phi$ does not contribute to non-Gaussianity
because non-linearity parameters are suppressed by the slow-roll of
$\phi$. Then, the curvature perturbation can be written as
\begin{eqnarray}
\label{eq:zetainflaton}
\zeta &=& \zeta^{(\phi)} + \zeta^{(\sigma)} \notag \\
&=& N_\phi \delta \phi_\ast 
+N_\sigma \delta \sigma_\ast 
+ \frac{1}{2} N_{\sigma\sigma} (\delta \sigma_\ast)^2 
+ \frac{1}{6} N_{\sigma\sigma\sigma} (\delta \sigma_\ast)^3,
\notag \\
\end{eqnarray}
where $\zeta^{(\phi)}$ and $\zeta^{(\sigma)}$ represent the
contributions to the curvature perturbation from $\delta \phi$ and
$\delta \sigma$, respectively\footnote{
  In some literature, the notation $\zeta_\phi$ indicates that the
  curvature perturbation on the slice where the energy density of $\phi$
  is uniform.  Here $\zeta^{(\phi)}$ just represents the contribution to
  $\zeta$ from $\delta \phi$. 
}.  In most cases, the inflaton and the other light scalar field can
be considered to be uncorrelated and non-linearity in the inflaton sector 
is generally suppressed by the slow-roll, here we set $f_{\rm NL}^{(\phi)} =
f_{\rm NL}^{(\phi\sigma)} = 0$, which indicates that $R_{f_{\rm NL}}
=0$ and $\overline{R} = R$.  Furthermore, in this case, we can also
set $g_{\rm NL}^{(\phi)} = g_{\rm NL}^{(\phi\phi\sigma)} = g_{\rm
  NL}^{(\phi\sigma\sigma)} = 0$.  Then we have the relation among
$f_{\rm NL}, \tau_{\rm NL}$ and $g_{\rm NL}$ as
\begin{equation}
\tau_{\rm NL}
= \left({1 + R \over R}\right) \left({6 \over 5}f_{\rm NL}\right)^2~,
\label{eq:mixedft}
\end{equation}
and
\begin{equation}
\label{eq:mixed_cur_fNL_gNL_tauNL}
{g_{\rm NL}\tau_{\rm NL}  \over f_{\rm NL}^3}
=
\left( \frac{6}{5} \right)^2 
\frac{g_{\rm NL}^{(\sigma)}}{f_{\rm NL}^{(\sigma)}}.
\end{equation}
This equation can be applied for a scenario where one of two sources
of fluctuations alone contributes to non-Gaussianity.  Also notice
that when $R$ is small, it corresponds to the situation where
$\tau_{\rm NL} \gg (6f_{\rm NL} / 5)^2$.  Thus large $\tau_{\rm NL} /
f_{\rm NL}^2$ may indicate that (almost) Gaussian field gives a large
contribution to the power spectrum, but non-Gaussianity comes from 
the other source\footnote{
  Here it should also be noted that, for certain parameter values of
  this kind of mixed models, the second order fluctuations give the
  leading contribution in $\zeta^{(\sigma)}$.  In this case, the model
  becomes like  ``ungaussiton'' model, categorized as
  ``constrained multi-field type,'' which will be discussed in the later
  section.
}.

In most models, the relation between $f_{\rm NL}^{(\sigma)}$ and
$g_{\rm NL}^{(\sigma)}$ for a single field sector can be given by the
following form:
\begin{equation}
\label{eq:gNL_fNL_general}
g_{\rm NL}^{(\sigma)} = C_{1\sigma} \left( f_{\rm NL}^{(\sigma)} \right)^p+ C_{2\sigma},
\end{equation}
where $C_{1\sigma}$ and $C_{2\sigma}$ are numerical constants and $p$ is
the power of $f_{\rm NL}^{(\sigma)}$.  For example, in the curvaton
model with a quadratic potential, these parameters are given as
$C_{1\sigma}= -(10/3), C_{2\sigma} = -(575/108)$ and $p=1$ as can be
read off from Eq.~\eqref{eq:fNL_gNL_pure_cur}. By assuming the form
of Eq.~\eqref{eq:gNL_fNL_general} for the $\sigma$ sector, we can
write the relation between $f_{\rm NL}$ and $g_{\rm NL}$ for the mixed
model with the inflaton as,
\begin{equation}
g_{\rm NL} = C_{1\sigma} \left( \frac{R}{1+R} \right)^{3-2p} f_{\rm NL}^p + C_{2\sigma}  \left( \frac{R}{1+R} \right)^{3}.
\end{equation}
We can also write this equation by using $\tau_{\rm NL}$ and
eliminating $R$ as,
\begin{equation}
\label{eq:gNL_fNL_multi}
g_{\rm NL} = \left( \frac{6}{5} \right)^{6-4p} C_{1\sigma} \frac{f_{\rm NL}^{6-3p}}{\tau_{\rm NL}^{3-2p}}
 +  \left( \frac{6}{5} \right)^{6}C_{2\sigma}  \frac{f_{\rm NL}^{6}}{\tau_{\rm NL}^{3} }.
\end{equation}

As an explicit example, first we give the formula for the case of the
mixed curvaton and inflaton fluctuations.  As mentioned above, even in
the curvaton scenario, the fluctuations from the inflaton can also
give some contribution to the curvature perturbations as well as those
from the curvaton. Here we denote $\sigma$ as the curvaton field.  As discussed in
Section~\ref{subsec:pure_curvaton}, assuming that the potential of the
curvaton is quadratic, $f_{\rm NL}^{(\sigma)}$ and $g_{\rm
  NL}^{(\sigma)}$ are related as Eq.~\eqref{eq:fNL_gNL_pure_cur} for
$r_{\rm dec} \ll 1$.  Putting the relation into
Eq.~\eqref{eq:mixed_cur_fNL_gNL_tauNL} or \eqref{eq:gNL_fNL_multi}, we
find the relation among the non-linearity parameters for the case with $r_{\rm dec} \ll 1$ as
\cite{Ichikawa:2008iq},
\begin{equation}
\label{eq:fNL_gNL_tauNL_mixed_cur}
g_{\rm NL} = -\frac{24}{5} \frac{f_{\rm NL}^3}{\tau_{\rm NL}} 
- \frac{9936}{625} \frac{f_{\rm NL}^6}{\tau_{\rm NL}^3}.
\end{equation}
This relation can be rewritten as
\begin{equation}
  g_{\rm NL} = -\frac{10}{3} \left( \frac{R}{1+R} \right) f_{\rm NL} - \frac{575}{108}\left( \frac{R}{1+R} \right)^3 ,
\end{equation}
which recovers the pure curvaton case Eq.~\eqref{eq:fNL_gNL_pure_cur} in
the limit of $R \rightarrow \infty$. When $R\sim 1$, $g_{\rm NL}$ is of the same order
of $f_{\rm NL}$, but when $R \ll 1$, $g_{\rm NL}$ is suppressed 
by the factor $R$.

A similar relation can be obtained for a mixed model of the modulated
reheating and the inflaton fluctuations.  As
discussed in Section~\ref{sec:pure_modulated}, when the term with
$\Gamma_{\sigma\sigma\sigma}$ is negligible, $f_{\rm NL}^{(\sigma)}$
is related to $g_{\rm NL}^{(\sigma)}$ as in
Eq.~\eqref{eq:fNL_gNL_modulated}.  By adopting the relation, we obtain
the corresponding equation in this model as \cite{Ichikawa:2008ne}
\begin{equation}
\label{eq:fNL_gNL_tauNL_mixed_mod}
g_{\rm NL} 
= \frac{72}{5} \frac{f_{\rm NL}^3}{\tau_{\rm NL}}
- 
\frac{31104}{625} \frac{f_{\rm NL}^6}{\tau_{\rm NL}^3} 
= 10 \left( \frac{R}{1+R} \right) f_{\rm NL} 
  - \frac{50}{3} \left( \frac{R}{1+R} \right)^3,
\end{equation}
which reduces to the pure modulated reheating case
(\ref{eq:fNL_gNL_modulated}) in the limit of $R \rightarrow \infty$.
Notice that $g_{\rm NL}$ is again suppressed by $R$ when $R \ll 1$.

Another example is the modulated trapping model with a sizable
contribution from the inflaton fluctuations.  Here $\sigma$ is assumed
to be modulaton.  For some functional forms of $\lambda$ and $m$, the
relation between $f_{\rm NL}^{(\sigma)}$ and $g_{\rm NL}^{(\sigma)}$
can be given as Eq.~\eqref{eq:mod_trap_relation}.  In this model, the
following relation holds:
\begin{equation}
g_{\rm NL} 
= \frac{2}{9} \left( \frac{1+R}{R} \right) f_{\rm NL}^2 
= \frac{25}{162} \tau_{\rm NL}.
\end{equation}

In general, the inflaton fluctuations can also contribute to the
primordial fluctuations even if we consider another source. 
The discussion above is applicable 
for other models  regarding $\sigma$ as the one in other single-source
models.  If the corresponding single-source model predicts the $f_{\rm
  NL}$--$g_{\rm NL}$ relation of the linear type, the counterpart of
its mixed model with the inflaton gives ``suppressed $g_{\rm NL}$''
type due to the ratio $R$ when $R \ll 1$, although $|f_{\rm NL}| \sim |g_{\rm NL}|$ 
when $R \sim 1$.

As a final remark, we comment on the spectral index $n_s$ and tensor-to-scalar ratio $r$
in this type of mixed models.
 $n_s$ and $r$ in this scenario are  generally given by
\begin{eqnarray}
n_s -1  & = &  -2 \epsilon - \frac{4 \epsilon -2 \eta}{1 +R}, \\
r  & = &  \frac{16 \epsilon}{1+R},
\end{eqnarray}
where $\epsilon$ and $\eta$ are slow-roll parameters for the inflaton defined as in 
Eq.~\eqref{eq:slow_roll}. 
The limit of $R \rightarrow 0$ corresponds to the 
case where only inflaton fluctuations are responsible for today's density 
fluctuations. 

On the other hand, the limit of $R \rightarrow \infty$ corresponds to a single-source model. 
As seen from the above equation, the tensor-to-scalar ratio $r$ 
becomes very small in this limit. Thus $r$ is considered to be generally very 
small in  models  with large non-Gaussianity. However, when $R\sim 1$, 
large $r$ is possible while non-Gaussianity can also be large.
Thus  a mixed model discussed in this section is interesting in this respect as well.

\subsubsection{Multi-curvaton model}\label{sec:multi_cur}

In a usual curvaton model, there exists only one curvaton
field. However, it may also be possible that multiple fields can play
a role of the curvaton.  Some authors have already considered this
kind of the scenario \cite{Choi:2007fya,Assadullahi:2007uw}, which is
called ``multi-curvaton'' model.  Here we discuss this model focusing
on its non-linearity parameters.
 
In the following, we basically follow the arguments in
\cite{Assadullahi:2007uw}. Here we adopt the sudden decay approximation and 
assume that two curvaton fields, denoted as $a$ and $b$, are
responsible for the present-day density fluctuations. We also assume
that the curvaton field $a$ decays first (i.e., the decay rates for
$a$ and $b$ are assumed to be $\Gamma_a > \Gamma_b$).  At an
infinitesimal time before and after the first curvaton decay, the total
curvature perturbation $\zeta_1$ at the first curvaton decay, the curvature perturbations on the
constant curvaton $a$ and $b$ hypersurfaces $\zeta_a$ and $\zeta_b$
are related as
\begin{eqnarray}
\label{eq:two_curvaton1}
(1 - \Omega_{a1}  - \Omega_{b1} ) e^{-4 \zeta_1}  
+ \Omega_{a1} e^{3 ( \zeta_a - \zeta_1)} + \Omega_{b1} e^{3 ( \zeta_b - \zeta_1)}& = & 1,\\
\label{eq:two_curvaton2}
(1   - \Omega_{b 1} ) e^{4 (\zeta_{\gamma 1} - \zeta_1)}  
 + \Omega_{b1} e^{3 ( \zeta_b - \zeta_1)}& = & 1,
\end{eqnarray}
where $\Omega_{a1} = \rho_{a1} / \rho_1$ and  $\Omega_{b1} = \rho_{b1} / \rho_1$
with $ \rho_{1}, \rho_{a1}$ and  $\rho_{b1}$ being energy densities of the total component, 
the curvaton $a$ and $b$ at the first curvaton decay, respectively.
At the time just before the second curvaton decay, the
following equation holds:
\begin{eqnarray}
\label{eq:two_curvaton3}
(1   - \Omega_{b 2} ) e^{4 (\zeta_{\gamma 1} - \zeta_2)}  
 + \Omega_{b2} e^{3 ( \zeta_b - \zeta_2)}& = & 1,
\end{eqnarray}
where $\Omega_{b2} = \rho_{b2} / \rho_2$ with $ \rho_{2}$ and
$\rho_{b2}$ being energy densities of the total component and the
curvaton $b$ at the second curvaton decay, respectively.  Here
$\zeta_{\gamma_ 1}$ is the curvature perturbation on the constant
radiation hypersurface at the first curvaton decay.  $\zeta_2$ is the
curvature perturbation after the second curvaton decay.  Here the
subscripts $1$ and $2$ indicate that the quantities are the ones at the
time of the first and the second curvatons (the curvatons $a$ and $b$)
decays, respectively.

Since we are interested in the final
curvature perturbation after the second curvaton decays, we evaluate
$\zeta_2$ up to third order, then find non-linearity parameters in the
model. The curvature perturbations $\zeta_a$ and $\zeta_b$ are related
to the field perturbations of the curvatons $a$ and $b$, denoted as $\delta a$ 
and $\delta b$,  by
\begin{eqnarray}
\label{eq:zeta_a}
\zeta_a & = & \frac{1}{3} \log ( 1 + \delta_a) 
= \frac{2}{3} \frac{\delta a}{a} - \frac{1}{3} \left( \frac{\delta a}{a} \right)^2
+ \frac{2}{9} \left( \frac{\delta a}{a} \right)^3,  \\
\label{eq:zeta_b}
\zeta_b & = & \frac{1}{3} \log ( 1 + \delta_b)
= \frac{2}{3} \frac{\delta b}{b} - \frac{1}{3} \left( \frac{\delta b}{b} \right)^2
+ \frac{2}{9} \left( \frac{\delta b}{b} \right)^3,
\end{eqnarray}
where $\delta_i = \delta \rho_i / \rho_i$ and we truncated at the third order. 
Here we have assumed that
the potentials for the curvatons are quadratic. Hence  the field values $a,b$ and 
their perturbations $\delta a, \delta  b$ are regarded as the ones evaluated at the time of 
horizon crossing.  

By using an iterative method, we can express
$\zeta_2$ as the series in $\zeta_a$ and $\zeta_b$ . 
Denoting the first order parts for the curvatons $a$ and $b$ in 
Eqs.~\eqref{eq:zeta_a} and \eqref{eq:zeta_b} as $\zeta_{a(1)}$ and $\zeta_{b(1)}$, respectively, 
we can generally write $\zeta_2$ as 
\begin{eqnarray}
\zeta_2 &=& C_{a}  \zeta_{a(1)} + C_{b} \zeta_{b(1)} 
+ C_{aa} \zeta_{a(1)}^2  +  C_{bb} \zeta_{b(1)}^2 + C_{ab} \zeta_{a(1)} \zeta_{b(1)}  \notag \\
&&+  C_{aaa} \zeta_{a(1)}^3 + C_{bbb}  \zeta_{b(1)}^3
+ C_{aab}   \zeta_{a(1)}^2 \zeta_{b(1)} + C_{abb}  \zeta_{a(1)} \zeta_{b(1)}^2.
\end{eqnarray} 
Here the coefficients such as $C_{a}$ and so on are functions of $f_{a1}, f_{b1}$ and 
$f_{b2}$ which are defined as
\begin{eqnarray}
f_{a1}  & = &  
\frac{3 \Omega_{a1}}{4  -  \Omega_{a1} - \Omega_{b1}}, \\
f_{b1}  & = &  
\frac{3 \Omega_{b1}}{4  -  \Omega_{a1}  - \Omega_{b1}}, \\
f_{b2}  & = &  
\frac{3 \Omega_{b2}}{4  - \Omega_{b2}}.
\end{eqnarray}
These quantities roughly represent the
ratio of the energy density of the curvatons $a$ and $b$ to the total
one at the first and second curvaton decays.
 
Since the expression for the  coefficients are very
complicated in general, here we consider only some limiting cases where
non-Gaussianity can be large.  We give the full expression of the
curvature perturbation in Appendix~\ref{app:multi_cur}.

\bigskip
\bigskip
\bigskip
\bigskip
\bigskip
\bigskip

\noindent 
$\bullet$ {\bf Case I: Both curvatons are subdominant at their decays} \\

When the curvatons are both subdominant at their decays, 
which corresponds to the case of $f_{a1}, f_{b2} \ll 1$  (and $ f_{b1} < f_{b2}$ being implicitly assumed),
the curvature perturbation after the second curvaton decay can be given by
\begin{eqnarray}
\zeta_2 &=& f_{a1} \zeta_{a(1)} + f_{b2} \zeta_{b(1)} 
+ \frac{3f_{a1}}{4} \zeta_{a(1)}^2  + \frac{3f_{b2}}{4} \zeta_{b(1)}^2 \notag \\
&&
- \frac{3f_{a1}^2}{2} \zeta_{a(1)}^3 - \frac{3f_{b2}^2}{2} \zeta_{b(1)}^3
-\frac{9}{4} f_{a1} f_{b2} \zeta_{a(1)}^2 \zeta_{b(1)} 
-\frac{9}{4} f_{a1} f_{b2} \zeta_{a(1)} \zeta_{b(1)}^2,
\end{eqnarray}
where we have kept $f_{a1}$ and $f_{b2}$ only at the leading order.

\begin{figure}[htbp]
  \begin{center}
    \resizebox{120mm}{!}{
    \includegraphics{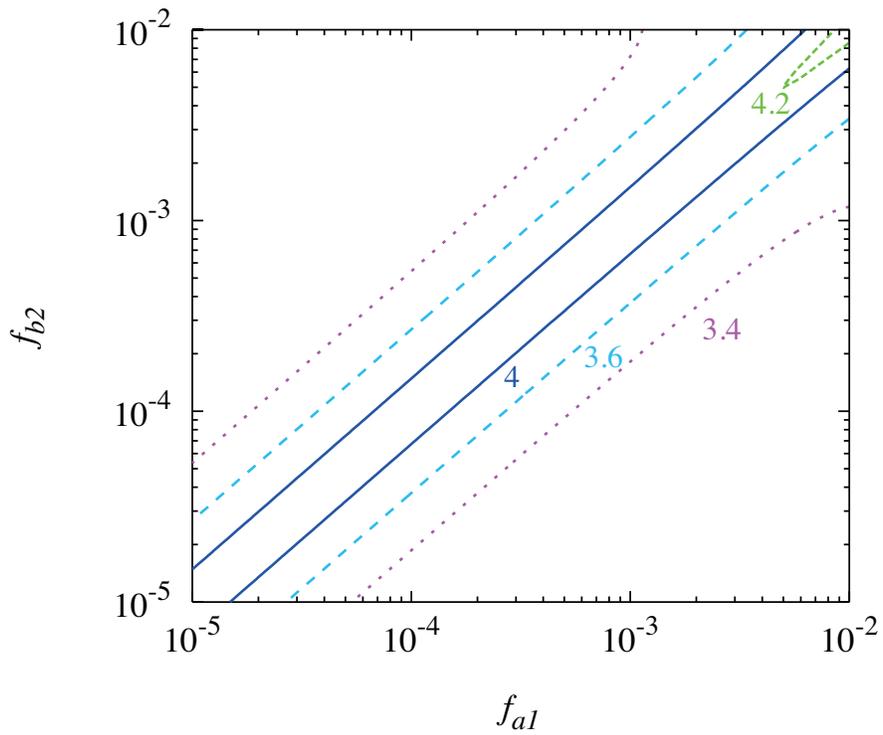}
    }
  \end{center}
  \caption{Contours of $-C_{\rm mc}$ in the $f_{a1}$--$f_{b2}$ plane. Here we take $K=1$.
  Notice that, in the lower right and upper left region, $C_{\rm mc} \rightarrow - (10/3)$, which 
  corresponds to the single curvaton case. }
  \label{fig:multi_curvaton}
\end{figure}

Then we can find the non-linearity parameters for this case as
\begin{eqnarray}
\frac65 f_{\rm NL} & \simeq& \frac{3}{2} \frac{f_{a1}^3 + f_{b2}^3 K^2}{(f_{a1}^2 + f_{b2}^2 K)^2}, \\
\tau_{\rm NL} & \simeq& \frac94 \frac{f_{a1}^4 + f_{b2}^4 K^3}{(f_{a1}^2 + f_{b2}^2 K)^3}, \\
\frac{54}{25}g_{\rm NL} & \simeq & 
- 9\, \frac{ f_{a1}^5 +  3 f_{a1}^3 f_{b2}^2 K +3 f_{a1}^2 f_{b2}^3 K^2 +  f_{b2}^5 K^3}{(f_{a1}^2 + f_{b2}^2 K)^3}, 
\end{eqnarray}
where $K$ is the ratio of the power spectra of the curvatons $a$ and $b$, 
denoted as $P_{\zeta_a}$ and $P_{\zeta_b}$, respectively:
\begin{equation}
K = \frac{P_{\zeta_b}}{P_{\zeta_a}}.
\end{equation}
With this definition, the ratio $R$ defined by Eq.~\eqref{eq:def_R} in Section~\ref{sec:multi_source} 
can be written as 
\begin{equation}
R \simeq \frac{f_{b2}^2}{f_{a1}^2} K.
\end{equation}
In the language of Section~\ref{sec:multi_source}, we have
\begin{eqnarray}
&&  \tau_{\rm NL} = \left( \frac{1+\overline{R}}{\overline{R}} \right)
                  \left( \frac65 f_{\rm NL} \right)^2, \qquad
  \overline{R} = \frac{1}{R} \left[
                 \frac{f_{a1}^3+f_{b2}^3 K^2}{f_{a1}^2 (f_{a1}-f_{b2} K)} 
                 \right]^2. 
 \end{eqnarray}        
Regarding the relation between $g_{\rm NL}$ and $f_{\rm NL}$, 
we find                 
\begin{equation}
 \frac{10}{3} f_{\rm NL} < - g_{\rm NL} < 10 f_{\rm NL}.
\end{equation}
Thus, $g_{\rm NL}$ is of the same order of $f_{\rm NL}$ with the
opposite sign in this case and we could write as
\begin{equation}
g_{\rm NL} =   C_{\rm mc} (f_{a1}, f_{b2}, K) f_{\rm NL}, 
\end{equation}
where $C_{\rm mc}$ is a negative coefficient of $\mathcal{O}(1)$, which
slightly depends on the parameters $f_{a1}, f_{b2}$ and $K$. In
Fig.~\ref{fig:multi_curvaton}, contours of $-C_{\rm mc}$ are plotted in the $f_{a1}$--$f_{b2}$ plane, from which
we can see that $C_{\rm mc}$ is generally $\mathcal{O}(1)$.

We also find the following inequalities for
three non-linearity parameters:
\begin{equation}
 - \frac{125}{27} \le \frac{f_{\rm NL} g_{\rm NL}}{\tau_{\rm NL}}  \le 0,
\end{equation}
\begin{equation}
\frac{\tau_{\rm NL} g_{\rm NL}}{f_{\rm NL}^3}  \le - \frac{24}{5},
\end{equation}
which may be useful for discriminating this case from other ones.

\bigskip
\bigskip
\bigskip
\bigskip

\noindent 

$\bullet$ {\bf Case II: Both curvatons are dominant at their decay} \\

It was first noted in \cite{Assadullahi:2007uw} that large
non-Gaussianity can be generated even if the curvatons are dominant when
it decays. This is possible only when there are two curvatons.  For this
case, we assume that $f_{a1} =1$ and $f_{b1}\ll 1$.  Although we also
assume that $f_{b2} \simeq 1$, we keep $f_{b2}$ up to the leading order,
i.e., we expand the expressions for $\zeta_2$ and non-linearity
parameters around $f_{b2}=1$. The limit $K \rightarrow \infty$
corresponds to the case where the first curvaton effectively homogeneous
and only the second curvaton fluctuates.  This situation is exactly the
same as the standard (one-field) curvaton case.  However, the case of $K
\rightarrow 0$ is quite different from the standard curvaton model,
which we are going to consider here.  Taking the limit $ K \rightarrow
0$, which corresponds to the case where the second curvaton is
effectively homogeneous, the curvature perturbation is given by
\begin{eqnarray}
\zeta_2 &=&  (1- f_{b2}) \zeta_{a(1)} + f_{b2} \zeta_{b(1)} 
+ \frac{5}{4(1- f_{b2} )}   [ (1- f_{b2}) \zeta_{a(1)} + f_{b2} \zeta_{b(1)} ]^2 \notag \\
&& \quad \quad \quad \quad \quad
+ \frac{5}{12(1-f_{b2} )^2} [ (1- f_{b2}) \zeta_{a(1)} + f_{b2} \zeta_{b(1)} ]^3,
\end{eqnarray}
from which we can find that this case effectively reduces to the single-source case
by regarding $ (1- f_{b2}) \zeta_{a(1)} + f_{b2} \zeta_{b(1)} $ being a Gaussian field.
The non-linearity parameters for this case can be evaluated as
\begin{eqnarray}
\frac65 f_{\rm NL} &=& \frac{5}{2 ( 1- f_{b2})},\\ \notag \\
\frac{54}{25}g_{\rm NL} &=& \frac{5}{2(1 - f_{b2} )^2 }.
\end{eqnarray}
Then we have the following consistency relations:
\begin{eqnarray}
\tau_{\rm NL} &=& \frac{36}{25} f_{\rm NL}^2, \\
g_{\rm NL}  &=& \frac{4}{15} f_{\rm NL}^2.
\end{eqnarray}
In this case,  $g_{\rm NL}$ is proportional to $f_{\rm NL}^2$ and
hence may become relatively large, which could be categorized as 
``enhanced $g_{\rm NL}$'' type.

\subsubsection{Multi-field inflation model} \label{subsec:multi_brid}

As is well known, the non-linearity parameters in the standard single
field inflation models are suppressed by the slow-roll parameters
evaluated at the time of horizon crossing. Due to the special property
of the single field model that the curvature perturbation is conserved
on super-horizon scales, the breakdown of the slow-roll conditions at
the end of inflation does not induce additional corrections to the
non-linearity parameters.  One may then expect that the situation
changes either if we go on to multi-field inflation models where
more than one field contribute to the inflation dynamics, or if more
than one field enter the game at the end of inflation where the
slow-roll conditions are violated\footnote{The special cases belonging
to this class have already been discussed in \ref{subsubsec:inhomo} and
\ref{subsubsec:thermal_inf}.}.  This is what we want to address in this
subsection.

We first briefly review the general argument which 
shows that, from a naive order counting, 
$f_{\rm NL}, \tau_{\rm NL}$ and $g_{\rm NL}$ generated during the slow-roll inflation are
suppressed by the slow-roll parameters. Then we consider a counter
example given in \cite{Byrnes:2008wi,Byrnes:2008zy,Byrnes:2010em}, where
the non-linearity parameters can be large during the slow-roll
inflation. Then, we also consider another multi-field inflation model
called multi-brid inflation model, where large non-Gaussianity is
generated at the end of inflation.

We set $M_{\rm Pl}$ to be unity only in this subsection to simplify the
equations.

\bigskip
\bigskip

\noindent
$\bullet$ {\bf General slow-roll multi-field inflation model} \label{subsec:multi_field} \\

Here we just present the formula for $f_{\rm NL}$ in a (multi-field)
slow-roll inflation model with canonical kinetic terms.  The formula for $f_{\rm NL}$
at some time $t_{\rm f}$ during inflation when the slow-roll
approximation is valid can be written in terms of the potential of scalar
fields $V$ as \cite{Yokoyama:2007uu,Tanaka:2010km}
\begin{eqnarray}
\frac{6}{5} f_{\rm NL} &= & (N^d_* N_{d*})^{-2}
\Big[ 
N_{ab}^{\rm f}\Theta^a(N_{\rm f})\Theta^b(N_{\rm f})
+\int^{N_{\rm f}}_{N_*} dN\, N_a(N) Q^a_{~bc}(N) \Theta^b(N)\Theta^c(N) \Big], \label{formula}
\end{eqnarray}
where $N^a_* = N^a(N_*)$ and
$N_{ab}^{\rm f}$ is given by
\begin{eqnarray}
N_{ab}^{\rm f} &=& \left({V^2 \over {V'}^2}\right)_{N=N_{\rm f}} 
\left[\eta_{ab} + 2 {V^cV^dV_aV_b \over \left({V'}^2\right)^2}\eta_{cd} + {V_a V_b \over V^2} 
- 4 {\eta_{c(a}V_{b)}V^c \over {V'}^2} \right]_{N=N_{\rm f}}~.
\end{eqnarray}
Here we introduced a notation for tensor indices as $t_{(ab)} = (1/2) (
t_{ab} + t_{ba})$ and $V_a = \partial V / \partial \phi^a$, $\eta_{ab} =
V_{ab} / V$ and $V'^2 = V_a V^a$.  All the quantities on the right hand
side in Eq.~(\ref{formula}) are evaluated on the background trajectory
in field space and the e-folding number is used as the time coordinate
($N_{\rm f}$ and $N_*$ are respectively the e-folding numbers
corresponding to $t_{\rm f}$ and $t_*$).  Here, we have taken the final
hypersurface at $N=N_{\rm f}$ to be the $V =$ constant one, which is
approximately corresponding to the uniform energy density one under the
slow-roll approximation.  The index is raised by the inverse of the
field space metric, which is assumed to be $\delta^{ab}$ here.  To
define $N_a$ and $\Theta^a$, it is convenient to introduce the
propagator
\begin{equation}
\Lambda^a_{~b}(N,N')=\left[T\exp\left(\int_{N'}^N P(N'')dN''\right)\right]^a_{~b},
\label{Tproduct}
\end{equation}
where $T$ means that the matrices $P^a_{~b}$ are ordered in time when the exponential is expanded in power of $P^a_{~b}$ and
\begin{equation}
P^a_{~~b}(N) \equiv -{V^a_{~~b}\over V}+{V^a V_b \over V^2}.
\end{equation}

Using this propagator, $N_a$ and $\Theta^a$ are defined as 
\begin{equation}
N_a(N)=N_b^{\rm f} \Lambda^b_{~a}(N_{\rm f},N), \hspace{5mm} \Theta^a(N)\equiv\Lambda^a_{~b}(N,N_*)N^b_*, 
\end{equation}
with
\begin{eqnarray}
N_b^{\rm f} = 
\left. {V V_b \over V_a V^a} \right|_{N=N_{\rm f}}~.
\label{eq:boundna}
\end{eqnarray}
$N_a(N)$ and $\Theta^a(N)$ respectively satisfy the following equation of motion:
\begin{eqnarray}
{d\over dN}N_a(N) &=& -P_a^{~b} N_b(N), \\
{d\over dN}\Theta^a(N) &=& P_b^{~a} \Theta^b(N).
\end{eqnarray}
The boundary conditions  are  given by 
$N_a(N_{\rm f})=N_a^{\rm f}$ and $\Theta^a(N_*)=N^a(N_*)$, respectively.
The three point interaction $Q^a_{~bc}$ is given by 
\begin{equation}
Q^a_{~bc}(N)\equiv -{V^a_{~~bc}\over V}+{V^a_{~~b}V_c\over V^2}+{V^a_{~c} V_b\over V^2}+{V^a V_{bc}\over V^2}-2{V^a V_b V_c\over V^3}. 
\end{equation}
In a naive sense, the slow-roll conditions require that the potential of
the inflaton is a smooth function of $\phi^a$ and hence higher order
differentiations with respect to $\phi^a$ are more suppressed\footnote{
The assumption of this kind of scaling would be natural to obtain an almost scale-invariant spectrum
although it is not strictly required~\cite{Stewart:2001cd}. 
}.  Using the similar slow-roll parameters as defined in
Eq.~\eqref{eq:slow_roll} and assuming those for every
field as of the same order $\mathcal{O}(\epsilon)$, $P^a_{~b}$ and
$Q^a_{~bc}$ can be estimated to be of $O(\epsilon)$ and
$O(\epsilon^{3/2})$, respectively.  Since the duration of the inflation
is roughly estimated as
\begin{equation}
N=O\left({V\over dV/dN}\right)=O\left( {H V\over V'\dot\phi}\right)=O\left(\epsilon^{-1}\right),
\end{equation}
the exponential factor in $\Lambda^a_{~b}$ defined in Eq.~\eqref{Tproduct} is $\mathcal{O}(1)$.
Furthermore, the value of $N_a^{\rm f}$ is estimated as
\begin{equation}
N_a^{\rm f}= {V V_a \over V^b V_b} =O(\epsilon^{-1/2}). 
\end{equation}
Hence, one can see that $N_a$ and $\Theta^a$ are $O(\epsilon^{-1/2})$. 
Substituting these estimates into Eq.~(\ref{formula}), we find that 
$
 f_{NL}=O(\epsilon).
$
This rough estimate  indicates that $f_{\rm NL}$ is typically smaller than ${\cal O}(1)$
within the range of validity of our present approximation.

However, when there are some large hierarchy between the slow-roll parameters, 
which is possible for a multi-field case, 
$f_{\rm NL}$ can be much larger than $\mathcal{O}(\epsilon)$.
For example, let us consider the case where
there is huge difference in size between $\epsilon_1$  and $\epsilon_2$ 
(which are slow-roll parameters for $\phi_1$ and $\phi_2$, respectively).
Assuming $\epsilon_1 \gg \epsilon_2$, $f_{\rm NL}$ can be 
estimated as $f_{\rm NL} \sim \mathcal{O}(\epsilon) \times (\epsilon_1 / \epsilon_2)$,
which can be large even though $\epsilon_1, \epsilon_2 < \mathcal{O}(1)$.
Hence, a multi-field inflation model can generate large non-Gaussianity 
although a typical estimate of $f_{\rm NL}$ is $\mathcal{O}(\epsilon)$.

Furthermore, the formulation given above is applicable only when 
the slow-roll condition is satisfied. However, when the slow-roll 
condition is violated, in particular, at the end of hybrid inflation, 
one could have large $f_{\rm NL}$. Some explicit example models of 
large $f_{\rm NL}$ will be discussed in the following.

The above discussion on $f_{\rm NL}$, of course, can be applied to the
higher order non-linearity parameters, i.e., $\tau_{\rm NL}$ and $g_{\rm
NL}$.  Following Ref.~\cite{Yokoyama:2008by}, $\tau_{\rm NL}$ and
$g_{\rm NL}$ are respectively given by
\begin{eqnarray}
\tau_{\rm NL} &=& \frac{1}{\left(N_*^bN_{*b}\right)^3}
\Omega_{a}(N_*)\Omega^a(N_*)~, \\
{54 \over 25}g_{\rm NL} &=& \frac{1}{\left(N_*^bN_{*b}\right)^3}
\biggl[
N_{abc}^{\rm f} \Theta(N_{\rm f})\Theta^b(N_{\rm f})\Theta^c(N_{\rm f}) + 
\int^{N_{\rm f}}_{N_*}dN N_{a}(N)S^{a}_{~bcd}(N)\Theta^b(N)\Theta^c(N)\Theta^d(N)
~\nonumber\\
&& \qquad\qquad\qquad\qquad\qquad
+3\int^{N_{}\rm f}_{N_*}dN\Omega_a(N)Q^a_{~bc}(N)\Theta^b(N)\Theta^c(N)
\biggr]~,
\end{eqnarray}
where $\Omega^a(N)$ is a new vector variable obtained by solving
\begin{eqnarray}
{d \over dN} \Omega_a(N) =  - \Omega_b(N)P^b_{~a}(N)-
 N_b(N)Q^b_{~ac}(N)\Theta^c(N)~,\label{eq:diffom}
\end{eqnarray}
with the boundary condition $\Omega_a(N_{\rm f}) = N_{ab}^{\rm f}\Theta^b(N_{\rm f})$.
$N_{abc}^{\rm f}$ is given by
\begin{eqnarray}
N_{abc}^{f} &=& \left({V^2 \over {V'}^2}\right)\Biggl[
\frac{V_{abc}}{V} + 6{V_aV_bV_c \over V^3} - 2 {V_aV_bV_cV^dV^eV^f \over \left({V'}^2\right)^3}{V_{def} \over V} - 4 \eta_{de} {V^d \over V}{V^e V_aV_bV_c \over \left({V'}^2\right)^2}  \nonumber\\
&&\qquad
- 4\eta_{de}{V^{ef}V^d V_fV_aV_bV_c \over \left({V'}^2\right)^3 }
+6{V_{de(a} \over V}{V^dV^e V_bV_{c)} \over \left({V'}^2\right)^2} 
+ 12 \eta^{de}{V_d V_{e(a}V_bV_{c)} \over \left({V'}^2\right)^2}\nonumber\\
&&\qquad
-6{V^d \over V}\eta_{d(a}{V_bV_{c)} \over {V'}^2}  -3{V^d V_{(a} \over {V'}^2}{V_{bc)d} \over V} 
- 6\eta_{d(a}{V_bV^d_{~c)} \over {V'}^2}
+6\eta_{de}{V^dV^e V_{(a} \over V {V'}^2}N_{bc)}^{\rm f} - 6 {V^d \over V}\eta_{d(a}N_{bc)}^{\rm f} \Biggr]~,
\nonumber\\
\end{eqnarray}
and
the four point interaction $S^a_{~bcd}(N)$ is given by
differentiating $Q^a_{~bc}$ with respect to $\phi^a$.
Then, based on the standard slow-roll approximation, where $V_{ab}/V = O(\epsilon)$,
$V_{abc}/V = O(\epsilon^{3/2})$ and $V_{abcd}/V = O(\epsilon^2)$, 
the orders of the non-linearity parameters $\tau_{\rm NL}$ and $g_{\rm NL}$ are respectively
estimated as
\begin{eqnarray}
\tau_{\rm NL} = O(\epsilon^2)\,,\quad g_{\rm NL} = O(\epsilon^2)\,.
\end{eqnarray}
From this rough estimation, we find that 
$\tau_{\rm NL}$ and $g_{\rm NL}$ are more suppressed by the slow-roll parameters
than $f_{\rm NL}$.

As in the case of $f_{\rm NL}$, the magnitudes of $\tau_{\rm NL}$
and $g_{\rm NL}$ are not necessarily suppressed like $O(\epsilon^2)$
and can be much larger than the rough estimation in some situations.
Furthermore, if slow-roll conditions are violated, we may have a chance
to generate large non-Gaussianity, which is not accommodated in this
formalism.  Indeed, some explicit multi-field inflation models have been
constructed that can produce large non-Gaussianity.  In the following,
we consider the generation of non-Gaussianity in those models.

\bigskip
\bigskip
\bigskip
\bigskip

\noindent 
$\bullet$ {\bf Two-field slow-roll inflation} \label{subsec:two_field} \\

In Ref.~\cite{Byrnes:2008wi,Byrnes:2008zy,Byrnes:2010em},
the authors have shown that the large non-Gaussianity 
can be generated even during the slow-roll inflation. 
As as example, we consider the potential of the form:
\begin{equation}
V (\phi_1, \phi_2) = V_{\rm inf}
\exp \left( {1 \over 2}\eta_1 \phi_1^2 + {1 \over 2}\eta_2 \phi_2^2 \right).
\end{equation}
In the following, we assume that $|\eta_1 \phi_1^2 | \ll 1$, $|\eta_2
\phi_2^2 | \ll 1$ and $V_{\rm inf}$ is constant.  In fact, the above
type of potential can lead to large {\it negative} $f_{\rm NL}$ in order
to have a red-tilted spectral index, which is required to be consistent
with current observations.  Since {\it positive} $f_{\rm NL}$ is favored
at this moment, this model might not be so attractive in this respect,
however, we briefly discuss this model as an example, in which
large non-Gaussianity may be generated even during a slow-roll inflation when two
fields exist.

By using slow-roll solutions, we can write the inflaton field values
during inflation as
\begin{equation}
\phi_1 = \phi_{1\ast} e^{- \eta_1 N},
\quad \quad
\phi_2 = \phi_{2\ast} e^{- \eta_2 N}.
\end{equation}
Here, we consider only the slowly-rolling phase and do not specify how
the end of inflation is triggered.  Hence we evaluate the curvature
perturbation on the uniform energy density hypersurface, which can be
determined by $V = $ constant hypersurface as in the
 ``general slow-roll multi-field inflation'' model discussed above.

The spectral index and the tensor-to-scalar ratio are calculated as
\begin{eqnarray}
n_s -1  
& = &   -2 \epsilon_\ast
+
2 \frac{ 
( \eta_1 - 2 \epsilon e^{2 \eta_1 N} ) \epsilon_1 e^{-2 \eta_1 N} 
+
( \eta_2 - 2 \epsilon e^{2 \eta_2 N} ) \epsilon_2 e^{-2 \eta_2 N} 
}{2 \epsilon_1 e^{-2 \eta_1 N} +2 \epsilon_2e^{-2 \eta_2 N} },
\\
r 
& = & 
\frac{16 \epsilon^2}{2 \epsilon_1 e^{-2 \eta_1 N} +2 \epsilon_2e^{-2 \eta_2 N} }.
\end{eqnarray}
The non-linearity parameter $f_{\rm NL}$ can be evaluated as
\begin{equation}
\frac65 f_{\rm NL} 
= 
\frac{ 
- \epsilon \left( \eta_1 \epsilon_1 + \eta_2 \epsilon_2 e^{4(\eta_1 - \eta_2) N} \right)
+ (2/\epsilon) \epsilon_1\epsilon_2 ( \eta_1\epsilon_2 +  \eta_2\epsilon_1)
( 1 - e^{2(\eta_1 - \eta_2) N})^2}{(\epsilon_1 + \epsilon_2 e^{2(\eta_1 - \eta_2) N})^2 },
\end{equation}
where the slow-roll parameters $\epsilon_1$ and $\epsilon_2$ are defined
for $\phi_1$ and $\phi_2$, respectively, in the same manner as in
Eq.~\eqref{eq:slow_roll} and $ \epsilon = \epsilon_1 + \epsilon_2 $. It
has been shown that large non-Gaussianity can be generated when the
following condition is met\footnote{
In fact, there is another case where large non-Gaussianity can be generated. However,
such another case follows  the same argument here by changing some parameters  
appropriately because of the symmetry of the potential \cite{Byrnes:2008zy}.
}:
\begin{equation}
 \frac{ \dot{\phi}_2^2 }{ \dot{\phi}_1^2 + \dot{\phi}_2^2}
\simeq  \frac{\epsilon_2}{ \epsilon_1 + \epsilon_2} \ll 1.
\end{equation}
In this case, we can have simpler formulae for $n_s$, $r$ and $f_{\rm NL}$ as follows, 
\begin{eqnarray}
n_s -1  & \simeq &  \frac{2 (\eta_1 + R \eta_2 )}{1 +R}, 
\\
r  & \simeq &  \frac{16 \epsilon_\ast}{ 1+R},
\\
\frac65 f_{\rm NL} &\simeq & \frac{R}{(1+R)^2} \eta_2 e^{2(\eta_1 - \eta_2)N},
\end{eqnarray}
where $R$ is defined as  $R  =  N_{\phi_2}^2 / N_{\phi_1}^2 = \left( \epsilon_2 / \epsilon_1\right) e^{2(\eta_1 - \eta_2)N}$(see Eq.~\eqref{eq:def_R}). 
From the above expressions, we find that $\eta_2 > 0$ should be satisfied 
to have large positive $f_{\rm NL}$.
However, large $f_{\rm NL}$ requires $\eta_1 > \eta_2$ to have a big factor 
from $e^{2(\eta_1 - \eta_2)N}$, which gives blue-tilted spectral index.
Thus, to have a consistent value of $n_s$ with current observations\footnote{
Current limit for the spectral index is $n_s = 0.963 \pm 0.012 ~(68 \% {\rm CL})$ \cite{Komatsu:2010fb}.
}, $f_{\rm NL}$ should be negative even though its size can be very
large.  Thus, in this sense, this model might not be a promising one 
from the viewpoint of current observations.  However, one could consider another
potential, which may give different predictions for $n_s, r$ and $f_{\rm
NL}$.

As a final remark, we mention the relation between $f_{\rm NL}$ and
$g_{\rm NL}$, which holds for the case considered here
\cite{Byrnes:2008zy}:
\begin{equation}
g_{\rm NL} = \frac{10}{3} \frac{R ( \eta_1 - 2 \eta_2) - \eta_2}{1+R} f_{\rm NL}.
\end{equation}
Thus, this model can be regarded as ``suppressed $g_{\rm NL}$'' type.
For $\tau_{\rm NL}$, we do not find anything more than the relation
given by Eq.~\eqref{eq:mixedft}.

\bigskip
\bigskip

\clearpage

\noindent 
$\bullet$ {\bf Multi-field hybrid (Multi-brid) inflation} \\

Here, we discuss a multi-field hybrid inflation (dubbed as multi-brid inflation
\cite{Sasaki:2008uc}) model.
The analysis of the non-Gaussianity in this model
is similar to that in the previous multi-field slow-roll inflation model, where
multiple (inflaton) fields can affect the dynamics of inflation. 
The difference between this multi-brid model and the previous multi-field slow-roll model
is mainly the assumption on how the inflation ends:
in the multi-brid model,  the end of inflation is characterized by an ellipse
in the field space\footnote{
There are several related studies~\cite{
Huang:2009vk, Alabidi:2006hg}.
This model, in fact, includes the
inhomogeneous end of hybrid inflation model, discussed in section
\ref{subsubsec:inhomo}, as a single source limit.
}.
In the following, we discuss two types of potential in the multi-brid inflation model.

\bigskip
\begin{description}
  \item[A.] {\bf Quadratic potential model}

Let us first consider the model whose potential is approximately
quadratic~\cite{Naruko:2008sq}:
\begin{eqnarray}
V &=& V_0 \exp \left( {1 \over 2}\eta_1 \phi_1^2 + {1 \over 2}\eta_2 \phi_2^2 \right), 
\end{eqnarray}
where
\begin{eqnarray}
V_0 &=& {1 \over 2} G(\phi_1, \phi_2)\chi^2 
+ {\lambda \over 4}\left( \chi^2 - {v^2 \over \lambda}\right)^2,\\
G(\phi_1, \phi_2) &=&
g_1^2(\phi_1 \cos \alpha + \phi_2 \sin \alpha )^2 + g_2^2 (-\phi_1 \sin \alpha + \phi_2 \cos \alpha)^2.
\label{eq:nsmodel}
\end{eqnarray}
Here $\eta_1$ and $\eta_2$ can be regarded as the masses squared for
the inflatons $\phi_1$ and $\phi_2$ normalized by $V_0$,
respectively. $\chi$ is a water-fall field and its VEV is
characterized by $v$. The angle $\alpha$ corresponds to the rotation
of the ellipse of $G(\phi_1, \phi_2) = {\rm const.}$ in the field space
relative to the $\phi_1$ axis. The slow-roll equation
of motion for $\phi_i$ is given by
\begin{eqnarray}
{d \phi_i \over dN} = \eta_i \phi_i,
\end{eqnarray}
where $N$ is the number of $e$-folds and $dN = -Hdt$, from which we
obtain
\begin{eqnarray}
N = {1 \over \eta_1} \ln \phi_1 - {1 \over \eta_1} \ln \phi_{1,f}.
\label{eq:efolds}
\end{eqnarray}
The inflation ends when the following condition is satisfied,
\begin{eqnarray}
v^2 = g_1^2(\phi_{1,f} \cos \alpha + \phi_{2,f} \sin \alpha )^2 + g_2^2 (-\phi_{1,f} \sin \alpha + \phi_{2,f} \cos \alpha)^2,
\end{eqnarray}
where $\phi_{i,f}$ denotes the field value at the end of inflation.
In order to parameterize $\phi_{i,f}$, we introduce an angular
parameter $\gamma$ defined as
\begin{eqnarray}
{v \over g_1} \cos \gamma &\equiv& \phi_{1,f} \cos \alpha + \phi_{2,f} \sin \alpha,\\
{v \over g_2} \sin \gamma &\equiv& - \phi_{1,f} \sin \alpha + \phi_{2,f} \cos \alpha.
\end{eqnarray}
These equations imply 
\begin{eqnarray}
\label{eq:phi1f_param}
\phi_{1, f}
 &=& 
 \frac{v}{g_1} \cos \alpha \cos \gamma  -  \frac{v}{g_2} \sin \alpha \sin \gamma, \\
 \label{eq:phi2f_param}
\phi_{2, f}
 &=& 
 \frac{v}{g_1} \sin \alpha \cos \gamma  +  \frac{v}{g_2} \cos \alpha \sin \gamma.
\end{eqnarray}
The angle $\gamma$ represents the position of the
inflationary trajectory at the end of inflation\footnote{
The definitions of $\alpha$ and $\gamma$ are shown schematically in Fig.~1 of \cite{Naruko:2008sq}. 
}.
From the equation of motion, we can show that $ \phi_1^{\eta_1} / \phi_2^{\eta_2}$ is a constant of motion, 
from which we can derive the following equation:
\begin{eqnarray}
{1 \over \eta_1} \ln \phi_1 - {1 \over \eta_2} \ln \phi_2 &=& {1 \over \eta_1} \ln \phi_{1,f} - {1 \over \eta_2} \ln \phi_{2,f} \nonumber\\
&=& {1 \over \eta_1} \ln \left[ {v \over g_1 g_2}\left( g_2 \cos \alpha \cos \gamma - g_1 \sin \alpha \sin \gamma \right) \right] \nonumber\\
&&\qquad
-  {1 \over \eta_2} \ln \left[ {v \over g_1 g_2}\left( g_2 \sin \alpha \cos \gamma + g_1 \cos \alpha \sin \gamma \right) \right]
.
\label{eq:constant-quad}
\end{eqnarray}
By expanding $N$ given in Eq.~(\ref{eq:efolds}), we can obtain $\delta
N$, up to the third order, as
\begin{eqnarray}
\delta N &=& {1 \over \eta_1} \biggl[
{\delta \phi_1 \over \phi_1} - {1 \over 2}\left( {\delta \phi_1 \over \phi_1 }\right)^2 
+ {1 \over 3} \left(  {\delta \phi_1 \over \phi_1 } \right)^3-  {\partial \ln \phi_{1,f} \over \partial \gamma} \left( \delta_{(1)} \gamma + \delta_{(2)} \gamma + \delta_{(3)} \gamma \right)~\nonumber\\
&&\qquad
- {1 \over 2} {\partial^2 \ln \phi_{1,f} \over \partial \gamma^2} \left( \delta_{(1)} \gamma + \delta_{(2)} \gamma\right)^2
- {1 \over 6} {\partial^3 \ln \phi_{1,f} \over \partial \gamma^3} \left( \delta_{(1)} \gamma \right)^3 \biggr],
\label{eq:multibriddeltaN}
\end{eqnarray}
where $\delta_{(1)} \gamma$, $\delta_{(2)} \gamma$ and $\delta_{(3)}
\gamma$ are respectively the first, the second and the third order
perturbations of $\gamma$. Since they can be described in terms of $\delta \phi_i$
by the relation given by Eq.~(\ref{eq:constant-quad}), the resultant
expression of $\delta N$ or the curvature perturbation $\zeta$
with respect to $\delta \phi_i$ is given by
\begin{eqnarray} 
\zeta = \delta N &=& 
{1 \over F(\gamma)}\left[ - A_2 {\delta \phi_1 \over \phi_1} + A_1 {\delta \phi_2 \over \phi_2} \right] \nonumber\\ 
&&+ {1 \over 2 F(\gamma)}\biggl[ 
A_2 \left({\delta \phi_1 \over \phi_1}\right)^2 - A_1 \left( {\delta \phi_2 \over \phi_2} \right)^2  +
{G(\gamma) \over F(\gamma)^2} \left(\eta_2 {\delta \phi_1 \over \phi_1} - \eta_1 {\delta \phi_2 \over \phi_2}  \right)^2 \biggr] \nonumber\\
&& + {1 \over 6 F(\gamma)}\biggl\{ -2 A_2 \left({\delta \phi_1 \over \phi_1}\right)^3 
+2 A_1 \left( {\delta \phi_2 \over \phi_2} \right)^3  \nonumber\\
&&\quad\quad -
3 {G(\gamma) \over F(\gamma)^2} \left(\eta_2 {\delta \phi_1 \over \phi_1} - \eta_1 {\delta \phi_2 \over \phi_2}  \right)
\left[ \eta_2 \left({\delta \phi_1 \over \phi_1}\right)^2 - \eta_1 \left({\delta \phi_2 \over \phi_2}\right)^2 \right] \nonumber\\
&&\quad\quad\quad
+ {1 \over F(\gamma)^4}\left( G'(\gamma)F(\gamma) - 3 G(\gamma)F'(\gamma) \right)
\left(\eta_2 {\delta \phi_1 \over \phi_1} - \eta_1 {\delta \phi_2 \over \phi_2}  \right)^3 \biggr\} ,
\label{eq:deltaN-quadratic}
\end{eqnarray}
where a prime denotes the derivative with respect to $\gamma$, and $A_1, A_2, F$ and $G$
are defined as
\begin{eqnarray}
A_1& =& {\partial \ln \phi_{1,f} \over \partial \gamma}, \quad \quad 
A_2 = {\partial \ln \phi_{2,f} \over \partial \gamma}, \\
F(\gamma) &=& \eta_2 A_1 - \eta_1A_2, \quad \quad
G(\gamma) =  A_1' A_2  - A_1 A_2'.
\end{eqnarray}
The spectral index and the tensor-to-scalar ratio are respectively given
by
\begin{eqnarray}
n_s - 1 &=& 2 {\eta_1 A_2^2 / \phi_1^2 + \eta_2 A_1^2 / \phi_2^2 \over A_2^2 / \phi_1^2 + A_1^2 / \phi_2^2}
- \left( \eta_1^2 \phi_1^2 + \eta_2^2 \phi_2^2\right),\\
r &=& {8 F^2 \over A_2^2 / \phi_1^2 + A_1^2 / \phi_2^2 }.
\end{eqnarray}
Here it is worth noting that the tensor-to-scalar ratio $r$ can be relatively 
large even when $f_{\rm NL}$ is large in this model \cite{Naruko:2008sq}. 
As mentioned in Section~\ref{sec:mixed_inf}, the tensor-to-scalar ratio 
$r$ tends to be very small 
in most models generating large non-Gaussianity. However, the multi-brid 
inflation model is one of those which can realize 
large tensor-to-scalar ratio and large non-Gaussianity simultaneously.

For the non-linearity parameters, we obtain
\begin{eqnarray}
{6 \over 5}f_{\rm NL} &=& {F \over \left( A_2^2 / \phi_1^2 + A_1^2 / \phi_2^2\right)^2} 
\left( {A_2^3 \over \phi_1^4} - {A_1^3 \over \phi_2^4}\right)
+ {G / F \over \left(A_2^2 / \phi_1^2 + A_1^2 / \phi_2^2\right)^2}\left( \eta_1 {A_1 \over \phi_2^2} + \eta_2 {A_2 \over \phi_1^2}\right)^2, \label{eq:multibridfNL} \notag \\ \\
\tau_{\rm NL} &=& {F^2 \over \left( A_2^2 / \phi_1^2 + A_1^2 / \phi_2^2\right)^3} 
\left( {A_2^4 \over \phi_1^6} + {A_1^4 \over \phi_2^6}\right) \nonumber\\ 
&&\qquad
+
{G^2 / F^2 \over \left( A_2^2 / \phi_1^2 + A_1^2 / \phi_2^2\right)^3}
\left( {\eta_2^2 \over \phi_1^2} + {\eta_1^2 \over \phi_2^2}\right)
\left( \eta_1 {A_1 \over \phi_2^2} + \eta_2 {A_2 \over \phi_1^2}\right)^2 \nonumber\\
&&\qquad\qquad
+ 2 {G \over  \left( A_2^2 / \phi_1^2 + A_1^2 / \phi_2^2\right)^3}
\left( \eta_1 {A_1 \over \phi_2^2} + \eta_2 {A_2 \over \phi_1^2}\right)
\left( \eta_2 {A_2^2 \over \phi_1^4} - \eta_1{A_1^2 \over \phi_2^4}\right),
\label{eq:multibridtauNL}
\end{eqnarray}
\begin{eqnarray}
{54 \over 25}g_{\rm NL} &=& 2
{F^2 \over \left( A_2^2 / \phi_1^2 + A_1^2 / \phi_2^2\right)^3} 
\left( {A_2^4 \over \phi_1^6} + {A_1^4 \over \phi_2^6}\right)
 \nonumber\\
&&\qquad
+
3 {G \over  \left( A_2^2 / \phi_1^2 + A_1^2 / \phi_2^2\right)^3}
\left( \eta_1 {A_1 \over \phi_2^2} + \eta_2 {A_2 \over \phi_1^2}\right)
\left( \eta_2 {A_2^2 \over \phi_1^4} - \eta_1{A_1^2 \over \phi_2^4}\right)
\nonumber\\
&&\qquad
-
{G / F \over \left( A_2^2 / \phi_1^2 + A_1^2 / \phi_2^2\right)^3}\left({G' \over G} - 3 {F' \over F} \right)
\left( \eta_1 {A_1 \over \phi_2^2} + \eta_2 {A_2 \over \phi_1^2}\right)^3.
\label{eq:multibridgNL}
\end{eqnarray}
Since these expressions are rather complicated in general, we consider
three limiting cases, which may generate large non-Gaussianity: 
Single-source case, equal mass case ($\eta_1 =\eta_2 =
\eta$), and large mass ratio case ($\eta_1 \gg \eta_2$).

\bigskip

\begin{description}
  \item[I.] {\bf Single source case} \\

    The setup of the multi-brid inflation model includes the
    inhomogeneous end of hybrid inflation discussed in section
    \ref{subsubsec:inhomo} as a limiting case where fluctuations of a
    single field, $\phi_1$ or $\phi_2$, only contribute to the curvature
    fluctuation and the angle $\alpha = 0$. Here we regard
    $\phi_1$ as the inflaton and $\phi_2$ as another light scalar
    field, denoted as $\sigma$ in Section \ref{subsubsec:inhomo}, in
    which the mass of the inflaton is much bigger than that of a light
    scalar field (i.e. $\eta_1 \gg \eta_2$). In this limit, we also
    assume that the inflaton does not contribute to the power spectrum,
    which is represented by the following condition:
\begin{equation}
  \frac{A_2^2}{\phi_1^2} \ll \frac{A_1^2}{\phi_2^2} \quad
 \Longleftrightarrow \quad
  g_1^2 \phi_{1,f}^2 \phi_2 \ll g_2^2 \phi_{2,f}^2 \phi_1
 \quad \biggl(  \Longleftrightarrow \quad
  g_1^2 \phi_{1,f} \ll g_2^2 \phi_{2,f} \biggr).
\end{equation} 
The last inequality applies only when $\eta_1 N \lesssim 1$, which leads
to the relation $\phi_{1,f}/\phi_1 \simeq \phi_{2,f}/\phi_2$. In
order to realize large non-Gaussianity, we further assume that $A_1^2
\ll A_2^2$, which is equivalent to consider the situation where $g_2^2
\phi_{2,f}^2 \ll g_1^2 \phi_{1,f}^2 \simeq v^2$. Putting these
conditions to the above general expressions with $\alpha = 0$, we obtain
the following formulae,
\begin{eqnarray} 
 && \frac65 f_{\rm NL} \simeq \eta_1 \frac{g_1^2 \phi_{1,f}^2}{g_2^2
  \phi_{2,f}^2} \simeq \eta_1 \frac{v^2}{g_2^2 \phi_{2,f}^2},
  \\
 && \frac{54}{25} g_{\rm NL} \simeq 6\eta_1(\eta_1-2\eta_2) \frac{g_1^2 \phi_{1,f}^2}{g_2^2
  \phi_{2,f}^2} \simeq \frac{36}{5} \eta_1 f_{\rm NL},
  \\
 && \tau_{\rm NL} \simeq \left(\frac65 f_{\rm NL} \right)^2,
\end{eqnarray}
where the following equations have been used:
\begin{eqnarray} 
 F(\gamma) \simeq -\eta_1 \frac{g_1 \phi_{1,f}}{g_2 \phi_{2,f}}\,,
  \qquad
 G(\gamma) \simeq -2 \frac{g_1 \phi_{1,f}}{g_2 \phi_{2,f}}\,.
\end{eqnarray}
These results reproduce  those obtained for the inhomogeneous end of
hybrid inflation model discussed in Section \ref{subsubsec:inhomo} and again 
notice that $g_{\rm NL} / f_{\rm NL}$ is suppressed by the slow-roll parameter
$\eta_1$. 

\bigskip

  \item[II.] {\bf Equal mass case ($\eta_1 = \eta_2 = \eta$)} \\
    
In the equal mass limit ($\eta_1 = \eta_2 = \eta$), $\alpha$-dependence
disappears~\cite{Naruko:2008sq} because of the symmetry.  Hence, we can
set $\alpha = 0$ without loss of generality. In such a case, we have
\begin{eqnarray}
&&A_1 = - {g_2 \phi_{2,f} \over g_1 \phi_{1,f}} = - \tan \gamma\,,\quad A_2 = {g_1 \phi_{1,f} \over g_2 \phi_{2,f}} = - {1 \over A_1}\,,\\
&& F = - {\eta \over \sin \gamma \cos \gamma}\,,\quad G = - {2 \over \sin \gamma \cos \gamma} = {2 \over \eta} F\,.
\end{eqnarray}
Then, the tensor-to-scalar ratio and the spectral index are reduced to
\begin{eqnarray}
r &=&  
{ 8 \eta^2 v^2 e^{2 \eta N_k} \over \left(g_1^2 \cos^2 \gamma + g_2^2 \sin^2 \gamma \right)}\,,\\
n_s - 1 &=& 2 \eta - {r \over 8 g_1^2 g_2^2}\left( g_1^2 \cos^2 \gamma + g_2^2 \sin^2 \gamma \right)
\left( g_1^2 \sin^2 \gamma + g_2^2 \cos^2 \gamma \right),
\end{eqnarray}
where we have used $\phi_i = \phi_{i,f} e^{\eta N_k}$. The non-linearity parameters 
in this limit are given by
\begin{eqnarray}
{6 \over 5}f_{\rm NL} = - { \eta \over \left( g_1^2 \cos^2 \gamma + g_2^2 \sin^2 \gamma \right)^{2}}
\left[ g_1^4 \cos^2 \gamma + g_2^4 \sin^2 \gamma  - 2 \left(g_1^2 - g_2^2 \right)^2 \sin^2 \gamma \cos^2 \gamma \right],
\notag \\ 
\end{eqnarray}
\begin{eqnarray}
\tau_{\rm NL} &=& { \eta^2 \over \left( g_2^2 \sin^2 \gamma + g_1^2 \cos^2 \gamma \right)^{3}}
\biggl[ g_1^6 \cos^2 \gamma + g_2^6 \sin^2 \gamma \nonumber\\
&&\qquad\qquad\qquad\qquad\quad
- 4 \left(g_1^2 - g_2^2 \right)^2 \left(g_1^2 \cos^2 \gamma + g_2^2 \sin^2 \gamma \right)
 \sin^2 \gamma \cos^2 \gamma \biggr], \notag \\
\end{eqnarray}
\begin{eqnarray}
{54 \over 25}g_{\rm NL} &=&  { 2 \eta^2 \over \left( g_2^2 \sin^2 \gamma + g_1^2 \cos^2 \gamma \right)^{3}}
\biggl\{ g_1^6 \cos^2 \gamma + g_2^6 \sin^2 \gamma \nonumber\\
&&\qquad\quad
- \left(g_1^2 - g_2^2 \right)^2  \sin^2 \gamma \cos^2 \gamma 
\left[3 \left(g_1^2 + g_2^2 \right)+ 2 \left( g_1^2 - g_2^2\right) \left( \cos^2 \gamma - \sin^2 \gamma \right) \right]
\biggr\}.\nonumber\\
\end{eqnarray}
After some algebra, we can find the simple relation between $f_{\rm
NL}$ and $g_{\rm NL}$:
\begin{eqnarray}
g_{\rm NL} = - {10 \over 3} \eta f_{\rm NL} - {50 \over 27} \eta^2.
\end{eqnarray}
As is the same with the single-source case, the ratio of the
non-linearity parameters $g_{\rm NL} / f_{\rm NL}$ is suppressed by the
slow-roll parameter $\eta$ in this case as well. We also obtain the
following relation,
\begin{eqnarray}
  \tau_{\rm NL} = \left( \frac{1+\overline{R}}{\overline{R}} \right)
                  \left( \frac65 f_{\rm NL} \right)^2,
\end{eqnarray}
where the ratio $\overline{R}$ is given by
\begin{eqnarray}
  \overline{R} = \left[ \frac{g_1^4 \cos^2 \gamma + g_2^4 \sin^2
		    \gamma - 2\left( g_1^2-g_2^2 \right) \sin^2 \gamma
                    \cos^2 \gamma}
                    {g_1 g_2 \left( g_1^2-g_2^2 \right) \sin \gamma \cos
		    \gamma} \right]^2,
\end{eqnarray}
from which we can calculate the size of $\tau_{\rm NL}$ relative to
$f_{\rm NL}^2$.

\begin{figure}[htbp]
  \begin{center}
  \resizebox{100mm}{!}{
    \includegraphics{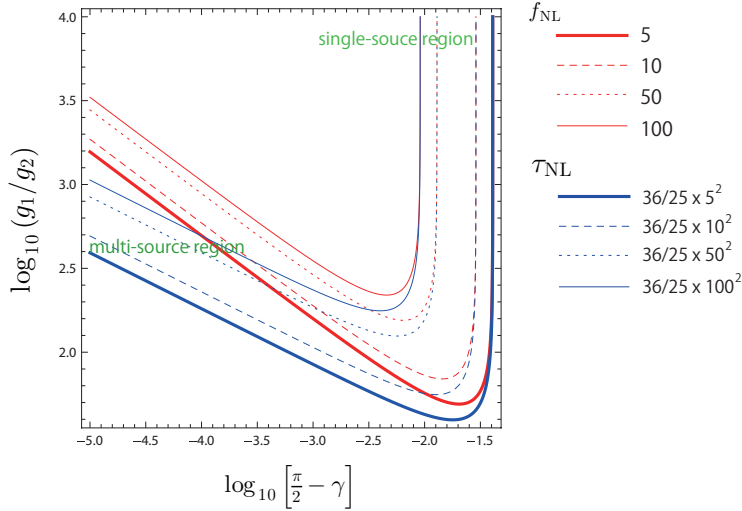}
    }
  \end{center}
  \caption{Contours of $f_{\rm NL}$ (red lines) and $\tau_{\rm NL}$
    (blue lines) on the $\gamma$-$g_1/g_2$ plane for the equal mass
    case.  The blue lines are for $f_{\rm NL} = 5$ (thick solid), $10$
    (dashed), $50$ (dotted) and $100$ (thin solid). The red lines are
    for $\tau_{\rm NL} = 36 / 25 \times 5^2 $ (thick solid), $36 / 25
    \times 10^2$ (dashed), $36 / 25 \times 50^2$ (dotted) and $36 / 25
    \times 100^2$ (thin solid).  We set $\eta_1 = \eta_2 = \eta = 0.01$
    here. From this figure, we can find that the upper right region on
    this plane corresponds to the single-source case because  the
    equality $\tau_{\rm NL} = 36/25 \times f_{\rm NL}^2$ is satisfied, which
    holds in the single source case. } \label{fig:fNL_tauNL_NS.eps}
\end{figure}
\begin{figure}[htbp]
  \begin{center}
  \resizebox{100mm}{!}{
    \includegraphics{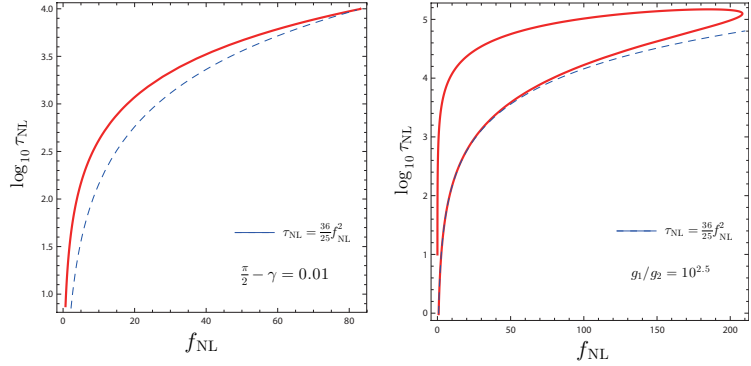}
    }
  \end{center}
  \caption{$f_{\rm NL}$--$\tau_{\rm NL}$ diagrams for the equal mass case. In the left panel,
    the red solid line is for the case with ${\pi \over 2}
    - \gamma = 0.01$ and $g_1/g_2$ being varied.  In the right panel, the red solid line is for
    the case with $g_1 / g_2 = 10^{2.5}$ and $\gamma$ being varied. In
    both the panels, the blue dashed lines indicate the equality $\tau_{\rm NL} = {36
      / 25} \times f_{\rm NL}^2$. We set $\eta_1 = \eta_2 = \eta = 0.01$ and 
    $\alpha =0$ here.}
  \label{fig:ft_NS.eps}
\end{figure}

In Fig.~\ref{fig:fNL_tauNL_NS.eps}, we depict contours of $f_{\rm
  NL}$ (red lines) and $\tau_{\rm NL}$ (blue lines) on the
$\gamma$-$g_1/g_2$ plane.  As seen from this figure, we can find that
the upper right region on this plane corresponds to the single source
case because the equality $\tau_{\rm NL} = 36/25 \times f_{\rm NL}^2$,
which holds in the single-source case, is approximated satisfied.

In Fig.~\ref{fig:ft_NS.eps}, we also show the $f_{\rm NL}$--$\tau_{\rm NL}$
diagrams in this limiting case.  In the left panel, the red solid line
is for the case with ${\pi \over 2} - \gamma = 0.01$.  In the right
panel, the red solid line is for the case with $g_1 / g_2 =
10^{2.5}$. In both the panels, the blue dashed lines show the equality
$\tau_{\rm NL} = {36 / 25} \times f_{\rm NL}^2$.

\bigskip

\item[III.] {\bf Large mass ratio case ($\eta_1 \gg \eta_2$) with $A_2 \ll 1$} \\

In order to study the case of large mass ratio, we have set $A_2 = 0$ in
the expressions for the non-linearity parameters
Eqs.~(\ref{eq:multibridfNL})--(\ref{eq:multibridgNL}). 
However, we would like to stress that, as already
pointed out in Ref.~\cite{Naruko:2008sq}, our expressions apply for
an arbitrary mass ratio as long as $A_2 = 0$.
  From the 
  expression for $\delta N$ given in
  Eq. (\ref{eq:multibriddeltaN}), we can easily find that the condition
  $A_2 = 0$ corresponds to neglecting the linear contribution from
  $\delta \gamma$.  In such a case, the non-linearity parameters are
  given by
\begin{eqnarray}
{6 \over 5}f_{\rm NL} &\! = \!&
\eta_2 \left[ -1 + \left({\eta_1 \over \eta_2}\right)^2 {1 \over A_1^2} \right], \\
\tau_{\rm NL} &\! = \!&
\left( {6 \over 5}f_{\rm NL}\right)^2 
+ {\eta_2^2 \phi_2^2 \over \eta_1^2 \phi_1^2} \eta_2^2 
\left({\eta_1 \over \eta_2}\right)^4 {1 \over A_1^4}\,,\\
{54 \over 25}g_{\rm NL} &\! = \!&
-3 \left( \eta_1 + \eta_2 \right){6 \over 5}f_{\rm NL}
- \eta_2^2 - 3 \eta_2 \eta_1 + 3 \left( 1 - {\eta_2 \over \eta_1}\right)\eta_2^2
\left({\eta_1 \over \eta_2}\right)^4 {1 \over A_1^4}\,.
\end{eqnarray}
From these equations, we find that, even if $A_1$ is order of unity,
large mass ratio $\eta_1 \gg \eta_2$ can generate the large
non-Gaussianity. For large $f_{\rm NL}$, we can approximately obtain
the following relations,
\begin{eqnarray}
{6 \over 5}f_{\rm NL} &\! \simeq \!& 
\eta_2\left({\eta_1 \over \eta_2}\right)^2 {1 \over A_1^2}\,,\\
\tau_{\rm NL} &\! \simeq \!&
\left( 1 +{\eta_2^2 \phi_2^2 \over \eta_1^2 \phi_1^2}\right) \left( {6 \over 5}f_{\rm NL}\right)^2 
,\\
{54 \over 25} g_{\rm NL}
&\! \simeq \!&
3 \left( 1 - {\eta_2 \over \eta_1}\right)
\left( {6 \over 5}f_{\rm NL}\right)^2
- \left( \eta_1 + \eta_2 \right)
{18 \over 5}f_{\rm NL}\,.
\end{eqnarray}
Here, it should be noticed that $g_{\rm NL}$ is of the order of $f_{\rm
NL}^2$ and can become relatively large. This is because both fields
significantly contribute to the non-linearity parameters as well
as the power spectrum despite their large mass ratio.
   
It should also be noted that, in the single source case discussed
earlier, we have assumed not only the large mass ratio but also the
large ratio of the field values, $\phi_1$ and $\phi_2$, in order to get
the large non-Gaussianity.  In such a case, the curvature perturbation
was effectively generated only from a single source at all orders.

On the other hand, here, we have assumed only $A_2 = 0$.  From
Eq.~\eqref{eq:deltaN-quadratic}, we can find that this case is different
from the single source case because the curvature perturbation can be
generated from multi-source at the second (and also the third) order.
Hence, the above expressions of non-linearity parameters are different
from those in the pure single source case.
   
\end{description}

\bigskip

\item[B.] {\bf Linear potential model} \\

Next let us briefly discuss the linear multi-brid model whose
potential is given by~\cite{Sasaki:2008uc, Huang:2009xa}
\begin{eqnarray}
V &=& V_0 \exp \left( m_1 \phi_1 + m_2 \phi_2 \right),
\end{eqnarray}
where
\begin{eqnarray}
V_0 &=& {1 \over 2}\left[ g_1^2\left( \phi_1 \cos \alpha + \phi_2 \sin \alpha\right)^2 
+ g_2^2 \left( - \phi_1 \sin \alpha + \phi_2 \cos \alpha \right)^2 \right]\chi^2 + {\lambda \over 4}
\left( \chi^2 - {v^2 \over \lambda}\right)^2.  \notag \\
\end{eqnarray}
For this potential, the slow-roll equation of motion is
\begin{eqnarray}
{d \phi_i \over dN} = m_i\,.
\end{eqnarray}
Hence the total $e$-folding number is evaluated
as
\begin{eqnarray}
N = {1 \over m_1} \left( \phi_1 - \phi_{1,f}\right),
\label{eq:efold}
\end{eqnarray}
where $\phi_{1,f}$ is the value of the scalar field at the end of
inflation.  As in a usual hybrid inflation model, the inflation ends
when the following relation is satisfied:
\begin{eqnarray}
v^2 = g_1^2\left( \phi_{1,f} \cos \alpha + \phi_{2,f} \sin \alpha\right)^2 
+ g_2^2 \left( - \phi_{1,f} \sin \alpha + \phi_{2,f} \cos \alpha \right)^2
.
\end{eqnarray}
We again parameterize $\phi_{1,f}$ and $\phi_{2,f}$
as in Eqs.~\eqref{eq:phi1f_param} and \eqref{eq:phi2f_param}. 
The field values at the end of inflation can be described in terms of
the values of $\phi_1$ and $\phi_2$ at some time during inflation as
\begin{eqnarray}
{1 \over m_1} \phi_1 - {1 \over m_2} \phi_2 = {1 \over m_1} \phi_{1,f} - {1 \over m_2} \phi_{2,f}\,.
\label{eq:constant}
\end{eqnarray}
Perturbing the above equation, we have the following relation,
\begin{eqnarray}
{1 \over m_1} \delta \phi_1 - {1 \over m_2} \delta \phi_2
&=& \left( {1 \over m_1} {\partial \phi_{1,f} \over \partial \gamma}
 - {1 \over m_2} {\partial \phi_{2,f} \over \partial \gamma} \right) \delta \gamma \nonumber\\
&&
 + {1 \over 2} \left( {1 \over m_1} {\partial^2 \phi_{1,f} \over \partial \gamma^2} 
 - {1 \over m_2} {\partial^2 \phi_{2,f} \over \partial \gamma^2} \right)  \delta \gamma^2
 + {1 \over 6} \left( {1 \over m_1} {\partial^3 \phi_{1,f} \over \partial \gamma^3}
 - {1 \over m_2} {\partial^3 \phi_{2,f} \over \partial \gamma^3} \right) \delta \gamma^3. \notag \\
\label{eq:perturbcons}
\end{eqnarray}
From Eq. (\ref{eq:efold}), $\delta N$ is given by
\begin{eqnarray}
\delta N = {1 \over m_1} \delta \phi_1 - {1 \over m_1} {\partial \phi_{1,f} \over \partial \gamma} \delta \gamma
- {1 \over 2} {1 \over m_1} {\partial^2 \phi_{1,f} \over \partial \gamma^2} \delta \gamma^2
- {1 \over 6} {1 \over m_1} {\partial^3 \phi_{1,f} \over \partial \gamma^3} \delta \gamma^3.
\end{eqnarray}
Substituting Eq. (\ref{eq:perturbcons}) into Eq. (\ref{eq:efold}), we
have the expression of $\delta N$ up to the third order,
\begin{eqnarray}
\delta N &=& 
\frac{1}{\widetilde{F}(\gamma)} 
\left( - \widetilde{A}_2(\gamma)  \delta \phi_1 + \widetilde{A}_1(\gamma) \delta \phi_2  \right)
+ \frac{\widetilde{G}(\gamma)}{ 2 \widetilde{F}(\gamma)^3}
\left( m_2 \delta \phi_1 - m_1 \delta \phi_2 \right)^2 \nonumber\\
&& \quad \quad \quad
- \frac{\widetilde{G}(\gamma)\widetilde{F}'(\gamma)}{2 \widetilde{F}(\gamma)^5 }  
\left( m_2 \delta \phi_1 - m_1 \delta \phi_2 \right)^3\, ,
\end{eqnarray}
 where
\begin{eqnarray}
&&\widetilde{A}_1(\gamma) = {\partial \phi_{1,f} \over \partial \gamma}\,,\quad
\widetilde{A}_2(\gamma) = {\partial \phi_{2,f} \over \partial \gamma}\,, \\
&& \widetilde{F}(\gamma) = m_2 \widetilde{A}_1(\gamma) - m_1 \widetilde{A}_2(\gamma)\,,\quad
\widetilde{G}(\gamma) = \widetilde{A}_1'(\gamma) \widetilde{A}_2(\gamma) 
- \widetilde{A}_2'(\gamma) \widetilde{A}_1(\gamma)\,,
\end{eqnarray}
and we have used
\begin{eqnarray}
\widetilde{A}_1'' = -\widetilde{A}_1,\quad \widetilde{A}_2'' = - \widetilde{A}_2.
\end{eqnarray}
From these equations, the non-linearity parameters can be calculated as
\begin{eqnarray}
{6 \over 5}f_{\rm NL} &=& {\widetilde{G} \over \widetilde{F}} {\left( m_1 \widetilde{A}_1 + m_2 \widetilde{A}_2 \right)^2 \over \left( \widetilde{A}_1^2 + \widetilde{A}_2^2\right)^2}, \label{eq:linfnl}\\
\tau_{\rm NL} &=& {\widetilde{G}^2 \over \widetilde{F}^2} {\left( m_1 \widetilde{A}_1 + m_2 \widetilde{A}_2 \right)^2 \over \left( \widetilde{A}_1^2 + \widetilde{A}_2^2\right)^3} \left( m_1^2 + m_2^2 \right), \label{eq:lintaunl} \\
{54 \over 25} g_{\rm NL} &=&  
\frac{3 \widetilde{F}'(\gamma) \widetilde{G} (\gamma)}{\widetilde{F}^2(\gamma)} 
\frac{ (m_1 \widetilde{A}_1 + m_2 \widetilde{A}_2 )^3}{(\widetilde{A}_1^2 + \widetilde{A}_2^2)^3}.
\end{eqnarray} 
By using the above expressions, we find that $g_{\rm NL}$ can be 
written with $f_{\rm NL}$:
\begin{eqnarray}
{54 \over 25}g_{\rm NL} = \left( {6 \over 5}f_{\rm NL} \right)^{\frac32}
      \frac{3( m_1\phi_{2,f}-m_2\phi_{1,f})}
           {\sqrt{\left( \widetilde{A}_1\phi_{2,f}-\widetilde{A}_2\phi_{1,f} \right)
                  \left(m_2 \widetilde{A}_1 - m_1 \widetilde{A}_2 \right) }}\,.
\label{eq:lingnl}
\end{eqnarray}
Thus, $g_{\rm NL}$ in the linear potential model is at least of the
order of $f_{\rm NL}^{3/2}$, and it can be relatively large compared
to $f_{\rm NL}$.

The spectral index and the tensor to
scalar ratio are given by
\begin{eqnarray}
n_s - 1 = - (m_1^2 + m_2^2)\,,\quad r = {8 \widetilde{F}^2(\gamma) \over \widetilde{A}_1^2 + \widetilde{A}_2^2}\,,
\end{eqnarray}
from which we find another simple relation among $f_{\rm NL}, n_s$ and $r$:
\begin{eqnarray}
{\left(6 f_{\rm NL} / 5\right)^2 \over \tau_{\rm NL}}
= 1 - {r \over 8 \left( 1 - n_s \right)}\,.
\label{eq:simplelinear}
\end{eqnarray}
 Although the expressions for the non-linearity parameters in this model are not so
complicated, but still it is a bit difficult to see its size
explicitly. Thus we look at some limiting cases in order.

\bigskip
\begin{description}
   
  \item[I.] {\bf Equal mass case ($m_1 = m_2 = M$)} \\
  
    Let us first consider the equal mass case.  The expression of
    $f_{\rm NL}$ in this case is given by
 \begin{eqnarray}
 \frac65 f_{\rm NL} &=& \frac{M}{v  \left(g_1^2 \cos
   ^2\gamma+g_2^2 \sin ^2\gamma\right)^2} \notag \\
   && \times\,g_1^2 g_2^2\,
   \frac{\bigl[
   - g_1 \cos \gamma (\cos \alpha - \sin \alpha) 
   + g_2 \sin \gamma (\cos \alpha +\sin \alpha)
   \bigr]^2}{
     g_1 \cos \gamma (\cos \alpha + \sin \alpha) 
   + g_2 \sin \gamma (\cos \alpha  -\sin \alpha)
    }\,.
    \label{eq:fNL_linear_equal_m}
\end{eqnarray}

Regarding $\tau_{\rm NL}$, from Eq.~(\ref{eq:simplelinear}) we obtain
the following relation
  \begin{eqnarray}
  \tau_{\rm NL} = \left( {1+\bar{R} \over \bar{R}}\right)
  \left( {6 \over 5}f_{\rm NL}\right)^2,
  \end{eqnarray}
  where the ratio $\bar{R}$ is given by
  \begin{eqnarray}
  \bar{R} = \left( {\widetilde{A}_1 + \widetilde{A}_2 \over \widetilde{A}_1 - \widetilde{A}_2}\right)^2.
  \end{eqnarray}
  We can also find the relation between $f_{\rm NL}$ and $g_{\rm NL}$ as
  \begin{eqnarray}
  g_{\rm NL} &=&
  2\, {g_1^2 \cos^2 \gamma + g_2^2 \sin^2 \gamma \over g_1 g_2}
  {g_1 \sin \gamma \left( \cos \alpha + \sin \alpha \right) 
  - g_2 \cos \gamma \left(\cos \alpha - \sin \alpha \right) \over
  g_2 \sin \gamma \left( \cos \alpha + \sin \alpha \right) 
  - g_1 \cos \gamma \left(\cos \alpha - \sin \alpha \right)}\,f_{\rm NL}^2 \cr\cr\cr
  &=&
  \left\{
\begin{array}{ll}
O(g_1/g_2)\times f_{\rm NL}^2,&~~g_1 \gg g_2 \cr\\
O(g_2/g_1)\times f_{\rm NL}^2,&~~g_2 \gg g_1
\end{array}
\right.
  \end{eqnarray}
In particular, $g_{\rm NL} = 2 f_{\rm NL}^2$ for the case with $g_1 = g_2$\,.

\bigskip

  \item[II.] {\bf Large mass ratio case ($m_1 \gg m_2$) with $\widetilde{A}_2 \ll 1$} \\
  
    Now we briefly discuss the large mass ratio case.  To study this
    case, we again set $\widetilde{A}_2 = 0$ as we did for the
    counterpart in the quadratic potential case.  Then the expressions
    for the non-linearity parameters become
  \begin{eqnarray}
  {6 \over 5}f_{\rm NL} &=& {m_1 \over m_2} {m_1 \phi_{2,f} \over \widetilde{A}_1^2}\,, \\
  \tau_{\rm NL} &=& \left( {1 + \bar{R} \over \bar{R}} \right) 
  \left( {6 \over 5}f_{\rm NL}\right)^2, \quad \bar{R} = {m_1^2 \over m_2^2}\,, \\
  g_{\rm NL} &=& 2 \left( 1 - {m_2 \phi_{1,f} \over m_1
		    \phi_{2,f}}\right) f_{\rm NL}^2\,.
  \end{eqnarray}
  
\end{description}

Thus, the linear potential model generically predicts
large $g_{\rm NL}$ relative to $f_{\rm NL}$ and in some limiting cases,
we have $g_{\rm NL} \sim f_{\rm NL}^2$ as we have shown above.

\end{description}

\subsection{Constrained multi-source model}
\label{subsec:constrained}

In this section, we discuss a class of ``constrained multi-source'' model. 
In fact, models of this category are particularly related to so-called loop 
contributions. Thus we start with a general discussion including loop 
terms.

\subsubsection{Expressions including loop contributions}

In the discussions up to the previous section, we have neglected loop 
contributions in the expressions for the power spectrum and
non-linearity parameters. However, if we include such loop terms, a
new type of model can appear.  The expression for $P_\zeta$ and the
non-linearity parameters, including one loop contributions, are
respectively given by
\begin{eqnarray}
\label{eq:P_zetaloop}
P_\zeta(k) &=& \left[ N_a N^a + N_{ab}N^{ab}  {\cal P}_\delta \ln (k L) \right] P_\delta (k)\,,
\\
{6 \over 5}f_{\rm NL}
&=& 
\frac{N_a N_b N^{ab}
+ N_a^{~b}N_b^{~c}N_c^{~a}{\cal P}_{\delta} \ln (k_{m1} L)}
{\left( N_a N^a  + N_{bc}N^{bc}{\cal P}_{\delta} \ln (k_i L) \right)^2}\,,
\label{eq:fNLloop} \\ \notag \\
\tau_{\rm NL}
 &=& 
 \frac{N_a N_{b} N^{ac} N_c^{~b} 
 + N_a^{~b}N_b^{~c}N_c^{~d} N_d^{~a}{\cal P}_{\delta} \ln (k_{m2} L) } 
{\left( N_a N^a  + N_{bc}N^{bc}{\cal P}_{\delta } \ln (k_i L) \right)^3}\,,
\label{eq:tauNLloop} 
\end{eqnarray}
where $L$ is the size of the box in which perturbations are defined
and we have introduced the notations $k_{m1} = \min \{ k_i \}$ and
$k_{m2}=\min \{|k_i + k_j|, k_l \}$.  In deriving the above equations,
we have truncated the expansion of Eq.~\eqref{eq:deltaN} at the second
order in the field fluctuations. Strictly speaking, higher order terms
contribute to the one-loop corrections.  The full expressions for the
one-loop correction are given in appendix \ref{app:1}.  In this
section, we do not take into account such higher order effects since
it turns out to be negligible in many cases.  Since the current observations indicate that the primordial
curvature perturbation is almost Gaussian,  the power spectrum
should not be dominated by the one-loop contributions, from which we
have the following constraint:
\begin{eqnarray}
1 > { N_{ab}N^{ab} \over N_c N^c} {\cal P}_\delta \ln (k L) 
\simeq { N_{ab}N^{ab} \over \left(N_c N^c\right)^2 } {\cal P}_\zeta \ln (k L).
\label{eq:powercons}
\end{eqnarray}
Now let us consider the relation between $f_{\rm NL}$ and $\tau_{\rm
  NL}$ in the case where the one-loop contributions dominate in the
non-linearity parameters.  In such a case, $f_{\rm NL}$ and $\tau_{\rm
  NL}$ are respectively given by
\begin{eqnarray}
{6 \over 5}f_{\rm NL} &=& P_{ab} M^{ab} {\cal P}_\zeta \ln (k_{m1}L), \\
\tau_{\rm NL} &=& M_{ab}M^{ab} {\cal P}_\zeta \ln (k_{m2}L),
\end{eqnarray} 
where
\begin{eqnarray}
P_{ab} \equiv \frac{N_{ab}}{N_dN^d}, \quad
M_{ab} \equiv \frac{N_{a}^{~c} N_{cb}}{\left( N_dN^d\right)^2}.
\end{eqnarray}
Using the Cauchy-Schwarz inequality, we find
\begin{eqnarray}
\left( {6 \over 5}f_{\rm NL}\right)^2 &=&
\left( P_{ab}M^{ab} \right)^2 {\cal P}_\zeta^2 \ln^2 (k_{m1} L) \nonumber\\
&\le & \left( P_{ab} P^{ab} \right) \left( M_{cd}M^{cd}\right) {\cal P}_\zeta^2 \ln^2 (k_{m1}L) \nonumber\\
& = & \tau_{\rm NL} {\ln^2 (k_{m1} L) \over \ln (k_{m2} L)} \left( P_{ab} P^{ab} \right) {\cal P}_\zeta \nonumber\\
& < & \tau_{\rm NL} {\ln^2 (k_{m1} L) \over \ln (k_{m2} L) \ln (kL)},
\end{eqnarray} 
where in the final inequality we have used Eq.~(\ref{eq:powercons}).
We have also assumed that $k$ is any wavenumber that satisfies $\ln (k
L) \simeq \ln (k_{m1} L) \simeq \ln (k_{m2} L) = O(1)$. Then we obtain
an approximate inequality given by
\begin{eqnarray}
\label{eq:local_ineq}
\tau_{\rm NL} \gtrsim \left( {6 \over 5}f_{\rm NL} \right)^2,
\end{eqnarray}
up to the logarithmic corrections. Hence, even in the case where the
one-loop contribution dominates in the non-linearity parameters, ``the
local-type inequality'' Eq.~(\ref{eq:inequality}) should be satisfied
under the condition where the contribution from the one-loop term should
be subdominant in power spectrum.

\subsubsection{Ungaussiton model}

In some cases, the first terms in the numerators of
Eqs.~\eqref{eq:fNLloop} and \eqref{eq:tauNLloop} are negligible
compared to the second order terms.  In such a case, the non-linear
parameters are dominated by the one loop contributions.  This kind of
model has been discussed in
\cite{Linde:1996gt,Boubekeur:2005fj,Suyama:2008nt,Hikage:2008sk} and
called ``ungaussiton'' in \cite{Suyama:2008nt} and ``quadratic model''
in \cite{Hikage:2008sk}.  Let us consider the simplest case where the
curvature perturbation is given by
\begin{eqnarray}
\zeta = N_\phi \delta \phi + {1 \over 2} N_{\sigma\sigma} \delta \sigma^2 + \cdots,
\end{eqnarray}
where $\phi$ is the inflaton and $\sigma$ is the so-called
``ungaussiton.''  In this model, the expression for the non-linearity
parameters given by Eqs.~\eqref{eq:fNL} and \eqref{eq:tauNL} cannot be
adopted in this case any more and $f_{\rm NL}, \tau_{\rm NL}$ and $g_{\rm NL}$ 
are given by
\begin{eqnarray}
{6 \over 5}f_{\rm NL} &=& {N_{\sigma\sigma}^3 
{\cal P}_{\delta \sigma} \ln \left( k_{m1} L\right) \over N_\phi^4}\,,\label{eq:fnlungauss} \\
\tau_{\rm NL} &=& {N_{\sigma\sigma}^4 {\cal P}_{\delta \sigma} \ln \left( k_{m2} L\right)\over N_\phi^6}\,,  \label{eq:taunlungauss} \\
\frac{54}{25} g_{\rm NL} &=&
\frac{ 
3 N_\sigma^2 N_{\sigma\sigma} N_{\sigma\sigma\sigma\sigma} \mathcal{P}_\sigma  \log (k_{m1} L) 
+
3 N_\sigma N_{\sigma\sigma}^2 N_{\sigma\sigma\sigma} \mathcal{P}_\sigma  \log (k_{m1} L) 
}
{N_\phi^6}.
\end{eqnarray}
 Since $N_\sigma$ and
$N_{\sigma\sigma\sigma}$ are strongly suppressed in this model,
the size of $g_{\rm NL}$ would be very small.  Thus the trispectrum is dominated
by $\tau_{\rm NL}$.  From the above expressions, we find the relation
between $f_{\rm NL}$ and $\tau_{\rm NL}$ as
\begin{equation}
\tau_{\rm NL} = C P_\zeta^{-1/3} f_{\rm NL}^{4/3} \sim 10^3  f_{\rm NL}^{4/3}\,,
\end{equation}
where we adopt the normalization for the power spectrum $P_\zeta \sim 2
\times 10^{-9}$ and $C$ is a constant defined as $C = (6/5)^{4/3} (\ln
(k_{m1} L) / [ \ln (k_{m2}) ]^{4/3} )$.  This equation clearly
contradicts with Eq.~\eqref{eq:local_ineq} if $f_{\rm NL}$ is
sufficiently large. However, it can be easily checked that, in such a
case, the power spectrum is also dominated by the loop contribution,
which does not satisfy our assumptions to derive
Eq.~\eqref{eq:local_ineq}.  Also notice that the curvature perturbation
induced from $\sigma$ starts from the second order.  Hence the
fluctuations from $\sigma$ are completely non-Gaussian, which cannot
explain the observed fluctuations.  Thus we need another source of
fluctuations to account for the observed almost Gaussian fluctuations,
which could be those from the inflaton.  In this sense, this model
requires multi-sources in nature. This motivates us to call this kind of
models ``constrained multi-source model.''

In fact, when the initial value (the field value during inflation) of
the scalar field $\sigma_\ast$ is less than its quantum fluctuation
during inflation $\delta \sigma_\ast \simeq H_{\rm inf} / (2\pi)$,
models of this category can be easily realized such as in the axion, the
curvaton, the modulated reheating and so on.

\subsection{Other models}

In this paper, we have discussed various models generating large
local-type non-Gaussianity focusing on models which can be
accommodated in the framework of inflationary universe (fluctuations
originating to quantum fluctuations during inflation) with adiabatic
mode.  However, there are other possibilities of generating large
non-Gaussianity in some other frameworks and/or models where 
a simple parametrization of the non-linearity parameters may not be adopted. 
Here we briefly mention such other models of the local type.

When one considers the generation of dark matter and baryon asymmetry
of the Universe, isocurvature perturbations can be generated in some
situations. Since the shapes of the bispectrum and the trispectrum are
different from the counterparts of the adiabatic ones, a simple
parametrization of $f_{\rm NL}, \tau_{\rm NL}$ and $g_{\rm NL}$ cannot
be easily compared with the ones defined for adiabatic fluctuations\footnote{
  However, we can have some approximate relation between $f_{\rm NL}$
  and $f_{\rm NL}^{\rm (iso)}$. While the former is defined in
  Eq.~\eqref{eq:def_tau_g_NL} and has been discussed in this paper,
  the latter is defined to parametrize non-Gaussianity in models with
  isocurvature mode \cite{Hikage:2008sk}. We can estimate how large 
  non-Gaussianity from isocurvature fluctuations can be compared to the 
  counterpart in adiabatic mode by using the relation between these non-linearity 
  parameters derived in \cite{Hikage:2008sk}.
}. In fact, isocurvature fluctuations have already been severely
constrained by cosmological observations such as CMB, however, the
non-linear effect can be large in some cases, in particular, where
fluctuations are dominated by the second order term.  In this case, the
situation is quite similar to the ungaussiton model discussed in the
previous section. The issue of non-Gaussianity in models with
isocurvature fluctuations has been investigated by several authors. We
refer the reader to Refs.~\cite{Langlois:2008vk,
Kawasaki:2008sn,Kawasaki:2008pa,Hikage:2008sk,Kawakami:2009iu,Langlois:2010dz}
for the detailed discussion on this issue.

Another local-type model is so-called the Ekpyrotic scenario
\cite{Lehners:2010fy,Koyama:2007if,Buchbinder:2007at}, in which the
scale-invariant curvature perturbation and large non-Gaussianity can be
generated.  The non-linearity parameters $f_{\rm NL}$ and $g_{\rm NL}$
in this models have been calculated and the relation between $f_{\rm
NL}$ and $g_{\rm NL}$ is obtained as $g_{\rm NL} \sim f_{\rm NL}^2$
\cite{Lehners:2010fy}.

There are also several studies on the possibility of generating large
non-Gaussianity during preheating phase~\cite{Kolb:2004jm,
Enqvist:2004ey, Enqvist:2005qu, Barnaby:2006km, Chambers:2007se,
Chambers:2008gu, Byrnes:2008zz, Kohri:2009ac, Bond:2009xx, Frolov:2010sz
}. In this case, the super-horizon scale fluctuations might be strongly
coupled to the evolution of small scale fluctuations.  In
Ref.~\cite{Bond:2009xx}, the authors have introduced a new parameter
$F_{\rm NL}$, which describes the non-linearity of the curvature
perturbations and claimed that during preheating era there seems to be a
possibility of generating large non-Gaussianity whose form is quite
different from the usual local-type non-Gaussianity.

Another possibility of local-type non-Gaussianity we would like to
mention is a model in which the $e$-folding number is quite sensitive to
the field value.  In such a case, the truncation at the leading order
or at the next leading order of Eq. (\ref{eq:deltaN}) will no longer
give a correct answer.  We then need to go to the higher order
calculations until the series converge or to resort to the full order
calculations.  In appendix \ref{app:2}, we give a simple toy model
belonging to this class and briefly discuss some interesting features
of the bispectrum and trispectrum.

\section{Discussion and Summary}
\label{sec:summary}

In this paper, we made a classification of models generating large
local-type non-Gaussianity by using some
consistency relations between the non-linearity parameters $f_{\rm NL},
\tau_{\rm NL}$ and $g_{\rm NL}$. The first key relation is the ratio
of $\tau_{\rm NL}/(6 f_{\rm NL}/5)^2$, by which we classify local-type
models into three categories:
\begin{itemize}
  \item  single-source model ($\tau_{\rm NL}/(6f_{\rm NL}/5)^2 =1$)
  \item  multi-source model ($\tau_{\rm NL}/(6f_{\rm NL}/5)^2 > 1$)
  \item constrained multi-source model ($\tau_{\rm NL} \propto f_{\rm NL}^n$)
\end{itemize}

We have also shown that the ``local-type inequality'' 
\begin{equation}
\tau_{\rm NL}  \gtrsim \left( \frac65 f_{\rm NL} \right)^2
\end{equation}
holds true not only for the case where the tree
level contribution dominates in the non-linearity parameters but also
for the case where the loop contribution dominates there.  This inequality has
been shown under the condition that a loop contribution is subdominant
in the power spectrum, which is required by current observations. To
our knowledge, since all models generating local-type large
non-Gaussianity known today should satisfy the ``local-type inequality,'' if
future observations confirm that this inequality does not hold, 
local-type models would be practically ruled out as a mechanism of
generating large non-Gaussian primordial fluctuations.

On the other hand, if future observations
find large $f_{\rm NL}$ of local type and probe the relation between
$\tau_{\rm NL}$ and $f_{\rm NL}$ with some accuracy, satisfying the local-type inequality,
we can see what category of models would be favored.  However, even if we can pick up
the one of these categories, as we have discussed, there still remain many possibilities for
each. Thus we need a further classification to pin down the
model of large non-Gaussianity. For this purpose, we can make use of the
relation between $f_{\rm NL}$ and $g_{\rm NL}$. We have shown that
models can be further divided into three types according to the relative
size of $g_{\rm NL}$ compared to that of $f_{\rm NL}$ as follows:
\begin{itemize}
  \item  Suppressed $g_{\rm NL}$  type 
  ($g_{\rm NL} \sim {\rm [suppression~factor]}\times f_{\rm NL}$ )
  \item  Linear $g_{\rm NL}$ type ($g_{\rm NL} \sim f_{\rm NL}$)
  \item  Enhanced $g_{\rm NL}$ type ($g_{\rm NL} \sim f_{\rm NL}^n$ with $n >1$ or $n=2$ for many models)
\end{itemize}
Thus if we further probe the relation $f_{\rm NL}$ and $g_{\rm NL}$ in
future observations, we may find that only a few models survive by using
the above categorizations.  Then we can figure out what type of models
are favored as a mechanism of the generation of primordial fluctuations.

We have also worked out the above mentioned relations for various concrete models in
this paper.  Although models can be categorized rather rigorously by
using the ratio $\tau_{\rm NL}/(6 f_{\rm NL} 5)^2$, the relation
between $f_{\rm NL}$ and $g_{\rm NL}$ can significantly differ
depending on some model parameters, in particular, in multi-source
models.  For example, let us  take the multi-brid inflation model with
quadratic potential, which was discussed in
Section~\ref{subsec:multi_brid}. This model predicts that $g_{\rm NL}
\sim \eta f_{\rm NL}$ with being $\eta$ a slow-roll parameter, which
is of the suppressed $g_{\rm NL}$ type, for the equal mass case, 
while the relation becomes $g_{\rm NL} \sim f_{\rm
  NL}^2$, which is of the enhanced $g_{\rm NL}$ type, for the large mass ratio case.  
  In other words, the relation between $f_{\rm
  NL}$ and $g_{\rm NL}$ should be carefully investigated to
discriminate a model because a model can predict quite different
relations depending on its model parameters. However, it also
means that the relation would be useful to explore the parameters of a
model.

In this paper, we have focused on non-Gaussianity in various models and
not discussed much the tensor-to-scalar ratio or gravitational waves. In most
models where large non-Gaussian primordial fluctuations are generated,
the tensor-to-scalar ratio is considered to be very small in general.
However, there are a few examples which can give a relatively large
tensor-to-scalar ratio as well as generating large non-Gaussianity. One
of such examples is a mixed model of inflaton fluctuations with some
other source discussed in Section~\ref{sec:mixed_inf}.  In this model,
the fluctuations from the inflaton can also be responsible for the
curvature perturbation, and in such a case, the tensor-to-scalar ratio
can be large, which could be detectable in near future
observations. Another example is multi-brid inflation model discussed in
Section~\ref{subsec:multi_brid}. If both the tensor-to-scalar ratio and
non-Gaussianity are found to be large in the future, above mentioned
models may become the target of detailed studies.  This illustrates that
a comprehensive investigation using non-Gaussianity and some other
information such as gravitational waves can give much insight into the pursuit
of the generation mechanism of primordial fluctuations.

If three non-linearity parameters $f_{\rm NL}, \tau_{\rm NL}$ and
$g_{\rm NL}$ are well determined in future observations, we may be
able to pin down the model of large non-Gaussianity and pick up a
right model of generating primordial fluctuations.  The classification
by using the consistency relation among the above three parameters
would be very useful to pursue the origin of the structure of the Universe
and give a deep understanding of the early Universe.

\section*{Acknowledgments}

T.~T, M.~Y. and S.~Y. thank the Yukawa Institute for Theoretical
Physics at Kyoto University, where this work was discussed during the
YITP-W-09-09 on ``The non-Gaussian universe,'' ``Gravity and
Cosmology 2010'' and the organizers and participants of these workshops 
for stimulating discussions. This work is partially supported by the Belgian
Federal Office for Scientific, Technical, and Cultural Affairs through
the Inter-University Attraction Pole Grant No. P6/11 (T.~S.); by the
Grant-in-Aid for Scientific research from the Ministry of Education,
Science, Sports, and Culture, Japan, Nos. 19740145 (T.~T.), 21740187
(M.~Y.) and 22340056 (S.~Y.). T.~S. is supported by a Grant-in-Aid for
JSPS Fellows.  S.~Y. acknowledges support from the Grant-in-Aid for
the Global COE Program ``Quest for Fundamental Principles in the
Universe: from Particles to the Solar System and the Cosmos'' from
MEXT, Japan.


\pagebreak

\appendix

\noindent
{\bf \LARGE Appendix}

\section{Full expression for  $\zeta$ in multi-curvaton model}
\label{app:multi_cur}

In Section~\ref{sec:multi_cur}, we have discussed the multi-curvaton model, but
investigated only some limiting cases. Here we give the full expression for the 
curvature perturbation in the model up to the third order.  In general, the curvature 
perturbation $\zeta$ can be written as 
\begin{eqnarray}
\zeta &=& C_{a}  \zeta_{a(1)} + C_{b} \zeta_{b(1)} 
+ C_{aa} \zeta_{a(1)}^2  +  C_{bb} \zeta_{b(1)}^2 + C_{ab} \zeta_{a(1)} \zeta_{b(1)}  \notag \\
&&+  C_{aaa} \zeta_{a(1)}^3 + C_{bbb}  \zeta_{b(1)}^3
+ C_{aab}   \zeta_{a(1)}^2 \zeta_{b(1)} + C_{abb}  \zeta_{a(1)} \zeta_{b(1)}^2.
\end{eqnarray} 
We list full expressions for the coefficients such as $C_a$ and so on in the following. 

For the first order, 
\begin{eqnarray}
C_a &=& -\frac{f_{a1} (f_{a1}+3) (f_{b2}-1)}{f_{a1} -3 f_{b1} + 3}, 
\\  \notag \\
C_b &=& \frac{f_{a1} (- f_{b1} f_{b2}+f_{b1}+f_{b2})-3 (f_{b1}-1) f_{b2}}{f_{a1}-3 f_{b1}+3}. 
\end{eqnarray}

For the second order,
\begin{eqnarray}
C_{aa} &=& \frac{1}{4 (f_{a1}-3 f_{b1}+3)^2}
\left[ f_{a1} (f_{b2}-1) \left(2 f_{a1}^4-2 f_{a1}^3 \left(2 f_{b1}+f_{b2}^2+3 f_{b2}-8\right)
\right.  \right.   \notag \\
&&
\left. \left. 
-3 f_{a1}^2 \left(2
   f_{b1}^2+4 f_{b2}^2+12 f_{b2}-13\right)+9 f_{a1} \left(5 f_{b1}-2 f_{b2}^2-6 f_{b2}+2\right)+27
   (f_{b1}-1)\right) \right],
   \\ \notag \\
 C_{bb} &=& 
 \frac{1}{4 (f_{a1}-3 f_{b1}+3)^2}
\left[ 2 f_{a1}^3 f_{b1}^2 (f_{b2}-1)+f_{a1}^2 \left(-4 f_{b1}^3 (f_{b2}-1)-2 f_{b1}^2 
\left(f_{b2}^3+2 f_{b2}^2-11 f_{b2}+8\right)  
   \right. \right. \notag \\
   &&
   \left. \left. 
   +f_{b1} \left(4 f_{b2}^3+8 f_{b2}^2-15 f_{b2}+3\right)+f_{b2}
    \left(-2 f_{b2}^2-4 f_{b2}+3\right)\right) 
    \right.  \notag \\
   &&
   \left. 
      -3 f_{a1} (f_{b1}-1) \left(2 f_{b1}^3 (f_{b2}-1)+2 f_{b1}^2 (f_{b2}-1)+f_{b1} \left(4
   f_{b2}^3+8 f_{b2}^2-15 f_{b2}+3\right)
     \right. \right. \notag \\
   &&
   \left. \left. 
   -2 f_{b2} \left(2 f_{b2}^2+4 f_{b2}-3\right)\right)-9 (f_{b1}-1)^2 f_{b2}
   \left(2 f_{b2}^2+4 f_{b2}-3\right) \right], 
\\   \notag \\
   C_{ab} &=& 
   \frac{1}{(f_{a1}-3 f_{b1}+3)^2}
    \left[ f_{a1} (f_{b2}-1) \left(\left(9-2 f_{a1}^2\right) f_{b1}^2+(f_{a1}+3) f_{b1} \left(f_{a1}^2-f_{a1}
   \left(f_{b2}^2+3 f_{b2}-5\right)
       \right. \right.  \right. \notag \\
   &&
   \left. \left. \left. 
   -3 \left(f_{b2}^2+3 f_{b2}+1\right)\right)-3 f_{a1} f_{b1}^3+(f_{a1}+3)^2 f_{b2}
   (f_{b2}+3)\right) \right]. 
   \end{eqnarray}

For the third order,
\begin{eqnarray}
   C_{aaa} &=&
   \frac{-1}{12 (f_{a1}-3 f_{b1}+3)^3} 
   \left[ f_{a1}^2 (f_{b2}-1) \left(6 f_{a1}^6-2 f_{a1}^5 \left(12 f_{b1}+3 f_{b2}^2+9 f_{b2}-37\right)
     \right. \right. \notag \\
   &&
   \left. \left. 
   +f_{a1}^4
   \left(-12 f_{b1}^2+2 f_{b1} \left(6 f_{b2}^2+18 f_{b2}-77\right)
   +6 f_{b2}^4+26 f_{b2}^3-48 f_{b2}^2-216    f_{b2}+343\right)
     \right. \right.   \notag \\
   &&
   \left. \left.    
   +3 f_{a1}^3 \left(24 f_{b1}^3+2 f_{b1}^2 \left(3 f_{b2}^2+9 f_{b2}-35\right)
   +f_{b1} \left(12 f_{b2}^2+36 f_{b2}-23\right) 
       \right. \right.  \right. \notag \\
   &&
   \left. \left. \left. 
   +3 \left(6 f_{b2}^4+26 f_{b2}^3-11 f_{b2}^2-105 f_{b2}+77\right)\right)+9 f_{a1}^2
   \left(6 f_{b1}^4+2 f_{b1}^3+f_{b1}^2 \left(6 f_{b2}^2+18 f_{b2}-55\right)
          \right. \right.  \right.  \notag \\
   &&
   \left. \left. \left. 
   +f_{b1} \left(-15 f_{b2}^2-45
   f_{b2}+128\right)+3 \left(6 f_{b2}^4+26 f_{b2}^3+3 f_{b2}^2-63 f_{b2}+15\right)\right)
     \right. \right. \notag \\
   &&
   \left. \left. 
   +27 f_{a1} \left(-9 f_{b1}^3+9
   f_{b1}^2-3 f_{b1} \left(6 f_{b2}^2+18 f_{b2}-23\right)+6 f_{b2}^4+26 f_{b2}^3+15 f_{b2}^2-27
   f_{b2}-17\right)
    \right.\right.  \notag \\
   &&
   \left.  \left. 
   -243 (f_{b1}-1) \left(-2 f_{b1}+f_{b2}^2+3 f_{b2}-2\right)\right) \right],  
      \end{eqnarray}
\begin{eqnarray}
   C_{bbb} &=& 
   \frac{-1}{24 (f_{a1}-3 f_{b1}+3)^3}
   \left[ 12 f_{a1}^5 f_{b1}^3 (f_{b2}-1)
     \right.  \notag \\
   &&
   \left.      
   +2 f_{a1}^4 f_{b1}^2 (f_{b2}-1) \left(-24 f_{b1}^2 
   -2 f_{b1} \left(3   f_{b2}^2+9 f_{b2}-37\right) 
   +6 f_{b2}^2+18 f_{b2}-9\right)
     \right. \notag \\
   &&
   \left.     
   +2 f_{a1}^3 \left(-12 f_{b1}^5 (f_{b2}-1) 
   +2 f_{b1}^4
   \left(6 f_{b2}^3+12 f_{b2}^2-95 f_{b2}+77\right)
    \right.\right.  \notag\\
   &&
   \left.  \left.    
   +f_{b1}^3 \left(6 f_{b2}^5+20 f_{b2}^4-86 f_{b2}^3-192
   f_{b2}^2+595 f_{b2}-343\right)
    \right.\right. \notag \\
   &&
   \left.  \left.       
   -3 f_{b1}^2 \left(6 f_{b2}^5+20 f_{b2}^4-33 f_{b2}^3-86 f_{b2}^2+126
   f_{b2}-33\right)
    \right.\right.  \notag \\
   &&
   \left.  \left.      
   +6 f_{b1} f_{b2} \left(3 f_{b2}^4+10 f_{b2}^3-4 f_{b2}^2-18 f_{b2}+9\right) 
   -f_{b2}^2 \left(6
   f_{b2}^3+20 f_{b2}^2+f_{b2}-18\right)\right)
   \right. \notag \\
   &&
    \left.  
   +6 f_{a1}^2 (f_{b1}-1) \left(24 f_{b1}^5 (f_{b2}-1)+2 f_{b1}^4
   \left(3 f_{b2}^3+6 f_{b2}^2-32 f_{b2}+23\right)
    \right.\right.  \notag \\
   &&
   \left.  \left.    
   +3 f_{b1}^3 \left(4 f_{b2}^3+8 f_{b2}^2-37 f_{b2}+25\right)+3
   f_{b1}^2 \left(6 f_{b2}^5+20 f_{b2}^4-29 f_{b2}^3-78 f_{b2}^2+108 f_{b2}-27\right)
    \right.\right.  \notag \\
   &&
   \left.  \left.    
   -12 f_{b1} f_{b2} \left(3
   f_{b2}^4+10 f_{b2}^3-4 f_{b2}^2-18 f_{b2}+9\right)+3 f_{b2}^2 \left(6 f_{b2}^3+20
   f_{b2}^2+f_{b2}-18\right)\right)
   \right.  \notag \\
   &&
    \left.    
   +9 f_{a1} (f_{b1}-1)^2 \left(12 f_{b1}^5 (f_{b2}-1)+28 f_{b1}^4 (f_{b2}-1)+6
   f_{b1}^3 (f_{b2}-1) \left(2 f_{b2}^2+6 f_{b2}-3\right)
    \right.\right.  \notag \\
   &&
   \left.  \left.   
   +6 f_{b1}^2 (f_{b2}-1) \left(2 f_{b2}^2+6
   f_{b2}-3\right)+12 f_{b1} (f_{b2}-1) f_{b2} \left(3 f_{b2}^3+13 f_{b2}^2+9 f_{b2}-9\right)
    \right.\right.  \notag \\
   &&
   \left.  \left.    
   -6 f_{b2}^2 \left(6
   f_{b2}^3+20 f_{b2}^2+f_{b2}-18\right)\right)+54 (f_{b1}-1)^3 f_{b2}^2 \left(6 f_{b2}^3+20
   f_{b2}^2+f_{b2}-18\right) \right],
      \end{eqnarray}
\begin{eqnarray}
   C_{aab} &=& 
   \frac{-1}{4 (f_{a1}-3
   f_{b1}+3)^3}
   \left [ 
   f_{a1} (f_{b2}-1) \left(54 f_{a1}^2 f_{b1}^5+9 f_{a1} \left(8 f_{a1}^2+2 f_{a1}-15\right)
   f_{b1}^4
    \right.\right.  \notag \\
   &&
   \left.  \left.   
   +(f_{a1}+3)^3 f_{b2} (f_{b2}+3) \left(2 f_{a1}^2-2 f_{a1} \left(3 f_{b2}^2+4 f_{b2}-5\right)-3\right)
    \right.\right.  \notag \\
   &&
   \left.  \left.    
   -3
   f_{b1}^3 \left(4 f_{a1}^4-2 f_{a1}^3 \left(3 f_{b2}^2+9 f_{b2}-35\right)-3 f_{a1}^2 \left(6 f_{b2}^2+18
   f_{b2}-43\right)-45 f_{a1}-27\right)
    \right.\right.  \notag \\
   &&
   \left.  \left.    
   -(f_{a1}+3) f_{b1}^2 \left(24 f_{a1}^4-2 f_{a1}^3 \left(6 f_{b2}^2+18
   f_{b2}-41\right)+3 f_{a1}^2 \left(2 f_{b2}^2+6 f_{b2}-55\right)
       \right.\right. \right. \notag \\
   &&
   \left.  \left. \left.
   +9 f_{a1} \left(9 f_{b2}^2+27 f_{b2}-25\right)+27
   \left(f_{b2}^2+3 f_{b2}+2\right)\right)+(f_{a1}+3)^2 f_{b1} \left(6 f_{a1}^4
       \right.\right. \right. \notag \\
   &&
   \left.  \left. \left.   
   -2 f_{a1}^3 \left(3 f_{b2}^2+9
   f_{b2}-19\right)+f_{a1}^2 \left(6 f_{b2}^4+26 f_{b2}^3-16 f_{b2}^2-120 f_{b2}+49\right)
   \right.\right. \right. \notag \\
   &&
   \left.  \left. \left.   
   +3 f_{a1} \left(6
   f_{b2}^4+26 f_{b2}^3+23 f_{b2}^2-3 f_{b2}-25\right)+9 \left(2 f_{b2}^2+6 f_{b2}+1\right)\right)\right)
   \right],         
   \end{eqnarray}
\begin{eqnarray}   
C_{abb} &=& 
   \frac{-1}{4 (f_{a1}-3 f_{b1}+3)^3}
   \left[ f_{a1} (f_{b2}-1) \left(18 \left(4 f_{a1}^2+f_{a1}-3\right) f_{b1}^5
    \right.\right.  \notag \\
   &&
   \left.  \left.   
   +(f_{a1}+3)^2 f_{b1} \left(f_{a1}^2
   \left(4 f_{b2}^2+12 f_{b2}-3\right)-f_{a1} \left(12 f_{b2}^4+52 f_{b2}^3+13 f_{b2}^2-105 f_{b2}+15\right)
   \right.\right. \right. \notag \\
   &&
   \left.  \left. \left.   
   -3 \left(12
   f_{b2}^4+52 f_{b2}^3+46 f_{b2}^2-6 f_{b2}-3\right)\right)-3 f_{b1}^4 \left(4 f_{a1}^3-2 f_{a1}^2 \left(3
   f_{b2}^2+9 f_{b2}-35\right)
   \right.\right. \right. \notag \\
   &&
   \left.  \left. \left.   
   -3 f_{a1} \left(6 f_{b2}^2+18 f_{b2}-35\right)-18\right)+f_{b1}^3 \left(-24 f_{a1}^4+2
   f_{a1}^3 \left(6 f_{b2}^2+18 f_{b2}-77\right)
   \right.\right. \right. \notag \\
   &&
   \left.  \left. \left.    
   +3 f_{a1}^2 \left(8 f_{b2}^2+24 f_{b2}-29\right)-9 f_{a1} \left(8
   f_{b2}^2+24 f_{b2}-45\right)-27 \left(4 f_{b2}^2+12 f_{b2}-5\right)\right)
    \right.\right.  \notag \\
   &&
   \left.  \left.    
   +(f_{a1}+3) f_{b1}^2 \left(6 f_{a1}^4-2
   f_{a1}^3 \left(3 f_{b2}^2+9 f_{b2}-28\right)+f_{a1}^2 \left(6 f_{b2}^4+26 f_{b2}^3-38 f_{b2}^2-186
   f_{b2}+163\right)
   \right.\right. \right. \notag \\
   &&
   \left.  \left. \left.   
   +3 f_{a1} \left(12 f_{b2}^4+52 f_{b2}^3+15 f_{b2}^2-99 f_{b2}-1\right)+9 \left(6 f_{b2}^4+26
   f_{b2}^3+29 f_{b2}^2+15 f_{b2}-8\right)\right)
    \right.\right.  \notag \\
   &&
   \left.  \left.    
   +54 f_{a1} f_{b1}^6+(f_{a1}+3)^3 f_{b2} \left(6 f_{b2}^3+26
   f_{b2}^2+21 f_{b2}-9\right)\right) \right]. 
   \end{eqnarray}

\section{Full expressions including one-loop corrections}
\label{app:1}

In subsection~\ref{subsec:constrained}, we have considered the
one-loop terms in the expression for the power spectrum and the
non-linear parameters, neglecting the third order terms in the $\delta
N$ formalism.  Here, we show the full expressions for the one-loop
correction terms, including higher order ones in the $\delta N$
formalism~\cite{Kawakami:2009iu}.

The expression for the power spectrum including the full one-loop
correction terms is given by
\begin{eqnarray}
P_\zeta ( k) = \left[ 
N_a N^a +
N_{ab}N^{ab}  {\cal P}_\delta \ln (kL)
+ N_a {N^{ab}}_b   {\cal P}_\delta \ln (k_{\rm max}L)
\right] P_\delta (k)\,,
\end{eqnarray}
where $k_{\rm max}$ is the cut-off scale of the power spectrum and the
last term in principle can be removed by shifting the homogeneous
value of the scalar field \cite{Byrnes:2007tm}.

The non-linearity parameters including the one-loop corrections 
are given as follows. For $f_{\rm NL}$, we obtain
\begin{eqnarray}
\frac{6}{5}f_{\rm NL} &=& 
\left[ 
N_a N^a +
N_{ab}N^{ab}  {\cal P}_\delta \ln (k_i L)
+ N_a {N^{ab}}_b   {\cal P}_\delta \ln (k_{\rm max}L)
\right]^{-2} \nonumber\\
&& \times
\biggl[
N_aN_bN^{ab} 
+ \Biggl( N_{ab} N^{bc} {N_c}^{a} 
+ 2N_a N_{bc}N^{abc}
\Biggr) {\cal P}_\delta \ln (k_{m1}L) 
\nonumber\\
&& \qquad\qquad\qquad
+\biggl(
N_a N^{ab}{N_{bc}}^c
+ {1 \over 2} N_a N_b {N^{abc}}_c   
\biggr) {\cal P}_\delta \ln (k_{\rm max}L)
\biggr],
\end{eqnarray}
where $k_{m1} = \min \{ k_i \}$. Please notice that a
term with $k_{\rm max}$ again can be removed by shifting the homogeneous
value of the scalar field.  Here we have assumed that $\ln (k_1 L) \sim
\ln (k_2 L) \sim \ln (k_3 L)$ and all of these are just represented as
$\ln (k_i L)$, which enables us to factor out the dependence on such a
factor in the power spectrum.  Other two non-linearity parameters are:
\begin{eqnarray}
\tau_{\rm NL} &=& 
\left[ 
N_a N^a +
N_{ab}N^{ab}  {\cal P}_\delta \ln (k_iL)
+ N_a {N^{ab}}_b   {\cal P}_\delta \ln (k_{\rm max}L)
\right]^{-3} \nonumber\\
&& \times
\biggl[
N_aN^{ab}N_{bc}N^c  
+ 2 N_a N^{ab}N^{cd} N_{bcd} {\cal P}_\delta \ln (k_{m1}L)
 \nonumber\\
&&\qquad
+\left( 
N_{ab}N^{bc}N_{cd}N^{da}
+ N_a N_{bc} N^{b}_{~d} N^{acd}
+ N_a N_b N^{acd} N^{b}_{~cd}
\right) {\cal P}_\delta \ln (k_{m2}L)
 \nonumber\\
&&\qquad
\biggl(
+N_a N^{ab} N_{bc} {N^{cd}}_{d}
+
N_a N^{ab} {N_{bc}}^{cd}N_d
\biggr)
{\cal P}_\delta \ln (k_{\rm max}L)
\biggr],
\end{eqnarray}
\begin{eqnarray}
\frac{54}{25}g_{\rm NL} &=&
\left[ 
N_a N^a +
N_{ab}N^{ab}  {\cal P}_\delta \ln (k_iL)
+ N_a {N^{ab}}_b   {\cal P}_\delta \ln (k_{\rm max}L)
\right]^{-3} \nonumber\\
&& \times
\biggl[
N_a N_b N_c N^{abc}  +
3 \left( 
N_a N_{bc} N^{b}_{~d} N^{acd}  
+N_a N_b N_{cd} N^{abcd} 
\right)
{\cal P}_\delta \ln (k_{m1}L)
 \nonumber\\
&&
\qquad
+
\left( 
{1 \over 2} N_a N_b N_c {N^{abcd}}_{d} 
+
{3 \over 2} N_a N_b N^{abc} {N_{cd}}^{d} 
\right){\cal P}_\delta \ln (k_{\rm max}L)
\biggl],
\end{eqnarray}
where $k_{m2} = \min \{ |k_i+k_j|, k_l \}$.

\section{Full order calculation} \label{app:2}

In this appendix, we briefly discuss the higher order perturbation
effects on the bispectrum and the trispectrum.  So far, our analysis has
been based on the perturbative expansion of the $e$-folding number in
terms of the field fluctuations.  In order to calculate the power spectrum,
the bispectrum, and the trispectrum, it was sufficient to consider the
lowest order contribution to those quantities (or the next lowest order
if the lowest one vanishes like in the ungaussiton case).  This
perturbative expansion is extremely accurate and powerful in many models
including all the models discussed  in this paper since the
higher order contributions are highly suppressed.

Yet, it is still logically possible to consider a model in which the
$e$-folding number is quite sensitive to the field value.  In such a
case, the truncation at the leading order or at the next leading order
will no longer give a correct answer.  We then need to go to the
higher order calculations until the series converge or to resort to
the full order calculations.

To see what happens if the higher order terms are taken into account,
we give a simple toy model that allows full order calculations at some
level.  Here, by at some level, it means that the final expressions of
the bispectrum and the trispectrum involve multidimensional integrals
that defeats analytic calculation.  Nevertheless, we can derive some
interesting consequences from these results, as we will see below.

Let us assume that the $e$-folding number depends on a field $\sigma$ by
\begin{equation}
N({\vec x})=\exp \left( -\frac{\sigma({\vec x})}{a} \right) \bigg/ \bigg\langle \exp \left( -\frac{\sigma({\vec x})}{a} \right) \bigg\rangle\,,
\end{equation}
where $\langle \cdots \rangle$ denotes the spatial average. The
magnitude of $\sigma$ field perturbation is not necessarily smaller
than $a$.  This type of model may be realized in the context of
modulated reheating scenario if the coupling constant for the inflaton
decay is multiplied by a factor like $\exp \left(-\frac{\sigma}{2a}
\right)$.  Then, using the $\delta N$ formula, the curvature
perturbation is given by
\begin{equation}
\zeta ({\vec x}) = \exp \left( -\frac{\delta \sigma ({\vec x})}{a} \right) \bigg/ \bigg\langle \exp \left( -\frac{\delta \sigma ({\vec x})}{a} \right) \bigg\rangle -1\,.
\end{equation}
The full order calculation was also done in \cite{Yamamoto:1992zb},
where the authors calculate the power spectrum of the field which
depends on the Gaussian field by the sine function rather than the
exponential one.

By assuming that $\delta \sigma$ is Gaussian, we write the two-point
function in Fourier space as
\begin{equation}
\langle \delta \sigma_{\vec k_1} \delta \sigma_{\vec k_2} \rangle = {(2\pi)}^3 P(k_1) \delta ({\vec k_1}+{\vec k_2})\,.
\end{equation}
We further assume the scale invariant power spectrum,
i.e. $P(k)=H^2_*/(2k^3)$.  Then, using the standard formula,
\begin{align}
&\bigg\langle \exp \left( -\int d^3x \ b({\vec x}) \delta \sigma ({\vec x}) \right) \bigg\rangle \nonumber \\
&= \exp \bigg[ \frac{1}{2} \int d^3 x_1 d^3 x_2 \int \frac{d^3q}{{(2\pi)}^3} P(q) e^{i {\vec q} \cdot ({\vec x_1}-{\vec x_2})} b({\vec x_1}) b({\vec x_2}) \bigg],
\end{align}
we find that the connected part of the three-point function of $\zeta$
is given by
\begin{align}
\langle \zeta_{\vec k_1} \zeta_{\vec k_2} \zeta_{\vec k_3} \rangle=\int &d^3x_1 d^3x_2 d^3x_3 \ e^{i {\vec k_1} \cdot {\vec x_1}+i {\vec k_2} \cdot {\vec x_2}+i {\vec k_3} \cdot {\vec x_3}} \nonumber \\
& \times \exp \bigg[ \frac{1}{a^2} \int \frac{d^3q}{{(2\pi)}^3} P(q) ( e^{i{\vec q} \cdot {\vec x_{12}} } +e^{i{\vec q} \cdot {\vec x_{23}} }+e^{i{\vec q} \cdot {\vec x_{31}} } )\bigg],
\end{align}
where ${\vec x_{ij}} \equiv {\vec x_i}-{\vec x_j}$.  This is
non-vanishing only when ${\vec k_1}+{\vec k_2}+{\vec k_3}=0$.  Then
after a suitable change of integration variables, we can show that the
bispectrum of $\zeta$ is given by
\begin{align}
B_\zeta({\vec k_1},{\vec k_2},{\vec k_3})= \int & d^3x_1 d^3x_2 \ e^{i{\vec k_1} \cdot {\vec x_1}+i{\vec k_2} \cdot {\vec x_2}} \nonumber \\
&\times \exp \bigg[ \frac{1}{a^2} \int \frac{d^3q}{{(2\pi)}^3} P(q) ( e^{i{\vec q} \cdot {\vec x_1} } +e^{i{\vec q} \cdot {\vec x_2} }+e^{i{\vec q} \cdot {\vec x_{12}} } )\bigg].
\end{align}
The momentum integration can be done analytically
\cite{Yamamoto:1992zb},
\begin{equation}
\exp \bigg[ \frac{1}{a^2} \int \frac{d^3q}{{(2\pi)}^3} P(q) e^{i{\vec q} \cdot {\vec x} } \bigg] = {\left( \frac{L}{x} \right)}^{2\beta},
\end{equation}
where $L$ is the size of the box and $\beta \equiv H^2_*/(8\pi^2
a^2)$.  Then the bispectrum becomes
\begin{align}
B_\zeta({\vec k_1},{\vec k_2},{\vec k_3})=
 \int & d^3x_1 d^3x_2 \ e^{i{\vec k_1} \cdot {\vec x_1}+i{\vec k_2} \cdot {\vec x_2}} {\left( \frac{L}{x_1} \right)}^{2\beta} {\left( \frac{L}{x_2} \right)}^{2\beta} {\left( \frac{L}{x_{12}} \right)}^{2\beta}. \label{full1}
\end{align}
It is not easy to do the integration over the real space coordinates.
However, we can extract the scale dependence of the bispectrum.
Indeed, under the scaling transformation ${\vec k_i} \to \lambda {\vec
  k_i}$ with $\lambda$ being constant, the bispectrum scales as
\begin{equation}
B_\zeta(\lambda {\vec k_1},\lambda {\vec k_2},\lambda {\vec k_3})=\lambda^{-6+6\beta} B_\zeta({\vec k_1},{\vec k_2},{\vec k_3})\,.
\end{equation}
Or, equivalently, the bispectrum runs with the scale as
\begin{equation}
\frac{d\log B_\zeta}{d \log k}=-6+6\beta\,. \label{run1}
\end{equation}
The derivative should be understood that the shape of the triangle is
kept unchanged.  Therefore, the larger $\beta$ is, the larger the
bispectrum is on small scales.  On the other hand, it is difficult
to extract the shape dependence from Eq.\ (\ref{full1}) and we stop
pursuing it further.

We can do the same for the four-point function.  The trispectrum is
given by
\begin{align}
T_\zeta({\vec k_1},{\vec k_2},{\vec k_3},{\vec k_4})=& 
 \int d^3x_1 d^3x_2 d^3x_3 \ e^{i{\vec k_1} \cdot {\vec x_1}+i{\vec k_2} \cdot {\vec x_2}+i{\vec k_3} \cdot {\vec x_3}} \nonumber \\
&\times {\left( \frac{L}{x_1} \right)}^{2\beta} {\left( \frac{L}{x_2} \right)}^{2\beta} {\left( \frac{L}{x_3} \right)}^{2\beta} {\left( \frac{L}{x_{12}} \right)}^{2\beta} {\left( \frac{L}{x_{23}} \right)}^{2\beta} {\left( \frac{L}{x_{31}} \right)}^{2\beta}.
\end{align}
Therefore, the trispectrum scales as
\begin{equation}
T_\zeta(\lambda {\vec k_1},\lambda {\vec k_2},\lambda {\vec k_3},\lambda {\vec k_4})=\lambda^{-9+12\beta} T_\zeta({\vec k_1},{\vec k_2},{\vec k_3},{\vec k_4})\,,
\end{equation}
and therefore, the trispectrum runs as
\begin{equation}
\frac{d\log T_\zeta}{d \log k}=-9+12\beta\,. \label{run2}
\end{equation}

Due to the same reason as the case of the ungaussiton, we assume that
the leading contribution to the power spectrum comes from inflaton
fluctuations. Therefore,
\begin{equation}
P_\zeta (k)=\frac{1}{4 \epsilon M_{\rm Pl}^2 k^3}\,,
\end{equation}
where $\epsilon$ is again the slow-roll parameter.  From (\ref{run1})
and (\ref{run2}), we find that a ratio
\begin{equation}
{\left( \frac{B_\zeta}{P_\zeta^2} \right)}^2 \bigg/ \left( \frac{T_\zeta}{P_\zeta^3} \right)=\frac{B_\zeta^2}{P_\zeta T_\zeta}\,,
\end{equation}
does not run with the scale.  But it will in general depend on the
shape of the triangle and the quadrilateral.

\end{document}